\documentclass[11pt, epsf]{article}
\usepackage{jheppub}
\usepackage{xcolor}
\usepackage{braket}
\usepackage{graphicx}
\usepackage{soul}
\usepackage{amssymb,amsmath}
\usepackage{bbm}
\usepackage{simpler-wick}

\newcommand{\pd}{\partial}
\newcommand{\nn}{\nonumber\\}

\DeclareMathOperator{\Tr}{Tr}
\usepackage{verbatim}
\usepackage{anyfontsize}
\usepackage{dsfont}
\usepackage{tikz}
\usepackage{caption}
\usepackage{subcaption}
\usepackage{sidecap}
\sidecaptionvpos{figure}{t}
\usepackage[english]{babel}
\usepackage{enumitem}  
\usepackage[export]{adjustbox}
\usepackage{url}

\newcommand{\ba}{\begin{align}}
\newcommand{\ea}{\end{align}}
\title{3d Gravity as a random ensemble}
\author[a]{Daniel L. Jafferis,}
\author[a]{Liza Rozenberg} 
\author[a,b]{and Gabriel Wong}
\affiliation[a]{Jefferson Physical Laboratory, Harvard University, Cambridge, MA 02138, USA}
\affiliation[b]{Mathematical Institute, University of Oxford,
Andrew Wiles Building, Radcliffe Observatory Quarter,
Woodstock Road, Oxford, OX2 6GG, U.K.}
\abstract{ We give further evidence that the matrix-tensor model studied in \cite{belin2023} is dual to AdS$_{3}$ gravity including the  sum over topologies. This provides a 3D version of the duality between JT gravity and an ensemble of random Hamiltonians,  in which the matrix and tensor provide random CFT$_2$ data subject to a potential that incorporates the bootstrap constraints.  We show how the Feynman rules of the ensemble produce a sum over all 3-manifolds and how surgery is implemented by the matrix integral.   The partition functions of the resulting 3d gravity theory agree with Virasoro TQFT (VTQFT) on a fixed, hyperbolic manifold.  However, on non-hyperbolic geometries, our 3d gravity theory differs from VTQFT, leading to a difference in the eigenvalue statistics of the associated ensemble.   As explained in \cite{belin2023}, the Schwinger-Dyson (SD) equations of the matrix-tensor integral play a crucial role in understanding how gravity emerges in the limit that the ensemble localizes to exact CFT's.  We show how the SD equations can be translated into a combinatorial problem about 3-manifolds. 

}
\emailAdd{jafferis@g.harvard.edu}
\emailAdd{erozenberg@g.harvard.edu}
\emailAdd{gabrielwon@gmail.com}

\begin{document}

\maketitle 
\section{Introduction}
\subsection{Motivations}
%random models describe chaotic dynamics
The idea that random Hamiltonians describe chaotic systems goes back to Wigner \cite{wigner55,wigner58}. In particular, he showed that ensembles of Hermitian matrices exhibit universal behavior associated to level repulsion.
This idea was extended to the study of other low-energy observables, where the eigenstate thermalization hypothesis (ETH) describes the matrix elements of an observable in the energy eigenbasis of an isolated quantum mechanical system  \cite{Deutsch91, Srednicki_1994}.  
More recently, the random statistics of OPE coefficients in chaotic CFT's were studied  in \cite{Chandra_2022, Collier_2020, jafferis2023jt, Cardy:2017qhl}.

The general paradigm is that in large entropy chaotic systems, the pseudo-random statistics of the micro-data, in an ensemble obtained by sampling over different precise energies, is the same as the maximum ignorance ensemble over theory data constrained by low-energy correlators. In this paper, following \cite{belin2023}, we also incorporate the constraints of microscopic consistency, such as locality, in an ensemble describing chaotic 2D CFT's. The implementation of these constraints introduces  non-gaussianities in the pseudo-random statistics: this is necessary to provide a complete description of the system, since independent gaussian random observables have exponentially suppressed out-of-time-order correlators \cite{PhysRevE.99.042139, Murthy:2019fgs, Jafferis:2022uhu}.

%gravity and random models
For holographic theories, expansions of the random models are related to the bulk.   For the case of random Hamiltonians, a particular double-scaled matrix integral was shown to be dual to Jackiw-Teitelboim (JT) gravity in a genus expansion over 2d spacetime topology \cite{saad2019jt}. This was extended in \cite{jafferis2023jt} to show that JT gravity with matter is dual to a non-gaussian ETH ensemble of matrices. In 
\cite{belin2023} it was proposed that pure AdS$_3$ gravity is dual to a  
tensor and matrix model that describes chaotic CFT$_2$'s.  In this work, we give a refinement of this model and give further evidence that its perturbation theory exactly matches the topological expansion of 3d gravity. 

%nod to the literature 
Our work touches upon many ideas that have been developed in the quest for a gravity interpretation of random, two dimensional CFT's \cite{DiUbaldo:2023hkc, DiUbaldo:2023qli,deBoer:2024kat,Anous:2021caj,Belin:2021ryy,collier2023,collier2023virasoro,Collier:2024mgv,Cotler_2021}.  The program was initiated in \cite{Cotler_2021}, where it was proposed that the low energy limit of the gravity path integral on torus wormholes describes the statistics of a double-scaled random matrix ensemble with Virasoro symmetry. Later, \cite{DiUbaldo:2023qli,Haehl:2023mhf,Haehl:2023tkr,Haehl:2023xys} showed that such ``off-shell" wormholes provide the minimal completion of the random matrix theory (RMT) correlators compatible with Virasoro symmetry and $SL(2,\mathbb{Z})$ invariance.  A check of classical 3d gravity calculations with a Gaussian ensemble of OPE's was performed in \cite{Belin:2020hea,Chandra_2022}.
In \cite{Belin:2021ryy, Anous:2021caj}, multi-boundary wormhole configurations were linked to non-Gaussianities in statistical distributions of heavy OPE coefficients of 2d CFT's.  In \cite{deBoer:2023vsm} the principle of maximum ignorance was applied to define a state averaging that agrees with the ensemble averaging over CFT's and the corresponding gravity calculations.
Finally, \cite{collier2023, Collier:2024mgv} applied the machinery of topological field theory to compute 3d gravity partition functions on hyperbolic geometries: We will make use of their results frequently to compare with the predictions of our model.

% 3D is different than 2D 
There is an important distinction between two dimensional JT gravity and AdS$_3$ gravity in their relation to random models.   In JT, a factorization puzzle appears because the bulk theory is UV complete. Then the bulk path integral together with the sum over topology provide a seemingly exact theory of gravity with a disordered dual, rather than merely an approximate description of certain averaged quantities that exhibited pseudo-random features in a fixed quantum system.

On the other hand, the boundary dual to AdS$_3$ must satisfy exact constraints of locality. These are incorporated as limits in the space of ensembles, where the potential becomes infinitely steep in certain directions\footnote{ Such phenomena also appear in the quantum mechanics dual to local bulk theories \cite{Jafferis:2022uhu, jafferis2023jt}, although in that case the constraints can never be exactly realized with a discrete spectrum.}.
In the limit where the constraints are exactly satisfied, we expect that the ensemble rigidifies into a particular quantum system, thereby circumventing the factorization puzzle.

A concrete realization of this idea was described in \cite{belin2023}. 
 Building on the works of \cite{Anous:2021caj,Belin:2021ryy}, they described how to use constraints of micro-locality via the crossing equations to build an ensemble of CFT data, which includes 3-point function structure constants and dimensions of primary operators graded by spin.  The resulting ensemble is highly non-Gaussian. In order to study the ensemble using the expansion of matrix and tensor integrals, the constraints need to be relaxed slightly to allow small deviations, which is a natural notion in holography since low-energy probes cannot distinguish such violations of crossing among black hole microstates. The expectation is that the limit of these approximate ensembles imposes the strict constraints and so restricts us to exact 2d CFTs, which are not expected to have the $e^{c}$ parameters of the ensemble. 
 
In this work, the focus is on a complementary perspective. The main lesson will be that the topological expansion of pure 3d gravity is exactly given by the expanding the dual CFT microdata around the Cardy density of states with a potential that imposes the constraints of locality. In that sense, pure gravity provides a solution to the bootstrap to all orders in $e^{-c}$, and its ultimate consistency is determined by the fine grained constraints of mutual locality of black hole microstate operators. 

\subsection{Summary of our work}
We elaborate on the tensor-matrix model of  \cite{belin2023}, which describes an ensemble of \emph{approximate} $\text{CFT}_2$.   The model is defined by an integral over the data of two dimensional conformal field theories, given by a matrix corresponding to the dilatation operator graded by integer spin, $\Delta_s$, and the tensor of the structure constants, $C_{ijk}$, acting on the Hilbert space of (non-identity) Virasoro primaries. Locality and conformal symmetry, and $CRT$ invariance, respectively determine the index symmetry and reality properties of the tensor. 

The tensor model potential depends on the central charge, $c$, and is defined for finite $c$. We will assume that the associated CFT is fully irrational, in the sense that no degenerate representations appear aside from the identity Virasoro module. The contributions of the identity operator are included explicitly. Conformal field theories with currents can be explored in an analogous way, but the associated tensor model potential will be different. The expansion of the ensemble integral that we will derive is an expansion around the Cardy density\footnote{We use this nomenclature to refer to exact spectral density obtained by $S$ transform of the identity block, not just its large weight asymptotics. This is equivalent to the BTZ spectral density.}, which has a gap to the black hole threshold of $\Delta_s = \frac{c-1}{12} + |s|$. However we don't strictly demand that no operators are present below the threshold.  Denoting by $V_{0}(\Delta_{s})$ the matrix model potential defined by the Cardy density, the partition function of the ensemble is given by
\begin{align}\label{Z0}
    \mathcal{Z}= \prod_{s \in \mathbb{Z}} \int D \Delta_{s} D   C_{ijk} e^{-V_{0}[\Delta_{s}]-\frac1 \hbar V[\Delta_{s},C_{ijk}] },
\end{align}
where $V[\Delta_{s},C_{ijk}]$ is a judiciously chosen ``constraint squared" potential that vanishes on the solutions of the modular bootstrap. The full expression for the potential is given below in \eqref{fullV}.
We will argue that the 't Hooft genus expansion of the matrix integral together with the Feynman diagram expansion of the tensors can be organized in a 3d topological expansion with parameter $e^{-c}$ that exactly matches the sum over topologies in pure 3d gravity. 
In the same way that the sum over $SL(2,\mathbb{Z})$ images of the BTZ saddle in 3d gravity imposes modular invariance of the dual CFT, we will see that the sum over hyperbolic 3-manifolds is the 4-point crossing symmetrizer. Moreover, the sum over all manifolds ensures that the result is modular invariant on any higher genus state cut, as required for consistency with the conformal block decomposition. 

The limit $\hbar \rightarrow 0$ produces a delta function of the bootstrap constraints, which are generated by four point crossing, and modular invariance of the torus partition function and one point functions. At nonzero $\hbar$ the model can be thought of as an ensemble of approximate CFT's\footnote{The axiom which they violate is locality, associated to the euclidean analyticity properties which imply the crossing equation. They are exactly conformally invariant by construction.}, in the sense that the crossing equations will only be obeyed approximately. We will be mostly interested in the limit $\hbar \to 0$, and the role of the parameter $\hbar$ is as a regulator that enables the perturbative expansion. 

The integral of a delta function of constraints can be understood as providing a precise definition of the maximum ignorance ensemble of CFT$_2$ data consistent with all exact bootstrap conditions \cite{deBoer:2023vsm}. As we explain in section \ref{ensembleDef}, the natural measure to define the delta function is provided by the Verlinde inner product on the space of conformal blocks\footnote{We will see that an important modification is required for the torus character.}. 
\paragraph{A diagrammatics for 3-manifolds }
The tensor integral can be expanded in triple line Feynman diagrams, which is an expansion in $\hbar$ around $C_{ijk}=0$. Note that the kinetic term has the wrong sign, since $C_{ijk}=0$ is not a solution of the bootstrap when the identity operator is present, and we are thus expanding around a local maximum of the potential. Usually in such circumstances, one finds the minimum of the potential and expands around that, but in this context that is tantamount to solving the bootstrap equations. Instead we reorganize the perturbation theory into an $e^{-c}$ expansion, and non-perturbatively resum the $\hbar$ expansion via Schwinger-Dyson equations, order by order in $e^{-c}$. 

The 't Hooft genus expansion of the matrix integral is an $e^{-c}$ expansion, since the Cardy density has infinite range, corresponding to a double scaled matrix model, and the density of eigenvalues at the edge of the cut scales as $e^c$. The parameter $c$ appears in the potential for the tensors, since it enters the Virasoro blocks that appear in the crossing equation. Each triple line diagram evaluates to a function of the weights of the operators that label every closed index line, and thus produces a multi-trace observable in the $\Delta_s$ matrix model. The index lines are then filled in with matrix model plaquets in the 't Hooft expansion. 

We will show that every such triple line Feynman diagram with index lines filled by a matrix 't Hooft diagram can be assigned to a 3d topology, as anticipated by \cite{belin2023}. This is done via a gluing and surgery construction. The tensor model vertices are assigned to simple smooth manifolds, detailed in section \ref{sec:gravity},  dressed with framed Wilson lines connected in the contraction pattern of the indexes and boundaries given by thrice punctured spheres, associated to a $C_{ijk}$. These boundaries are glued together by the tensor propagators. In section \ref{sec:checks}, we show how these gluing rules result in a type of connect sum of 3-manifolds that reproduces the partition functions of Virasoro TQFT \cite{collier2023,collier2023virasoro}. 

Finally, tubular neighborhoods of the Wilson lines are excised, and a manifold resulting from the matrix integral, with toroidal boundaries, is glued in. The simplest possibility is to glue in the matrix model disk on each Wilson line, associated to the leading spectral density given by the Cardy density; this performs an $S$ toroidal surgery on each Wilson line, corresponding to gluing in the euclidean BTZ topology. 

\paragraph{The matrix model and the random statistics of 3d gravity} 
The basic building blocks of the matrix expansion are the disk and annulus diagrams, corresponding to the BTZ topology and the torus wormhole $T^2 \times I$.  The disk is designed to match the BTZ spectrum by choice of the potential.  The annulus diagram, which is the standard random matrix theory two point function of the spectral density,
equals the $T^2 \times I$ partition function computed in \cite{Cotler_2021}, up to a factor of 2 that we will explain in section \ref{sec:gravityMM}. The annulus determines the random statistics of 3d gravity, since it is dictated by the Vandermonde potential which captures the eigenvalue statistics of ensemble.
%The annulus diagram determines the random statistics of 3d gravity: it is the standard random matrix theory two point function of the spectral density, which is determined by the Vandermonde potential that captures the eigenvalue statistics of the random matrices.   
%The annulus diagram, which is the standard random matrix theory two point function of the spectral density,
%equals the $T^2 \times I$ partition function computed in [], up to a factor of 2 that we will explain in section \ref{sec:gravityMM}.   

An important aspect of the ``gravitational" statistics obeyed by the matrix model arises from the fact that the $T^2 \times I$ wormhole has a non-trivial bulk mapping class group, $\mathbb{Z} \times \mathbb{Z}$, associated to large diffeomorphisms that are trivial at the boundaries, but shift one boundary relative to the other by a lattice translation (regarding $T^2$ as $R^2/(\mathbb{Z}\times \mathbb{Z})$)\footnote{This is distinct from the boundary mapping class group, which in this case is the $PSL(2,\mathbb{Z})$ modular group of the torus. In manifolds of the form $M \times I$, part of the bulk mapping class group is given by $\pi_1({\mathit{Diff}.}(M))$, which here is $\mathbb{Z} \times \mathbb{Z}$. The boundary mapping class group is $\pi_0({\mathit{Diff}}(M))$.}. This leads to an important distinction between 3d gravity and Virasoro TQFT, since the latter is defined without imposing the mapping class group identifications on the configuration space.  Indeed 
the VTQFT annulus produces an ensemble with no level repulsion, which does not belong to the standard symmetry classes of random matrix theory. 
Note that this issue really pertains to the non-hyperbolic, off shell manifolds: hyperbolic 3-manifolds have trivial or finite order bulk mapping class groups, which are generated by isometries of the hyperbolic metric, and thus lead only to overall symmetry factors.  

The upshot is that in 3d gravity, the gauging of the bulk mapping class group is entirely responsible for the level repulsion that is characteristic of RMT statistics.  Note that a similar distinction exists between 2d JT gravity and $SL(2,\mathbb{R})$ gauge theory \cite{Blommaert:2018iqz,Mertens:2020hbs,Fan:2021bwt}.  Here the analogue of the torus wormhole is the double trumpet, which has a $\mathbb{Z}$ mapping class group.  It is well known that gauging by $\mathbb{Z}$ changes the gluing measure used to produce the double trumpet from the gluing of two single trumpets along a bulk geodesic circle \cite{Mertens:2020hbs}.  This difference in gluing measures is the origin of the different statistics associated to gauge theory and gravity. 
  
Going beyond the disk and annulus, our matrix model also includes a double trace potential, associated to the square of $S$ modular invariance. We show that its effect, combined with the underlying definition of the model with integer spin, is to include a sum over $SL(2,\mathbb{Z})$ Dehn twists on every cycle of $\Sigma_{g,n} \times S^1$ where $\Sigma$ is a genus $g$ 't Hooft diagram with $n$ punctures.  Thus the full one and two point functions of the spectral density at genus 0 precise match the $SL(2,\mathbb{Z})$ sum of euclidean BTZ and the torus wormhole answer \cite{Cotler_2021, yan2023torus}, respectively. We leave for future work the determination of the off-shell 3d gravity calculation at higher genus and more boundaries, which we expect to match the associated matrix model results. 
\paragraph{The Schwinger Dyson equation and sum over 3-manifolds}

A special feature of the tensor model potential is a kind of integrability. It is associated to the identities obeyed by the 6J symbols and modular crossing kernels, such as the hexagon and pentagon identities, shown by Moore-Seiberg to lead to consistent rules of 3d TQFT. Fixing one index line reduces some of these tensor identities to the familiar Yang-Baxter equation of matrices, so they are an uplift of that notion of integrability.  A closely related integrability appeared in the matrix model for JT gravity with matter \cite{jafferis2023jt}.

A consequence of the integrability of the tensor model is that every diagram in the same 3-topology class evaluates to exactly the same function of $c$, up to overall factors and powers of $\hbar$. Moreover, we argue that this function is precisely the pure 3d gravity partition function on that manifold. This can be established for hyperbolic manifolds resulting from tensor model diagrams, by directly relating them to the gluing rules of Virasoro TQFT \cite{collier2023}. 

As a consequence, expectation values in the tensor/matrix ensemble takes the form:
\begin{equation} \label{ensembleexp}
\langle Z_{CFT}(\Sigma_{1}(q_{1})) \cdots Z_{CFT}(\Sigma_{n} (q_{n}))\rangle = \sum_{\substack{\textrm{connected} M_3, \\ \partial M_3 = \Sigma_{1}\cup\cdots \cup \Sigma_{n
}}} Z_\textrm{3d gravity}(M_3 ; c, \vec{q}) f(M_3 ; \hbar),
\end{equation}
where $Z_{CFT}(\Sigma(q))$  is CFT$_2$ partition function on a Riemann surface $\Sigma$ written in a specific conformal block channel, with $q$ labelling the moduli.  This should be viewed as a boundary  observable that inserts the boundary $\Sigma$ into the bulk gravitational theory. 

On manifolds for which $Z_{3\text{d gravity}}$ is well defined, which includes hyperbolic manifolds and ``matrix model manifolds",  we conjecture that $f \rightarrow 1$ in the limit $\hbar \rightarrow 0$.
It is in this limit, in which exact local CFT's are produced, that the partition function of 3d gravity with given asymptotically AdS boundary is recovered.
Building on the observations in \cite{belin2023}, we show how the conjectural limit cited above can be checked using the Schwinger Dyson equations of the matrix-tensor model.   We explain in more detail  how the SD equations reduce to a combinatorial question about 3-manifolds in the large $c$ limit.  
Finally, we show that all manifolds $M_{3}$ with a given boundary is produced on the RHS of \eqref{ensembleexp}.  
 \subsection{Outline}
 
Here is a brief outline of our paper.  In section \ref{ensembleDef} we define the matrix-tensor model and explain the associated triple line Feynman rules. In particular, we specify the symmetry class for the matrix $\Delta_{s}$ and construct the constraint squared potential that approximately implements the constraints of the modular bootstrap. 

In section \ref{sec:gravity}, we give a gravity interpretation of the ensemble, including an extended discussion of surgery and how it is implemented in our model.    In section \ref{sec:checks}, we provide evidence for the proposal.  In particular, we explain in detail how the Feynman diagrams of the tensor model map to the partition functions of VTQFT, and show that all manifolds are produced by the ensemble.   In section \ref{SchwingerDyson}, we translate the Schwinger-Dyson equations into a combinatorial problem whose solution is tantamount to proving that 3d gravity partitions are exactly produced by our model. 

\section{Definition of the ensemble} \label{ensembleDef}

We define an ensemble of approximate CFT data, given by a set of random matrices $\Delta_{s}$ and a tensor $C_{ijk}$.  These corresponds to the Dilation operator graded by spin s,  and the OPE coefficients.   We consider a model with only Virasoro symmetry,  possessing an infinite number of primaries in non-degenerate representations in each sector of fixed spin\footnote{  The results of \cite{Mukhametzhanov:2019pzy,Mukhametzhanov:2020swe} imply that an exact CFT spectrum has an infinite number of primaries for each spin $s$. This follows from the precise bounds they derived on the deviation of the (appropriately smeared) spectrum from the  Cardy density. }.  The  partition function for this ensemble can be defined as a finite matrix-tensor integral by truncating the number of primaries to $N<\infty$.  
Assuming the spins take all integer value, the ensemble partition function is given by
\begin{align}\label{Z}
    \mathcal{Z}= \prod_{s \in \mathbb{Z}} \int D \Delta_{s} D   C_{ijk} e^{-V_{0}(\Delta_{s})-\frac1 \hbar V_{\varepsilon}(\Delta_{s},C_{ijk}) }.
\end{align}
Here, $V_{0}(\Delta_{s})$ is the single-trace potential that produces the Cardy density of states for a fixed spin $s$, while  $V_{\varepsilon}(\Delta,C_{ijk}) $ is a ``constraint squared" potential that is minimized on the solutions to the bootstrap constraints, with $\varepsilon$ a suitably chosen regulator.  $\hbar$ is a small parameter that controls the deviation from these constraints, with $\hbar \to 0$ corresponding to the exact implementation of the bootstrap.  We will consider a diagrammatic expansion of this integral in the limit $N\to \infty$, and $\hbar \to 0$.  A priori, the central charge $c$ is a fixed parameter entering the potential.  However, in order to make a connection  to 3d  gravity, we will eventually re-organize the perturbation theory into an asymptotic expansion  in $e^{-c}$  about the Cardy density, and then take the $\varepsilon, \hbar \to 0 $ limit term by term in $e^{-c}$.

\subsection{The GOE ensemble and symmetries of $C_{ijk}$'s }
 In addition to rotational symmetry, the Dilatation operator must also commute with CRT symmetry by the general axioms of relativistic QFT\footnote{R refers to the reflection of a single spatial coordinate.
 }.   For the bosonic theories with integer spins, this anti-unitary symmetry squares to 1, implying that the random matrices $\Delta_{s}$ belong to the GOE ensemble  \cite{Jensen:wip2024, yan2023torus}.  In this ensemble, we can simultaneously diagonalize CRT and the Dilatation operator.   However, due to the anti-unitary nature of CRT, this diagonalization is not preserved by a general unitary change of basis.   Instead, one has to restrict to orthogonal changes of basis.  The matrices $\Delta_{s}$ can then be taken to be real and symmetric. 
 
 The choice of a GOE ensemble for $\Delta_{s}$ is particularly relevant for the reality condition\footnote{We review the derivation of this formula in appendix \ref{app:real}.} that we impose on the OPE coefficients:
\begin{align}
     C^{*}_{ijk} 
 &= \exp \left(i\pi (s_{i} +s_{j} +s_{k} )\right) C_{ijk} 
\end{align}
It implies that $C_{ijk}$ is real when the sum of the spins is even and purely imaginary when the sum of the spins is odd.  Note that these condition are not preserved by a general unitary change of basis. Instead, they are  preserved by orthogonal change of basis that commutes with CRT, consistent with the GOE ensemble.

To fully specify the model, we must determine the symmetry properties of the tensor $C_{ijk}$. Recall that these OPE coefficients are defined by the 3 point function
\begin{align}\label{3pt1}
\braket{O_{i}(z_{1}) O_{j}(z_{2}) O_{k}(z_{3})} &=\frac{C_{ijk}}{z_{12}^{h_{i}+h_{j}-h_{k}} z_{23}^{h_{j}+h_{k}-h_{i}} z_{31}^{h_{i}+h_{k}-h_{j}} \times (\text{anti-holomorphic}) } \nn
z_{ij} &=z_{i} -z_{j} 
\end{align} 
where we have introduced the left and right moving dimensions
$h= \frac{1}{2} \left( \Delta +s \right)$, $\bar{h} = \frac{1}{2} \left( \Delta -s \right)$. The integrality of spin is essential for the locality of this 3 point function: otherwise branch cuts develop in this expression\footnote{The issue is that when  $z_{i}$ is braided around $z_{j}$, a phase is produced multiplying the relative distance:
\begin{align}
    z_{ij} \to e^{2 \pi i (h_{i}-\bar{h}_{i})}z_{ij}
\end{align}
the total 3 point function gains a phase that can't be absorbed into the transformation of the tensor $C_{ijk}$.}.  
For integer spins, these branch cuts disappear provided 
$C_{ijk}$ satisfies the symmetry property (see appendix \ref{app:sym}):
\begin{align}\label{sym2}
 C_{ijk} &=
 \begin{cases}
  C_{\sigma(i) \sigma(j)\sigma (k) } \exp(i \pi ( s_{i} +s_{j} +s_{k}) ) & \text{for odd permutations } \sigma \in S_{3}  \\
  C_{\sigma(i)\sigma(j)\sigma(k) } & \text{for even permutations } \sigma \in S_{3}
 \end{cases}
 \end{align}

\subsection{The constraint-squared potential }
Correlations functions of a two dimensional CFT are constrained by crossing equations, which are generated by $4$-point sphere crossing, and modular invariance of the torus 1-point function.   These constraints ensure that correlation functions can consistently computed via different slicings of the same manifold, which correspond to different channels in the conformal block decomposition of the correlators.     

To define a potential $V$ for $(C_{ijk},\Delta_{s} )$ that vanishes on the solutions of the crossing equations, we need a positive quadratic form $|\cdots |^{2}$ on the vector space of  conformal blocks with which we can define the sum of squares \cite{belin2023}:
\begin{align}
     V \sim \sum |\text{constraint}|^2
\end{align}

 To obtain this quadratic form, we make use of the following fact.  On a surface $\Sigma_{g,n}$ of genus $g$ and $n$ punctures, the space $\mathcal{H}_{g,n}$ of conformal blocks forms a Hilbert space spanned by a linear combination of blocks with the same external weights at the $n$ punctures, but with arbitrary  internal weights.  $\mathcal{H}_{g,n} $  is endowed with an inner product that was defined by Verlinde in \cite{Verlinde:1989ua}.
Explicitly,  for a pair of chiral conformal blocks $\ket{\mathcal{F}_{1}} ,\ket{\mathcal{F}_{2}} \in \mathcal{H}_{g,n} $,
their inner product is given by a string theory-like  path integral \cite{collier2023}:
\begin{align}\label{Verlinde}
\langle \mathcal{F}_1 | \mathcal{F}_2 \rangle =\int_{\mathcal{T}_{g,n} }Z_\text{bc} \, Z_\text{timelike Liouville} \, \mathcal{F}^{*}_{1} \, \mathcal{F}_2 ,
\end{align}
where $\mathcal{T}_{g,n}$ is the Teichmuller space of the surface $
\Sigma_{g,n}$, $Z_\text{bc}$ is a ghost partition function and $Z_\text{timelike Liouville}$ is the partition function of time- like Liouville theory with central charge $26-c$. The \emph{Liouville} conformal blocks, which have internal weights above $(c-1)/24$, form a complete basis with respect to this inner product.  The full Hilbert space is a tensor product of the above with the Hilbert space of anti-chiral blocks.

For four-point crossing on the sphere and one-point crossing on the torus, we will define the square of the constraints using the Verlinde inner product.  However, compatibility with the 3d gravity interpretation will require us to choose a different norm for the zero point torus constraint. 
We will find it useful at times to label the conformal primaries by their left and right moving Liouville momenta $(P,\bar{P})$. These are related to the left and right moving conformal dimensions
\begin{align} 
h= \frac{1}{2}\left(\Delta +s \right), \quad  \bar{h}= \frac{1}{2} \left( \Delta- s \right)
\end{align}
by the relations
\begin{align}
    h=\frac{Q^2}{4} + P^2,\qquad \bar{h} = \frac{Q^2}{4} + \bar{P}^2,\qquad Q=(b+b^{-1}), \qquad  c=1+6Q^2
\end{align}
\paragraph{Regularizing the inner product}
Due to the continuous weights, the Liouville conformal blocks are delta function normalized with respect to the  Verlinde inner product.  The general formula is given in eq. 2.21  of  \cite{collier2023}.  To give a non-singular definition for the the tensor and matrix model potential, we will regularize by smearing this delta function over a width of order $\varepsilon$.   We will define a diagrammatic expansion of the ensemble in which each diagram has a smooth $\varepsilon \to 0 $ limit, so this regulator can be safely removed at the end of the computation. 
\subsubsection{The tensor model and 4-point crossing}
 The 4 point crossing equation comes from identifying two ways of time slicing the 4 punctured sphere:
\begin{align}\label{4ptcross}
    \vcenter{\hbox{\includegraphics[scale=.5]{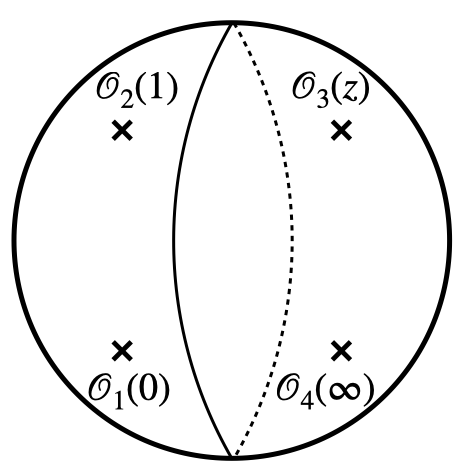}}} \qquad &= \qquad \vcenter{\hbox{\includegraphics[scale=.5]{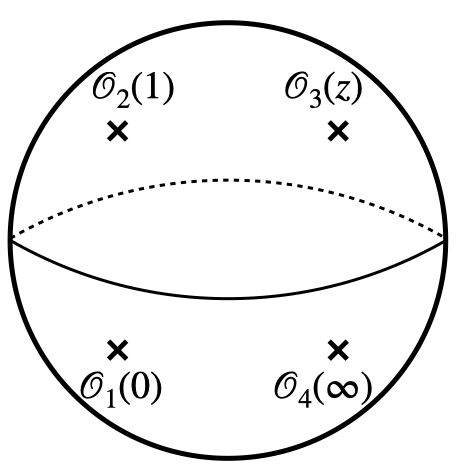}}} \\
    s-\text{channel} \quad \qquad \qquad & \qquad \qquad \qquad t-\text{channel} \notag
\end{align}
Equating these gives 
\begin{align} \label{cross}
    \sum_p C_{12p} C_{p34} \left| \vcenter{\hbox{\includegraphics[scale=0.22]{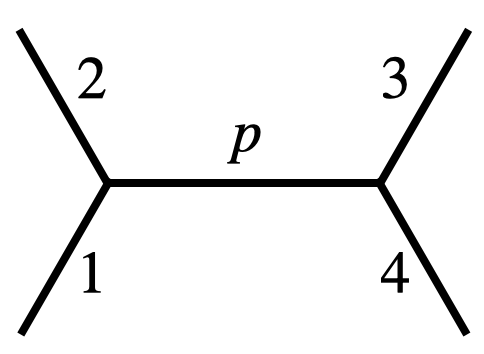}}} \right\rangle \left| \vcenter{\hbox{\includegraphics[scale=0.22]{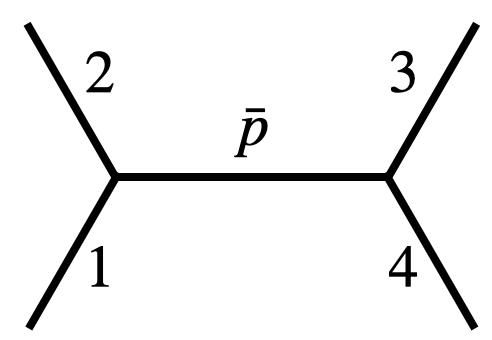}}} \right\rangle - \sum_q C_{23q} C_{q41} \Bigg| \left. \vcenter{\hbox{\includegraphics[scale=0.2]{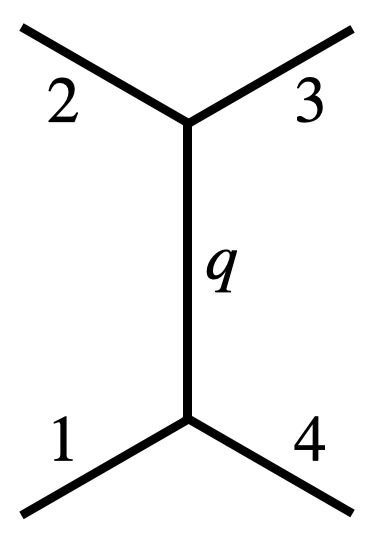}}} \right\rangle \Bigg| \left. \vcenter{\hbox{\includegraphics[scale=0.2]{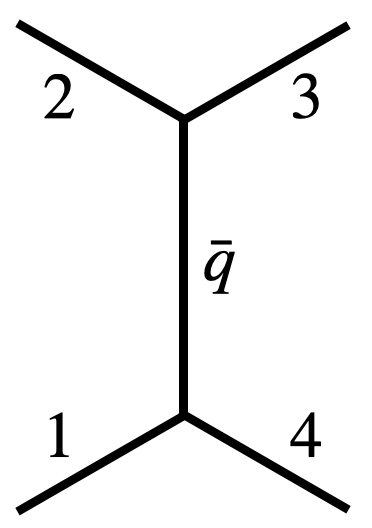}}} \right\rangle = 0
\end{align}
where we represented the Virasoro conformal blocks as vectors in a Hilbert space $\mathcal{H}_{\Sigma_{0,4}}$ associated to the 4 punctured sphere.   In terms of the Liouville momenta $(P_{i},\bar{P}_{i})$ the inner product \eqref{Verlinde} on $\mathcal{H}_{\Sigma_{0,4}}$ is  \cite{collier2023virasoro}
\begin{align}\label{iprod}
    \left\langle \vcenter{\hbox{\includegraphics[scale=0.22]{figures/fin_blockp.png}}} \right| \left. \vcenter{\hbox{\includegraphics[scale=0.178]{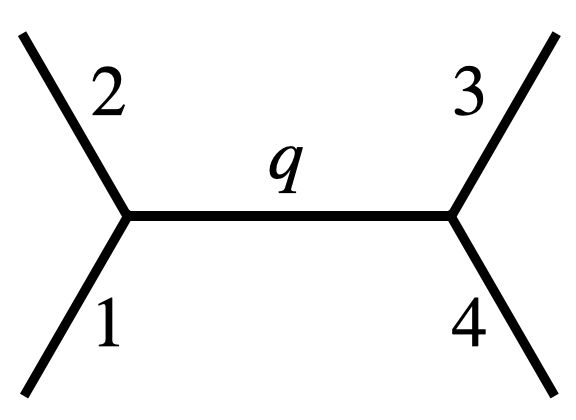}}} \right\rangle = \frac{\delta^{(2)}(P_q - P_q)}{\rho_0(P_q) C_0(P_1,P_2,P_q) C_0(P_3,P_4,P_p)},
\end{align}
where $C_{0}(P_{1},P_{2},P_p)$ is the Liouville 3-point function in a particular normalization. 

We define a potential given by the norm square of the crossing constraint \eqref{cross}, summed over the external operators: 
\begin{align}\label{V4*}
V_{4} = \sum_{i_1 \cdots i_4}{\!\!}' \sum_{p,q} & \left( C_{i_1 i_2 p} C_{p i_3 i_4 } C^*_{i_1 i_2 q } C^*_{q i_3 i_4 } + cc. \right) 
\left| \left\langle \vcenter{\hbox{\includegraphics[scale=0.22]{figures/fin_blockp.png}}} \right| \left. \vcenter{\hbox{\includegraphics[scale=0.178]{figures/fin_blockq2.png}}} \right\rangle \right|^2 \\
& - \left( C_{i_1 i_2 p} C_{p i_3 i_4 } C^*_{q i_4 i_1 } C^*_{i_2 i_3 q } + cc. \right) \left| \left\langle \vcenter{\hbox{\includegraphics[scale=0.22]{figures/fin_blockp.png}}} \right| \left. \vcenter{\hbox{\includegraphics[scale=0.2]{figures/fin_blockq.png}}} \right\rangle \right|^2
\end{align}
Here the sum  $\sum_{i_1\cdots i_4}{\!\!}'$  over external operators excludes the identity. Note that we have used a condensed notation above in which the absolute valued squared denotes the product of the overlaps for conformal blocks in the holomorphic sector with those in the anti-holomophic sector.  However, the overlaps in these sectors are not complex conjugates of each other, and we are not assuming the holomorphic and anti holomorphic weights are equal. 

To compute the overlaps we observe that the Liouville blocks in the s and t-channels are related by the Virasoro fusion kernel  $\mathbb{F}_{P_p P_q}$  defined by Ponsot and Teschner \cite{Ponsot_2001}.
\begin{align}
\left| \vcenter{\hbox{\includegraphics[scale=0.22]{figures/fin_blockp.png}}} \right\rangle = \int_0^\infty d P_q \, \mathbb{F}_{P_p P_q} \,
\begin{bmatrix}
    P_2 & P_3 \\
    P_1 & P_4
\end{bmatrix}
\Bigg| \left. \vcenter{\hbox{\includegraphics[scale=0.2]{figures/fin_blockq.png}}} \right\rangle
\end{align}
Accounting for the normalization \eqref{iprod},  we can compute the overlap for these blocks to be
\begin{align}
\left\langle \vcenter{\hbox{\includegraphics[scale=0.22]{figures/fin_blockp.png}}} \right| \left. \vcenter{\hbox{\includegraphics[scale=0.2]{figures/fin_blockq.png}}} \right\rangle
&= \frac{\mathbb{F}_{P_q P_p} \,
\begin{bmatrix}
    P_3 & P_4 \\
    P_2 & P_1
\end{bmatrix}}{\rho_0(P_p) C_0(P_1,P_2,P_p) C_0(P_3,P_4,P_p)} \\
&= \frac{\begin{Bmatrix}
    q & 4 & 1 \\
    p & 2 & 3
\end{Bmatrix}}{C_0(P_1, P_2, P_p) C_0 (P_3, P_4, P_p) C_0(P_2, P_3, P_q) C_0 (P_1, P_4, P_q)}
\end{align}

In the last equality, we made use of the tetrahedrally symmetric Virasoro 6J symbol $\begin{Bmatrix}
    q & 4 & 1 \\
    p & 2 & 3
\end{Bmatrix}$, so that the total expression has manifest tetrahedral symmetry\footnote{The Virasoro 6J symbol is invariant under interchanging any pair of columns, or interchanging elements in two columm simultaneously.  It can be related to the fusion kernel $\mathbb{F}$ in two ways \cite{collier2023}:
\begin{equation}\label{eq.6jDefition}
   \left| \begin{Bmatrix}  q & 4 & 1 \\ s & 2 & 3\end{Bmatrix} \right|^2 = \left\{ \begin{array}{c} \left|\rho_0(P_s)^{-1}C_0 (P_1, P_4, P_q) C_0 (P_2, P_3, P_q) \mathbb{F}_{P_q P_s} \left[ \begin{array}{c c} P_3 & P_4 \\ P_2 & P_1 \end{array}\right] \right|^2  \\[25pt]
   \left|\rho_0(P_q)^{-1}C_0 (P_1, P_2, P_s) C_0 (P_3, P_4, P_s) \mathbb{F}_{P_s P_q} \left[ \begin{array}{c c} P_3 & P_2 \\ P_4 & P_1 \end{array}\right] \right|^2
    \end{array}
    \right. \,.
\end{equation}}.

Similarly, the identity block can be expanded in terms of the Liouville blocks: 
\begin{align}
    \left| \vcenter{\hbox{\includegraphics[scale=0.22]{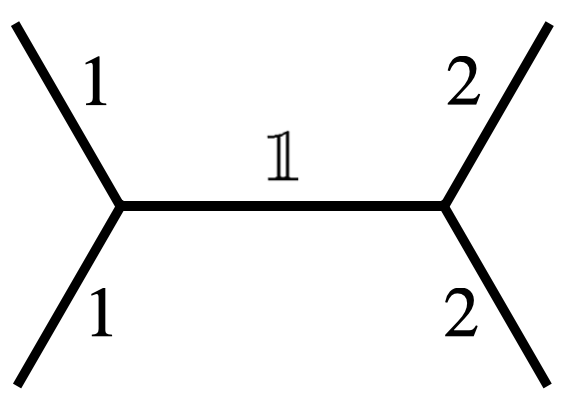}}} \right\rangle &= \int_0^\infty d P_q \, \mathbb{F}_{\mathbbm{1}P_q} \,
\begin{bmatrix}
    P_1 & P_2 \\
    P_1 & P_2
\end{bmatrix}
\Bigg| \left. \vcenter{\hbox{\includegraphics[scale=0.2]{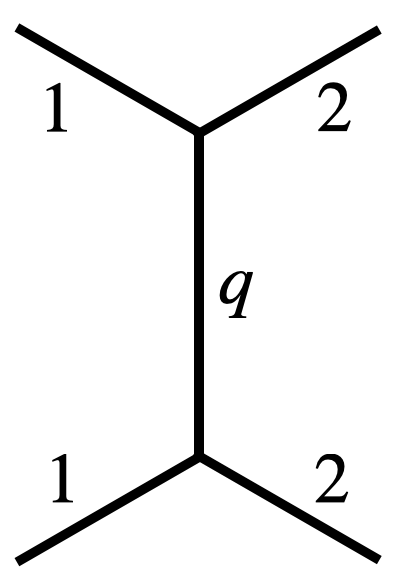}}} \right\rangle \\
&= \int d P_q \, \rho_0(P_q) C_0(P_1,P_2,P_q) \, \Bigg| \left. \vcenter{\hbox{\includegraphics[scale=0.2]{figures/fin_blockq3.png}}} \right\rangle 
\end{align}

where $\rho_0(P_q)$ is the Cardy density of states\footnote{ The Cardy density \cite{Cardy:1986} was originally defined as the $b \to \infty$ limit of this density, while $\rho_{0}(q)$ is sometimes referred to as the Plancherel measure associated to Virasoro representations. Here we keep to the Cardy terminology for convenience. }
\begin{align}
    \rho_0(P_q) = \sinh{(2\pi b P_q)} \sinh {(\frac{2 \pi}{b} P_q)} 
\end{align}
This gives the overlap:
\begin{align}
    \left\langle \vcenter{\hbox{\includegraphics[scale=0.22]{figures/fin_blockq3.png}}} \right| \left. \vcenter{\hbox{\includegraphics[scale=0.22]{figures/fin_blockid.png}}} \right\rangle = \frac{1}{C_{0}(P_{1},P_{2},P_q)} \equiv \begin{Bmatrix}
    q & 2 & 1 \\
    p & 1 & 2
\end{Bmatrix}
\end{align}
Finally, using the relation $C_{ijk}^*= C_{kji}$, we can write the quartic vertex as\footnote{Here we have used the fact that the inner product is invariant under crossing transformations, so that the diagonal terms are the same.}:
\begin{eqnarray}\label{eq.V4}
V_4 &=& 2 \sum_{i_1\cdots i_4}{\!\!}' \sum_{p,q}\left( \frac{C_{i_1 i_2 p} C_{p i_3 i_4 } C_{q i_2 i_1  } C_{ i_4 i_3 q }}{|\rho_0(p)C_0(12p)C_0(34p)|^2} \delta^{(2)}_{\varepsilon} \left(P_p - P_q \right) - \right. \nonumber \\ &&  \hspace{3.5cm}\left. \frac{ C_{i_1 i_2 p} C_{p i_3 i_4 } C_{i_1 i_4 q } C_{q i_3 i_2 }  }{|C_0(12p) C_0 (34p) C_0(23q) C_0 (14q)|^2 } \left|\begin{Bmatrix}
    q & 4 & 1 \\
    p & 2 & 3
\end{Bmatrix}\right|^{2} \right)\,.
\end{eqnarray}
This is the same tensor model potential written in \cite{belin2023}. 
In the first term, we introduced a regulator $\varepsilon$ that produces a smearing of the delta function in anticipation of the fact that the width has to be greater than the average inter-eigenvalue spacing dictated by the matrix model ensemble, which scales like $e^{-c}$.   Note also that in writing this potential, we have used a short-hand notation where we only keep track of the indices which label the momenta in the functions $C_0, \rho_0$, and operator indices in the 6J or $\mathbb{F}$ symbols. Indices will always be lower-case letters, while momentum will always be denoted by capital $P$.  We will adopt this convention for the rest of the paper. 

Note that each term in \eqref{eq.V4} has symmetries that correspond to the symmetries of the pillow and tetrahedron graphs which arise from gluing together together two s-channel blocks or an s and t-channel block respectively:
\begin{align}\label{graphs}
 \vcenter{\hbox{ \includegraphics[scale=.3]{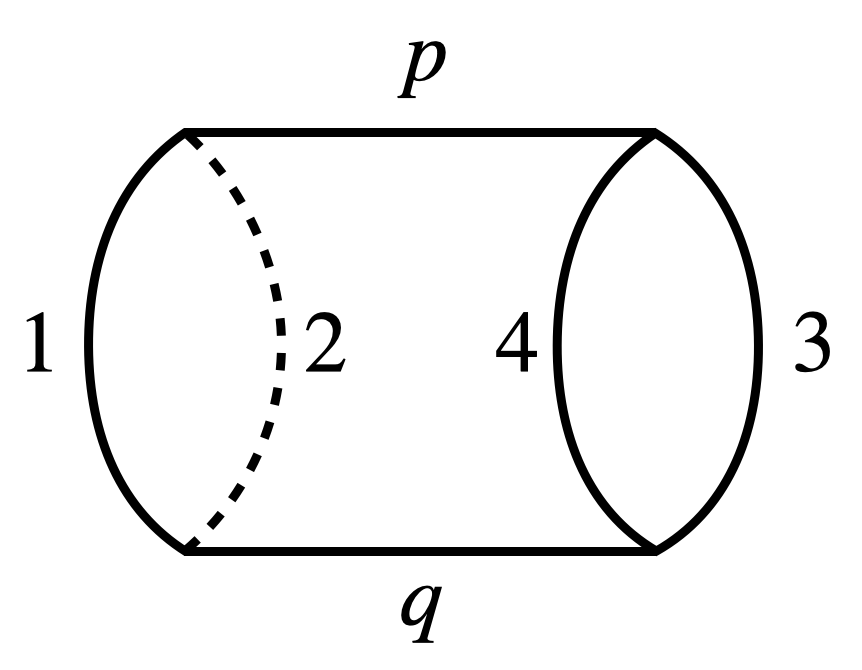}}} ,\qquad    \vcenter{\hbox{\includegraphics[scale=.3]{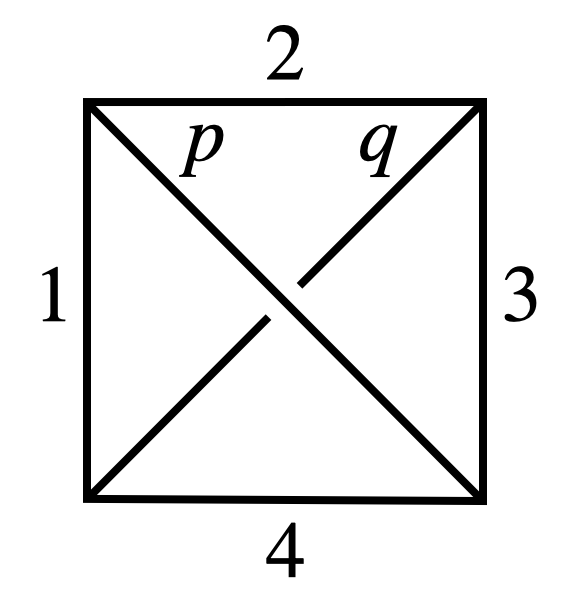}}}
\end{align}
Each vertex of these graphs corresponds to an OPE coefficient,  and each edge indicates a contraction of indices.    The cyclic ordering of the indices in the OPE coefficients is indicated in these figures by the clockwise orientation around each vertex when viewing the graphs from the ``outside".

\subsubsection{Tensor model propagator and Triple line diagrams} 
The propagator for the tensor model is determined by the quadratic terms in $V_4$ that appear because $C_{\mathbbm{1} ii} =1$. These originate from terms in \eqref{eq.V4} with $p =\mathbbm{1}$, $ i_{1}=i_{2}$, $i_{3}=i_{4}$  and similarly for $q=\mathbbm{1}, \, i_{1}=i_{4},i_{2}=i_{3}$.  These give rise to the kinetic term:
\begin{align}\label{K2}
    V_{4} \supset K_{2} = -2 \sum_{i_{2} i_{3}} {\!\!}' \sum_{q} \frac{C_{i_{2}i_{3}q} C_{q i_{3} i_{2}}}{|C_{0}( i_{2}i_{3}q)|^2}    
\end{align} 
Notice that this has the wrong sign, signaling that we are expanding around an unstable maximum as we alluded to in the introduction.  Due to the symmetry property \eqref{sym2} of the OPE coefficients,  the corresponding propagator will contract any pair of $C_{ijk}$'s whose indices differ by a permutation, with a nontrivial phase factor for even permutations\footnote{The propagator which involves $C^{*}_{ijk}$ follows from rewriting it in terms of the $C_{ijk}$ using the symmetry property \eqref{sym}.  This gives:
\begin{align} \label{Prop*}
\wick{ \c1{C^{*}_{ijk}} \c1{C^{*}_{\sigma(i) \sigma(j) \sigma(k)} } }  &=
\begin{cases}
-\hbar \, e^{+i \pi ( s_{i} +s_{j} +s_{k}) } |C_{0}(ijk)|^{2}  &\text{for even permutations } \sigma \in S_{3}  \\
-\hbar \, | C_{0}(ijk)|^{2} &\text{for odd permutations } \sigma \in S_{3},
\end{cases}
\end{align} 
and
\begin{align} \label{C*C}
\wick{ \c1{C^{*} _{ijk}} \c1{C_{\sigma(i) \sigma(j) \sigma (k) }}} = - \hbar\, e^{i \pi ( s_{i}+s_{j}+s_{k})} \wick{ \c1{C_{ijk}} \c1{C_{\sigma(i) \sigma(j) \sigma(k) }}} &=
\begin{cases}
 |C_{0}(ijk)|^{2}  &\text{for even permutations } \sigma \\
 -\hbar \,e^{ + i \pi (s_{i}+s_{j}+s_{k})} |C_{0}(ijk)|^{2} &\text{for odd permutations } \sigma
\end{cases}
\end{align} }.
\begin{align} \label{Prop} 
\wick{ \c1{C_{ijk}} \c1{C_{\sigma(i) \sigma(j) \sigma(k)} } }  &= \begin{cases} -\hbar \, e^{-i \pi ( s_{i} +s_{j} +s_{k}) } |C_{0}(ijk)|^{2}  &\text{  for even permutations } \sigma \in S_{3}  \\
 -\hbar \,| C_{0}(ijk)|^{2}   &\text{  for odd permutations } \sigma \in S_{3}
\end{cases}
\end{align}
The minus sign above arises because the propagator is the inverse of the kinetic term.
\paragraph{Triple line Feynman diagrams}
The contraction rules \eqref{Prop} define the triple line propagator: 
\begin{align}\label{3lines}
   \vcenter{\hbox{\includegraphics[scale=.3]{figures/fin_3lines}}} = (\text{phase}) \times \hbar \,| C_0(ijk)|^2
\end{align}
Similarly, the quartic vertices \eqref{eq.V4} corresponding to the 6J and pillow graph  have a triple line representation given by \cite{Regge:1961px,Sasakura:1990fs,Ambjorn:1990ge,Williams:1991cd,Gross:1991hx,BOULATOV_1992,Rivasseau:2016zco,Gurau:2016cjo,hartle2022simplicialquantumgravity}
\begin{align}\label{quarticv}
    \vcenter{\hbox{\includegraphics[scale=.3]{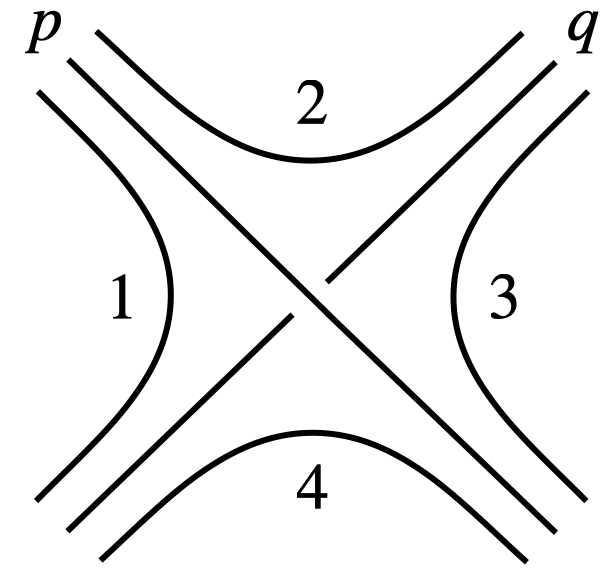}}} &= \frac{1}{\hbar} \frac{1}{\left|C_0(12p) C_0(34p) C_0(23q) C_0(14q)\right|^2} 
   \left| \begin{Bmatrix}
        q & 4 & 1 \\
        p & 2 & 3
    \end{Bmatrix} \right|^2 \\
    \vcenter{\hbox{\includegraphics[scale=.52]{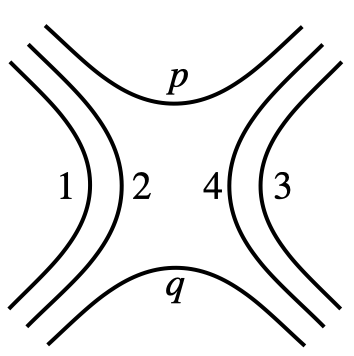}}} &= \frac{1}{\hbar} \frac{\delta^{(2)}(P_p - P_q)}{\left|\rho_0(p) C_0(12p) C_0(34p)\right|^2}
\end{align}
 These diagrams are obtained from the 6J and pillow graphs by removing a neighborhood of the junctions.  This allows them to be glued together by the triple line propagator to create a general tensor model diagram.   However, these diagrams do not capture the cyclic order of the indices\footnote{This is the information that was captured by a choice of orientation each vertex of the pillow and 6J graphs in \eqref{graphs}. } in each $C_{ijk}$, which leads to ambiguities associated to the phases in \eqref{sym2} when we introduce non-integer spins in section \ref{sec:gravity}.   We will explain how to incorporate these phases into the diagrammatics in section \ref{sec:checks}.  
 
 With these Feynman rules we can immediately compute some simple observables, which are products of the OPE coefficients with an index structure compatible with the propagator and the quartic vertices: 
\begin{align}
    \braket{C_{ijk} C_{ijk}^* } &= -\hbar  |C_0(ijk)|^2 +\cdots \nn
     \braket{C_{12p} C_{34p}C_{41q} C_{32q} } &= - \hbar^3 \left| \begin{Bmatrix}
    q & 4 & 1 \\
    p & 2 & 3
\end{Bmatrix} \right|^2 +\cdots
\nn
 \braket{C_{12p} C_{34p}C_{12q} C_{34q} } &=\hbar^3 \frac{\delta^{(2)} (P_p-P_q)}{|\rho_{0}(p)|^2}+\cdots
\end{align}
More generally, observables involve contraction of these indices.  Since these produce insertions in the matrix integral, we will discuss these after defining the matrix model.

\subsubsection{The matrix model and torus modular invariance}

 Our model \eqref{Z} is defined so that in each sector of fixed, quantized spin $s$, the Dilatation operator $\Delta_{s}$ is a double-scaled random matrix  with the  spectral density given by the \emph{integer} Cardy formula
\begin{align}\label{Icardy}
    \rho_{0}(\Delta,s) &= \braket{\rho(\Delta,s)}_{\text{disk}} \nn
    &=  \mu(\Delta,s)  \sinh \left( 2 \pi b \sqrt{ \frac{\Delta +s}{2} - \frac{c-1}{24} }\right) \sinh \left( 2 \pi b^{-1} (\sqrt{ \frac{\Delta +s}{2}- \frac{c-1}{24} }\right)\times \nonumber \\    
    & \qquad \qquad \, \sinh \left( 2 \pi b \sqrt{ \frac{\Delta -s }{2}- \frac{c-1}{24} }\right) \sinh \left( 2 \pi b^{-1} (\sqrt{ \frac{ \Delta -s}{2}- \frac{c-1}{24} }\right) , \nn
    \mu(\Delta,s) &\equiv \frac{1}{8} \frac{1}{\sqrt{ \frac{\Delta+s}{2} - \frac{c-1}{24}}}  \frac{1}{\sqrt{ \frac{\Delta-s}{2} - \frac{c-1}{24}}}, \qquad  s \in  \mathbb{Z}, \,\,\Delta \geq  |s| +\frac{c-1}{12} 
\end{align}
In the standard  matrix model interpretation, this is obtained diagrammatically by filling in a boundary circle representing the observable $\rho(\Delta, s)$ with  't Hooft diagrams compatible with a disk topology\footnote{More precisely, this is the inverse Laplace transform of the observable \par $\braket{Z(\beta)}=\braket{\int d \Delta \,\rho(\Delta, s) \left(e^{-\beta \Delta} + \text{descendants} \right).}$ that is normally associated to a boundary circle of length $\beta$. }. In addition to the spectral density,  another important observable is given by the density-density correlator, whose leading contribution is given by the 't Hooft diagrams with the annulus topology:
\begin{align} 
    C(\Delta_{1},s_{1}; \Delta_{2},s_{2}) & \equiv
    \braket{\rho(\Delta_{1},s_{1}), \rho(\Delta_{2},s_{2})}_{\text{annulus}} 
\end{align}
In each spin sector, this correlator is universal and dictated by the symmetry class specifed by Vandermonde.   For GOE, the Vandermonde is
\begin{align}\label{matrix_measure}
K(\Delta_{1},s_{1}; \Delta_{2},s_{2} ) =  \delta_{s_1,s_2} \log |\Delta_1 - \Delta_2|,
\end{align}
The annulus $C$ is then given by the functional inverse of $K$, because $K$ is the quadratic kernel for the density $\rho(\Delta, s)$ in the matrix model effective action.  It is important to note that the functional inverse depends on the particular shape of the cut for $\Delta$ as a function of $s$.

We find that 
\begin{align}\label{annulus}
     C(\Delta_{1},s_{1}; \Delta_{2},s_{2}) 
    &= 
    2 \frac{\Delta_{1}+\Delta_{2} -2s - \frac{c-1}{6}}{(\Delta_{1}-\Delta_{2})^2 \sqrt{( \Delta_{1} + \Delta_{2} -2s -\frac{c-1}{6} )^2 - (\Delta_1 - \Delta_2)^2} }\delta_{s_{1},s_{2}}  
\end{align}
for $\Delta_1, \, \Delta_2 \geq s+ \frac{c-1}{12}$
where we denote $\frac{|s_1+s_2|}{2} = s$.
We interpret  $C=K^{-1}$ as the matrix model propagator.  

In addition to the {\it a priori} potential $V_{0}(\Delta_{s})$ defined by the spectral density \eqref{Icardy}, we now introduce the constraint-squared potentials that impose $S$ modular invariance on the torus.  We will treat modular invariance for zero point and one point functions separately.  
\paragraph{$S$ modular invariance on the torus}
Consider the torus partition function of an approximate CFT, expressed in terms of the Virasoro characters 
\begin{align}
        Z(\tau , \bar{\tau}) &=  \chi_{\mathbbm{1}}(\tau) \chi_{\mathbbm{1}} (\bar{\tau}) +   \sum_{i, \bar{i} }  \chi_{P_{i}}(\tau) \chi_{\bar{P}_{\bar{i} }}(\tau)  \nn
        \chi_{P}(\tau) &=  \frac{q^{P^{2} }}{\eta(\tau) } , \qquad  \chi_{\mathbbm{1}} (\tau)= \frac{1-q}{\eta(\tau)} q^{-\frac{c-1}{24}} ,\qquad q =e^{2 \pi i \tau}
\end{align}
Here, we have used Liouville notation to label the conformal dimenions.
$S$-modular invariance is the statement that 
\begin{align}
    Z(\tau,\bar{\tau}) -Z( -1/\tau, -1/\bar{\tau}) =0.
\end{align}
With respect to the Verlinde inner product  on the Hilbert space  $\mathcal{H}_{T^2}$ of torus conformal blocks,  the Virasoro characters form a delta function normalized basis which we denote by $\ket{P_{i}}$: 
\begin{align}
    \braket{P_{i}|P_{j}} = \delta(P_{i}-P_{j}) 
\end{align}

The $S$ modular transformation is represented by an operator $S$ on $\mathcal{H}_{T^2}$ with matrix elements
\begin{align} \label{s_mat_def}
S_{P_{i} P_{j} } &= \cos (4 \pi  P_{i} P_{j}) \nn
S_{\mathbbm{1}P_{i}} &= \sinh 2 \pi b P_i \sinh 2 \pi b^{-1} P_i \nn
S_{P_{i}\mathbbm{1} } &= S_{ \mathbbm{1} \mathbbm{1}}=0 .
\end{align}

As in the case of 4 point crossing, 
we can write the modular invariance constraint as the vanishing of a vector:
\begin{align}\label{S}
 \ket{\mathbbm{1} } \ket{ \mathbbm{1}} + \sum_{i ,\bar{i} }  \ket{ P_{i}}   \ket{ \bar{P}_{\bar{i}}}  -  \int dP dP' S_{\mathbbm{1}P}S_{\mathbbm{1}P'} \ket{P}\ket{P'}  - \sum_{i ,\bar{i} } \int d P  dP'   S_{P_{i} P} S_{\bar{P}_{\bar{i}} P'} \ket{P} \ket{P'}  &=0.
\end{align} 
To construct a potential for $S$ modular invariance that is compatible with 3d gravity, we will take the square of the constraint using an inner product that is different than the Verlinde inner product. We find that the Vandermonde $K$ defines the appropriate gravitational inner product on the space of Virasoro characters. This leads us to define the matrix model potential\footnote{The braket notation is potentially confusing here: we are really just writing down the matrix element of a quadratic form, and we should not read $\bra{\Delta_{i},s_{i}}$ as an operation that uses an inner product to turn a ket in to a bra.  }:
\begin{align}\label{VsMatrix}
V_{S} &\equiv \sum_{i,j} \braket{ \Delta_{i},s_{i} | \hat{V}_{S} | \Delta_{j} ,s_{j} } \nn
    \hat{V}_{S} &\equiv 
  (\mathbbm{1} - \hat{S}) \hat{K} (\mathbbm{1} - \hat{S})  
\end{align}
The positivity of $K$ ensures that this potential is also minimized when $1-S=0$.  Some care must be taken to interpret the operator $\hat{V}_{S}$, since $S$ takes us outside the space of quantized spins.  In particular, we will define a regulated potential by replacing the Kronecker delta in definition \eqref{matrix_measure} of $K$ with a smeared delta function:  
\begin{align}\label{Keps}
\braket{\Delta_{1},s_{1}|\hat{K}_{\varepsilon}|\Delta_{2},s_{2}} \equiv \delta_{\varepsilon} (s_{1}-s_{2}) \log |\Delta_1 - \Delta_2|
\end{align}

We will address the subtleties related to introduction of  continuous spins in section \ref{sec:gravity}.

\paragraph{One point $S$ modular invariance on the torus}
The $S$ modular invariance of the one point function on the torus is given by
\begin{align}\label{1ptT}
    0=\braket{O_{i}}_{\tau} -  \tau^{h_{i}} \bar{\tau}^{\bar{h}_{i}} \braket{O_{i}}_{-1/\tau} &= \sum_{j} C_{ijj} |\mathcal{F}_{P_{j}} (P_{i};\tau)|^{2}-C_{ijj}|\mathcal{F}_{P_{j}} (P_{i};-1/\tau)|^{2}\nn
    &=\sum_{j}C_{ijj}\left( |\mathcal{F}_{P_{j}} (P_{i};\tau)|^{2} -\sum_{k} S_{P_{j}P_{k}}[{P_i}] |\mathcal{F}_{P_{k}} (P_{i};\tau)|^{2}\right)
\end{align}
In the first equality, we introduced the one point torus conformal block $\mathcal{F}_{P_{j}} (P_{i};\tau)$ (see \eqref{fnorm} below for their graphical representation).
In the second equality we used the one-point modular $S$ matrix $S_{P_{j}P_{k}}[P_{i}]$ to implement an $S$ transform on the conformal blocks. We define the corresponding constraint squared potential by taking the norm-square of \eqref{1ptT} using the Verlinde inner product.  The conformal blocks themselves have the norm
\begin{align}\label{fnorm}
    \braket{\mathcal{F}_{P_{j}} (P_{i};\tau)|\mathcal{F}_{P_{j}} (P_{i};\tau)} = \left\langle \vcenter{\hbox{\includegraphics[scale=0.3]{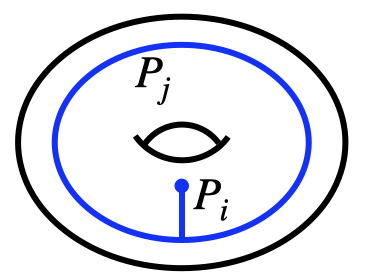}}} \right. \left| \vcenter{\hbox{\includegraphics[scale=0.3]{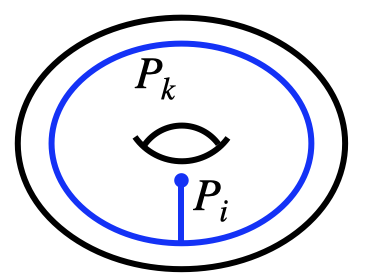}}}\right\rangle =\frac{\delta(P_{j}-P_{k})}{\rho_{0}(P_{j}) C_{0}(ijj)} 
\end{align}
where we have used the notation $\rho_{0}(P_{j}) \equiv S_{\mathbbm{1}P_j} $.   The constraint square potential obtained from the norm of \eqref{1ptT}is then given by\footnote{This differs from the potential written in \cite{belin2023}, where $|\rho_{0}(P_{j}) C_{0}(ijj)|^2$ was in the numerator instead of the denominator.}
\begin{align}\label{VS}
   V^{1\text{pt}}_{S}  &=\sum_{i,j,k} C_{ijj}C_{ikk} \left( 
    \left| \left\langle \vcenter{\hbox{\includegraphics[scale=0.3]{figures/fin_torus1.png}}} \right. \left|\vcenter{\hbox{\includegraphics[scale=0.3]{figures/fin_torus2.png}}} \right\rangle \right|^2 - \left| \left\langle \vcenter{\hbox{\includegraphics[scale=0.3]{figures/fin_torus1.png}}} \right. \left| \vcenter{\hbox{\includegraphics[scale=0.3]{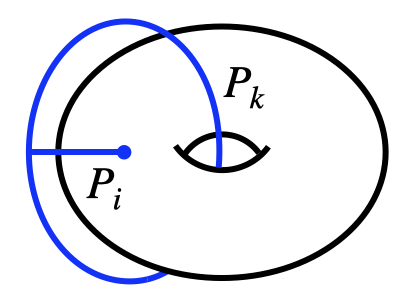}}}\right\rangle \right|^2 \right)   \nn
   &=\sum_{i,j,k} \frac{C_{ijj}C_{ikk}}{|\rho_{0}(P_{j}) C_{0}(ijj)|^2}  \left(\delta^{2}_{\varepsilon}(P_{j}-P_{k}) -|S_{P_j P_k}[P_{i}]|^2 \right),
\end{align}
where we have once again introduced a smearing of the delta function.
This gives two quadratic vertices 
with  the triple line diagrams 
\begin{align}
\vcenter{\hbox{\includegraphics[scale=0.4]{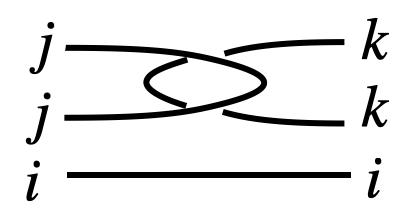}}} &=  \frac{C_{ijj}C_{ikk}}{|\rho_{0}(P_{j}) C_{0}(ijj)|^2}  \delta^{2}_{\varepsilon}(P_{j}-P_{k}) \\
\vcenter{\hbox{\includegraphics[scale=0.4]{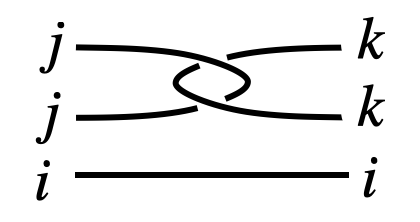}}}&=\frac{C_{ijj}C_{ikk}}{|\rho_{0}(P_{j}) C_{0}(ijj)|^2}  |S_{P_j P_k}[P_{i}]|^2 
\end{align}
\paragraph{Matrix model observables}
The standard observables of the matrix model consists of products of CFT partition functions:
\begin{align}
    \braket{ \Tr q^{L_{0} }\bar{q}^{\bar{L}_{0}}  \cdots \Tr q^{L_{0}} \bar{q}^{\bar{L}_{0} }}.
\end{align}

When coupled to the tensor model, there are more matrix model observables that arise from sums over the indices of $C_{ijk}$, e.g. 
\begin{align}
    \sum_{k} \braket{C_{ijk} C_{klm} }  
\end{align}
In terms of 't Hooft diagrams, the $k$ index here behaves like a quark line which is attached to the double line 't Hooft graphs of the matrix model. 
Finally, there are bubble diagrams of the tensor model lines which are also inserted into the matrix integral. We will elaborate on the diagrammatic description of these observables in section \ref{sec:gravity}.

\paragraph{The total potential}
We summarize the full constraint-squared potential as
\begin{align}\label{fullV}
    V_{\varepsilon} = V_S + V^{1\text{pt}}_{S} + V_4\,,
\end{align}
where the pieces $V_S, V^{1\text{pt}}_{S}$ and $V_4$ are written in \eqref{VsMatrix}, \eqref{VS} and \eqref{eq.V4} respectively. Note that $V_4$ includes the propagator term, which results from setting some indices to identity.

\section{Gravity from the CFT ensemble } \label{sec:gravity}
\subsection{Quantizing spins by summing over T transforms } 
Our ensemble of approximate CFT's was defined with quantized spins. Indeed this is important for the $C_{ijk}$'s to define a local theory.  
However, to relate the perturbative expansion of the ensemble \eqref{Z} to the topological expansion of 3d gravity, it will be useful to introduce  \emph{continuous} spins\footnote{For example,  this matches with the spectrum of BTZ black holes.}  in the matrix model expansion.

For example, consider the disk density, given by the integer Cardy formula.  We can trivially re-write this as a sum over delta functions:
\begin{align}
    \braket{\rho(\Delta, s)}_{\text{disk}} = \sum_{n \in \mathbb{Z}} \delta(s-n) \rho_{0}(\Delta, s)
\end{align}
which produces a density that can be integrated against functions of a continuous spins $s$.   This rewriting has a nice interpretation when we write the  Dirac comb  in terms of its Fourier transform:
 \begin{align}
 \braket{\rho(\Delta, s)}_{\text{disk} } = \sum_{n \in \mathbb{Z}} e^{2\pi i n s} \rho_{0}(\Delta, s) 
\end{align} 
Consider the Laplace transform of the disk density, which produces the averaged partition function:
\begin{align}
    \braket{Z(\tau, \bar{\tau})}_{\text{disk} } &= \int ds \int d \Delta \braket{\rho(\Delta, s)}_{\text{disk} } \chi_{\frac{\Delta+s}{2}} (\tau) \bar{\chi}_{\frac{\Delta-s}{2}}(\bar{\tau})\nn
    &= \sum_{n \in \mathbb{Z}}   \int ds \int d \Delta   \, \rho_{0}(\Delta, s) e^{2\pi i n s}   \chi_{\frac{\Delta+s}{2}} (\tau) \bar{\chi}_{\frac{\Delta-s}{2}}(\bar{\tau}) 
    \nn 
    &= \sum_{n \in \mathbb{Z}}   \int ds \int d \Delta   \, \rho_{0}(\Delta, s)   \chi_{\frac{\Delta+s}{2}} (\tau+n) \bar{\chi}_{\frac{\Delta-s}{2}}(\bar{\tau}+n) 
\end{align}
In the second line, we assume that given some appropriate regularization of the Dirac delta comb, we can interchange the order of summation and integration.  In the last line, we observed that for fixed $n$, the phase $e^{2\pi i n s}$ describes the effect of applying a  $T$-transformation to the Virasoro characters $n$ times.   This produces a $T$ symmetrization of partition functions for BTZ black holes.  The same procedure can be applied to the insertions of the annulus diagram $\braket{\rho(\Delta_{1},s_{1}) \rho(\Delta_{2},s_{2})}_{\text{annulus}}$, where we replace the Kronecker delta with an appropriately smeared Dirac delta function  $\delta_{\varepsilon}(s_{1}-s_{2})$ as in \eqref{Keps}.

Notice that once we have interchanged the order of summation and integration in the Dirac delta comb, we must analytically continue functions of integer spin $s$ to non-integer spins.    For example, the 
%To do so, we will  spins $s\in \mathbb{Z}$ as an integral over continuous spins, times a Dirac-delta comb: 
%\begin{align}
    %\sum_{s \in \mathbb{Z}} \to \int ds %\sum_{n=-\infty}^{\infty} e^{2\pi i n s}.
%\end{align}
integer Cardy density \eqref{Icardy} is continued into a   holomorphically factorized Cardy density:
\begin{align}\label{cardy}
    \rho_{0} (\Delta,s) \to \rho_{0}(P) \rho_{0} (\bar{P}),\qquad P,\bar{P} \geq 0
\end{align}
which allows for all real and positive spins.

In the tensor model, allowing for non integer spin introduces non-local braiding phases. These arise from the analytical continuation of the symmetry properties of the OPE coefficients: 
\begin{align}\label{sym2*}
 C^{*}_{ijk} &= e^{ \left(i\pi (s_{i} +s_{j} +s_{k} )\right)} C_{ijk} ,\nn
 C_{ijk} &= 
    \begin{cases} C_{\sigma(i) \sigma(j)\sigma (k) } e^{i \pi ( s_{i} +s_{j} +s_{k})} & \text{for odd permutations } \sigma \in S_{3}  \\
    C_{\sigma(i)\sigma(j)\sigma(k) } & \text{for even permutations } \sigma \in S_{3}
    \end{cases}
 \end{align} 
When the spins are continued to  non integer values, the sign factors $\exp(i \pi ( s_{i} +s_{j} +s_{k}) )$ become general braiding phases.   The definition \eqref{3pt1} of $C_{ijk}$ in terms of three point functions then breaks down due to the appearance of branch cuts.  Thus for continuous spins, formulating a path integral over $C_{ijk}$ is a subtle problem that we will not address in this paper.  Instead we observe that the diagrammatic rules for the tensor model remain well defined for non integer spin:  each line in a Feynman diagram is thickened into a ribbon whose twisting captures the braiding phases. We will explain these phases in more detail in section \ref{sec:checks}.   From an operational point of view, we can simply define the tensor model perturbative expansion using the same contraction rules in \eqref{Prop} and with the same vertices, except that $C_{ijk}$ is treated as a formal variable.

Note that  the analytic continuation of spin to non-integer values only occurs at the intermediate stages in the perturbative expansion of the  matrix-tensor model, i.e. before summing over $T$ transforms.   The full sum over $T$ transforms restores the integrality of spin, consistent with the original definition of the ensemble.
\begin{comment} 
Thus, we can use the diagrammatic expansion as a \emph{definition} of \eqref{Z'}, in which the analytic continuation of spin to non-integer values only occurs at intermediate stages of a calculation. Indeed the 
sum over phases $e^{2\pi i n s}$ enforces integrality of spin in the final answer for any observable.   This sum has a natural interpretation as a sum over T transforms which acts on a partition function according to
\begin{align}
    T^{n} :Z(\tau,\bar{\tau}) \to Z(\tau +n,\bar{\tau}+n ) = \sum_{i} \chi_{h_{i}} (\tau+n) \bar{\chi}_{\bar{h}_{i}}(\tau+n)= \sum_{i}  e^{2\pi i n s_{i}} \chi_{h_{i}} (\tau) \bar{\chi}_{\bar{h}_{i}}(\tau)
\end{align}
Summing over $n$ symmetrizes over T transforms and project onto integer spin.  At the level of the leading spectral density, this has the effect of mutliplying the holomorphically factorized Cardy density \eqref{cardy} with a Dirac delta-comb, restoring the integrality of spin. 
\end{comment} 
\subsection{Implementing $S$-modular invariance}
Consider the perturbative expansion of the potential for zero point modular invariance.  We now interpret the objects $C, S, K$ in \eqref{annulus}, \eqref{s_mat_def}, \eqref{matrix_measure} as operators $\hat{C},\hat{S},\hat{K}$ acting on the enlarged space of holomorphically factorized torus blocks $\mathcal{H}_{T^2} \otimes \mathcal{H}_{T^2}$, where the spin is allowed to be continuous.  But this means that the operator associated to the matrix model propagator (i.e. the annulus contribution to the density-density correlator)  
is not just given by $\hat{C}$. Instead it must be attached to a Dirac-delta comb $\hat{D}$ that projects onto integer spin.  Thus, on the enlarged space of continuous spins we should write the annulus operator as  
\begin{align}
    \hat{A}= \hat{C} \hat{D},\qquad  \hat{D}=\sum_{n=-\infty}^{\infty} \hat{T}^{n} 
\end{align}    
We identify this with the annulus diagram in the expansion of the matrix model.

One can check that the operators $\hat{K}$ and $\hat{S}$ commute (note that we do not attach a Dirac comb to $\hat{K}$ so it remains the inverse to $\hat{C}$).  Using this fact, we can write 
\begin{align}
    \hat{V}_{S} =2 (\hat{1}-\hat{S}) \hat{K}
\end{align}
and the corresponding potential is \begin{align}
    V_{S} &= V_{S,1}+V_{S,2} \nn
    V_{S,1} &= 2 \sum_{j} \left \langle \mathbbm{1}, \bar{\mathbbm{1}} \right| 2 (\hat{1}-\hat{S})\hat{K} \left| P_j, \bar{P}_j \right \rangle\nn
    V_{S,2}&=\sum_{i,j} \left \langle P_{i}, \bar{P}_{i} \right| 2 (\hat{1}-\hat{S})\hat{K}\left| P_j, \bar{P}_j \right \rangle 
\end{align}
with no restrictions on the momenta $(P,\bar{P})$.  Here we have explicitly separated out the single trace contribution $V_{S,1}$, coming from the terms where the initial or final dimension are set to identity, from the double trace contribution $V_{S,2}$.
The single trace potential has only one index sum and therefor acts as a one point vertex, while the double trace part acts as a quadratic vertex.

The perturbative expansion of the matrix model corresponds to the expansion of $e^{-\frac{1}{\hbar} V_{S}}$ in the matrix model integral.  In these perturbative computations, both the single and double trace potentials are Wick contracted with the matrix model propagator $\hat{A}$.   In the operator language, the Wick contraction is given by the composition of linear maps on the Hilbert space $\mathcal{H}_{T^2} \otimes \mathcal{H}_{T^2}$. Diagrammatically we represent these propagators and vertices as follows  
\begin{align}
    V_{S,1} &= \vcenter{\hbox{\includegraphics[scale=0.35]{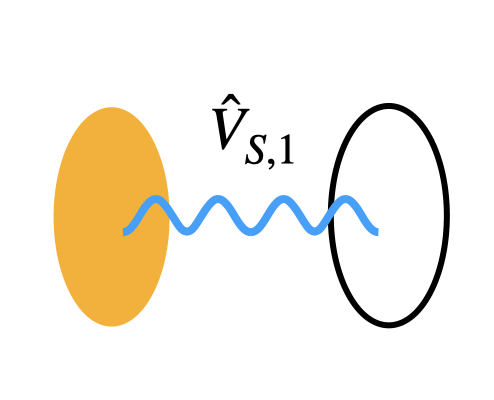}}} \\
    V_{S,2} &= \vcenter{\hbox{\includegraphics[scale=0.35]{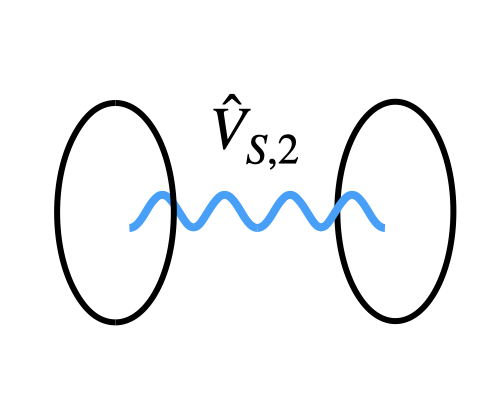}}} \\
    \hat C &= \vcenter{\hbox{\includegraphics[scale=0.35]{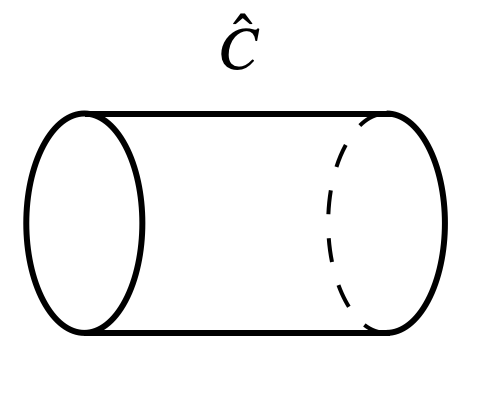}}}
\end{align}
The circles correspond to index sum, while the solid disk corresponds to identity. As we can see, $V_{S,1}$ has one circle and one solid disk, which corresponds to it being a single trace term. $V_{S,2}$ has two index sums, hence has two circles joined by a wavy line to schematically indicate the action of operator $\hat{V}_{S,2}$.

If we restrict to the sector where $n=0$ (so no $T$ transforms have been performed), introducing the potential $V_{S}$ does not change the leading spectral density with the disk topology.  This is because the contributions between 
$V_{S,1}$ and $V_{S,2}$ cancel.  Indeed, the double trace part glued to a disk is 
 \begin{align}
     2\sum_{j} \langle \mathbbm{1}, \bar{\mathbbm{1}} | \hat{S} \, 2(\hat{1}-\hat{S}) \hat{K} | P_j, \bar{P}_j \rangle &= 2 \sum_{j} \langle \mathbbm{1}, \bar{\mathbbm{1}} | 2(\hat{S}- \hat{1}) \hat{K} | P_j, \bar{P}_j \rangle \\
     &= -V_{S,1},
 \end{align}
 which cancels the single trace contribution.  
However, for $n\neq 0$, the disk density does change, and in fact will be replaced with a sum over  $SL(2,\mathbb{Z})$ transformations.

To see how the $SL(2,\mathbb{Z})$ sum comes about, observe that since the vertices are at most quadratic, at $n$-th order in the perturbation series, we get a sequence 
\begin{align} 
\frac{(\frac{1}{\hbar})^n}{n!}
\cdots \hat{A} \hat{V} \hat{A} \hat{V} \hat{A}  \cdots = \frac{(\frac{2}{\hbar})^n}{n!}\cdots \hat{C} \hat{D} (\hat{1}-\hat{S})\hat{K}\hat{C} \hat{D} (\hat{1}-\hat{S})\hat{K}\hat{C} \hat{D} \cdots
\end{align} 
which can be contracted on both sides with matrix model observable. We can represent this diagrammatically as 
\begin{align}\label{string}
    \sum_{n=0}^{\infty} \frac{(\frac{1}{\hbar^n})}{n!} \vcenter{\hbox{\includegraphics[scale=0.35]{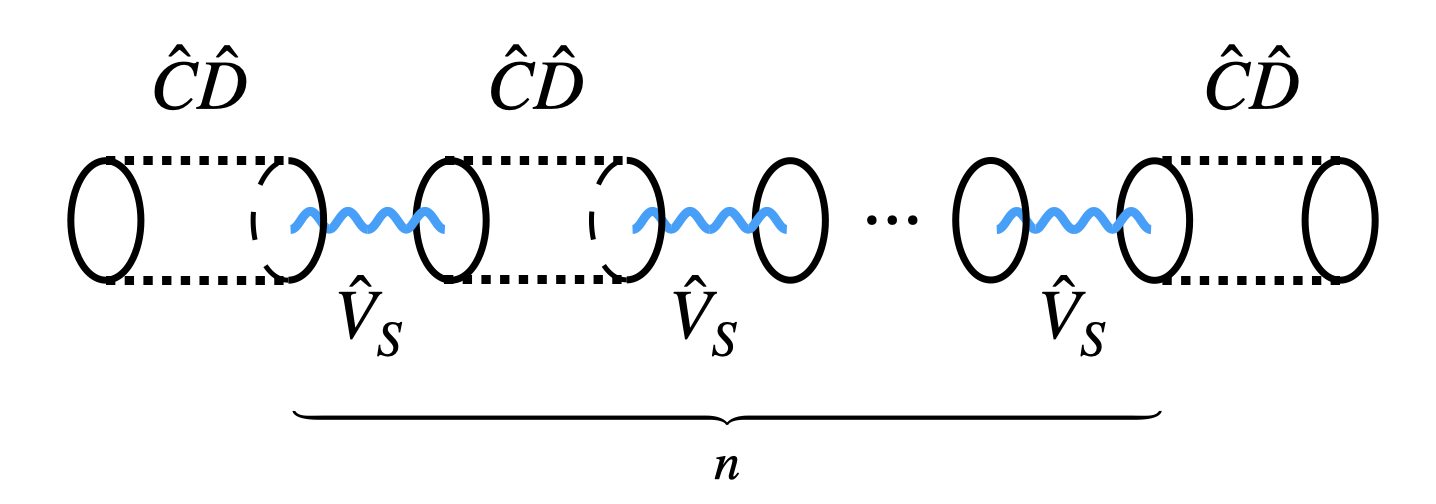}}}
\end{align}
Here, the dotted line on the $\hat C \hat D$ cylinders represents the fact that $\hat D$ implements a sum over $T$-transforms that implements the restriction  to integer spin. 
Note that since $\hat{K}\hat{C}=  \mathbbm{1}$, so these operators disappear from the string. The perturbative series \eqref{string} can then be re-written as 
\begin{align} \label{string2}
   \sum_{n} \frac{(\frac{2}{\hbar})^n}{n!} \cdots \hat{D}(\hat{1}-\hat{S})\hat{D}(\hat{1}-\hat{S}) \cdots
\end{align}
This operator is an  $SL(2,\mathbb{Z})$ symmetrizer that projects onto $SL(2,\mathbb{Z})$ invariant states on the torus.  To see this, consider the simplest case, where we restrict to the identity part $\hat{T}^{0}$
of $\hat{D}$.  Because the symmetry factors cancel the $\frac{1}{n!}$, we get a formal geometric series
\begin{align}
    \sum_{n}
    (\frac{2}{\hbar})^n  (\hat{1}-\hat{S})^n \sim \frac{1}{1- \frac{2}{\hbar}(\hat{1}-\hat{S})}.
\end{align}
In the $\hbar \to 0$ limit, this gives zero unless $\hat{S}=1$. (Note that since $\hat{S}$ is Hermitian and $\hat{S}^2=1$, it has eigenvales $\pm 1$). Thus formally the perturbative sum gives a projector 
\begin{align}
\frac{1+\hat S}{2}
\end{align}
onto $S$-invariant states on $\mathcal{H}_{T^{2}}\otimes \mathcal{H}_{T^{2}}$.  When we incorporate the effects of the non trivial $T$ transforms in \eqref{string2}, the combination of $S$ and $T$ generates a full $SL(2,\mathbb{Z})$ symmetrizer in the limit that $\hbar \to 0$.   We explain  this works in appendix \ref{app:SL2}.

As an application of this result, we can consider the computation of the disk density in which we incorporate the full effect of the perturbative expansion \eqref{string}.    Diagrammatically this involves a sum over an annulus connected to a string of the form \eqref{string} of arbitrary length, which is capped off by the insertion of the single trace potential $V_{S,1}$. 
\begin{align}
    \vcenter{\hbox{{ \includegraphics[scale=.4]{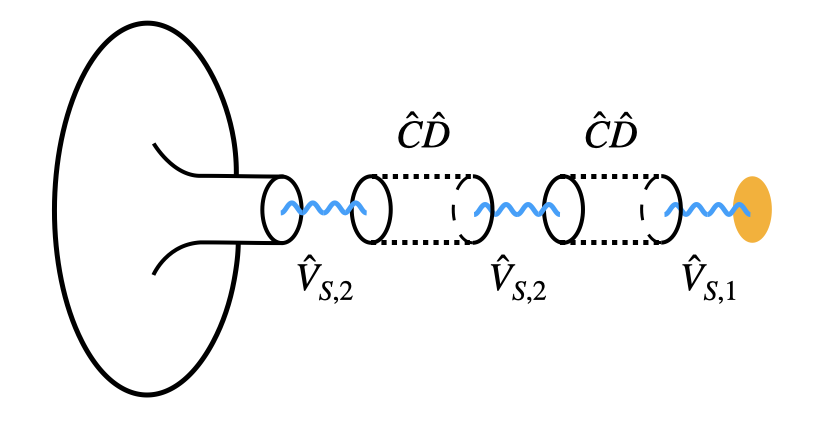}}}}
\end{align}

The resulting symmetrization over $SL(2,\mathbb{Z})$ transforms produces a modular invariant partition function: In the gravity interpretation, this  corresponds to the  $SL(2,\mathbb{Z})$  sum over BTZ black holes.

\subsection{Summary of the proposal}
\label{sec:proposal}
\subsubsection{Gravity interpretation of the tensor model}
Due to its relation to Chern Simons theory, 3d gravity has many aspects in common with TQFT\footnote{The precise relation is to Virasoro TQFT \cite{collier2023,collier2023virasoro}, which we review in section \ref{sec:checks}.}. In particular, it makes sense to insert Wilson lines in the gravity theory.  The representation labels $(P,\bar{P})$   of the Wilson lines give the mass and spin of the associated gravitational object: for dimensions below the black hole threshold, these are conical defects, while for dimensions above the threshold, these are microcanonical black hole states\footnote{We emphasize that these black hole states correspond to geometries that are not required to satisfy the Hawking mass-temperature relation. In pure gravity, these are always averaged over a microcanonical energy window.}.  

We now describe the 3d gravity interpretation of the ensemble in which the triple line Feynman diagrams are mapped to building blocks of 3-manifolds with Wilson lines inserted.   These are multi-boundary wormhole topologies in which each line of the Feynman diagram is mapped to a Wilson line, and each triple endpoint (corresponding to a $C_{ijk}$ in the potential) is mapped to a boundary sphere with 3 punctures.  

For example, in the tensor model,  the triple line propagator is mapped to $S^2 \times I$ with three Wilson line inserted:  this is a two-boundary wormhole that we refer to as the $C_{0}$ manifold.   The quartic vertices are mapped to 4-boundary wormholes, with a pillow or 6J pattern of Wilson lines inserted.  These manifolds are obtained by placing the  Wilson line network corresponding to the graphs \eqref{graphs} inside  $S^3$ and then removing a 3 ball around the junctions.

Similarly,  the triple line diagrams associated to the potential for one-point modular invariance is mapped to a two-boundary wormhole whose sphere boundaries have punctures corresponding to the indices of $C_{ijj}$.    The precise geometry is determined by  gluing the tori with Wilson lines that appear in the overlaps in \eqref{VS}, and then removing solid balls around the triple line junctions.     

We summarize the 3-manifold -- Feynman graph correspondence below:

\paragraph{Propagator} We refer to this as the $C_{0}$ manifold. 
\begin{align}
   -\hbar  |C_0(ijk)|^2 \to \vcenter{\hbox{\includegraphics[scale=0.35]{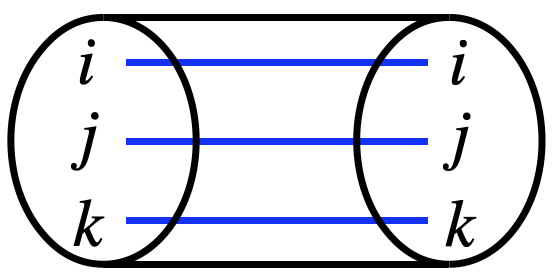}}}
\end{align}

\paragraph{Quadratic vertex}
There are two terms which are quadratic in $C_{ijk}$, and they are assigned different diagrams. 
\begin{align}
   \frac{1}{\hbar} \frac{\delta^{(2)}(P_i - P_j)}{|\rho_0(i) C_0(iik)|^2 } &\to \vcenter{\hbox{\includegraphics[scale=0.3]{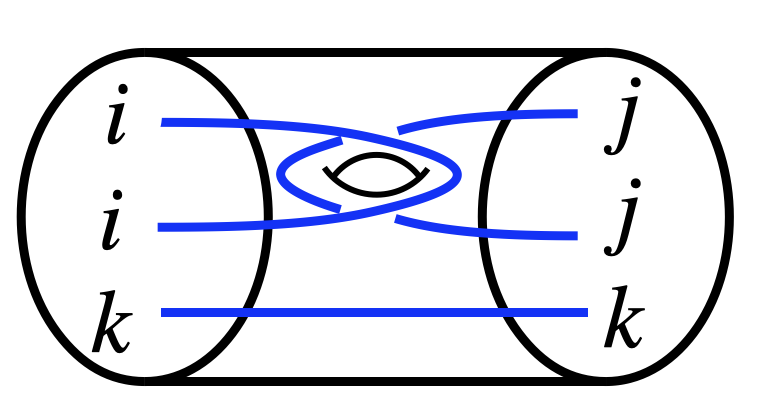}}} \\
   -\frac{1}{\hbar}\frac{|\mathbb{S}[P_k]_{P_i P_j}|^2}{|\rho_0(i) C_0(iik)|^2 } &\to \vcenter{\hbox{\includegraphics[scale=0.4]{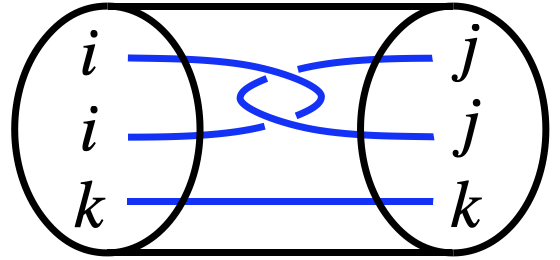}}}
\end{align}

\paragraph{Quartic vertex}
There are two terms at the quartic order in $C_{ijk}$. We refer to these as the Pillow and 6J manifolds.
\begin{align}
   -\frac{1}{\hbar} \frac{\delta^{(2)} (P_p-P_q)}{|\rho_0(p) C_0(12p) C_0(34p)|^2} &\to \vcenter{\hbox{\includegraphics[scale=0.4]{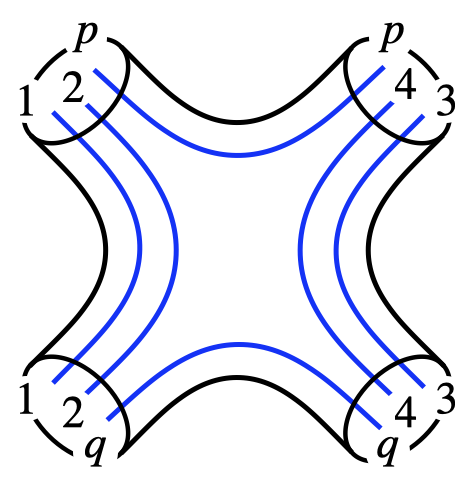}}} \\
     \frac{1}{\hbar} \frac{\left|\begin{Bmatrix}
    q & 4 & 1 \\
    p & 2 & 3
\end{Bmatrix}\right|^2}{|C_0(12p) C_0(34p) C_0(23q) C_0(14q)|^2} &\to \vcenter{\hbox{\includegraphics[scale=0.4]{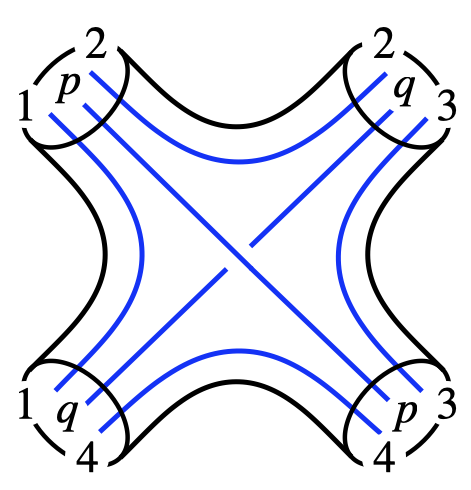}}}
\label{6Jman}
\end{align}
  Thus, the tensor model generates a sum over 3-manifolds by gluing together these basic building blocks, each of which are assigned to functions of weights given by the Virasoro crossing kernels. We will show in section \ref{sec:checks} that on a fixed manifold hyperbolic manifold, these Feynman rules agree with the 3d gravity partition functions as defined by Virasoro TQFT.

\subsubsection{Gravity interpretation of the matrix model}\label{sec:gravityMM}
To give a gravity interpretation of the matrix model, let us first consider the genus expansion of the doubled scaled random matrix $\Delta_{s}$ at fixed spin s, prior to the introduction of the constraint squared potential.  The genus expansion arise from the sum over 2D topologies determined by double line 't Hooft diagrams: in our ensemble this corresponds an $e^{-c}$ expansion.  This is because the double scaling limit produces an expansion in the level spacing near the edge of spectral density.   In this region, the spectral density \eqref{Icardy} is given by the Cardy formula $\rho_{0}= e^{\sqrt{\frac{c}{6} L_{0}}}$, and the edge of the spectrum starts at black hole threshold $L_{0}= \frac{c-1}{24}$.  This gives the $e^c$ scaling of the eigenvalue density, leading to an level spacing of $e^{-c}$. Note the since $\Delta_{s}$ is in the GOE ensemble, the genus expansion includes 2D non-orientable surfaces such as the cross cap. 

To incorporate the different spin sectors, we lift the genus expansion to a sum over 3-manifolds by attaching a circle over the 2D geometries.  For orientable 2D surfaces, we just attach the circle as a direct product, while for non-orientable surfaces, we fiber the circle in a way that makes the 3-manifold orientable.  This latter procedure was explained in \cite{yan2023torus}. 
Thus, the disk representing the spectral density becomes a solid torus $D^{2} \times S^1$ and annulus describing the density-density correlator becomes a $T^2 \times I$  wormhole:

\begin{align} \label{T2I}
\braket{\rho}_{\text{disk}} = \vcenter{\hbox{\includegraphics[scale=0.2]{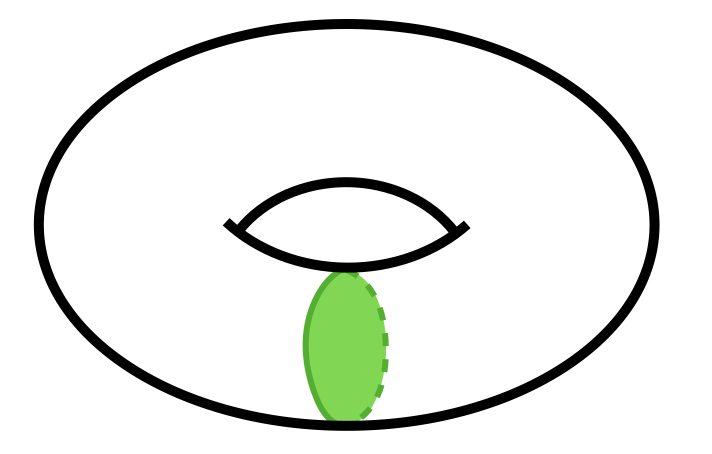}}},\qquad 
C =\braket{\rho \rho}_{\text{annulus}} = \vcenter{\hbox{   \includegraphics[scale=.25]{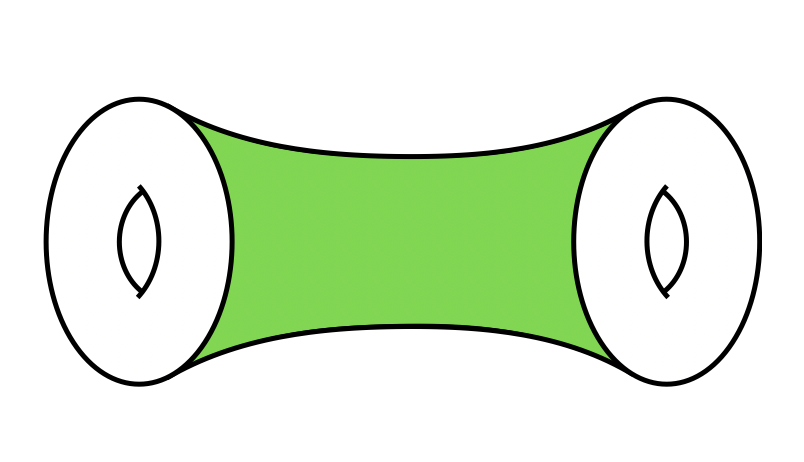}}},  
\end{align}
The disk density is the density of states for BTZ black holes, summed over $T$ transformations.   
Remarkably, the annulus \eqref{annulus} agrees with the 3d gravity partition function on the torus wormhole as computed by Cotler and Jensen\footnote{The agreement here is with Cotler Jensen's partition function without the sum over modular images.  This was computed in \cite{Cotler_2021} as a function of the modular parameters $\tau_{1},\tau_{2}$ on the boundary.  We checked that the Laplace transform agrees with \eqref{annulus}.}, with an extra factor of 2 due to the orientable lift of 2D nonorientable cylinders as described in \cite{yan2023torus}.  Note that while the annulus is determined on the matrix model side simply by the Vandermonde associated to the GOE ensemble, together with the fact that the different spin sectors decouple,  the 3D gravity computation of  $T^2 \times I$ partition function is highly nontrivial. This is due to the  fact that it is an off-shell geometry which does not admit a classical solution, and because of the presence of a  bulk mapping class group 
that must be gauged \cite{Cotler_2021} - we elaborate on the latter point in section \eqref{sec:mappingclass}.    
``Higher genus" $e^{-c}$ corrections to these manifolds are obtained by attaching $T^2 \times I$ handles to the geometry.  For the solid torus, this is illustrated below. 
\begin{align}
\braket{
\rho} =
\vcenter{\hbox{\includegraphics[scale=0.23]{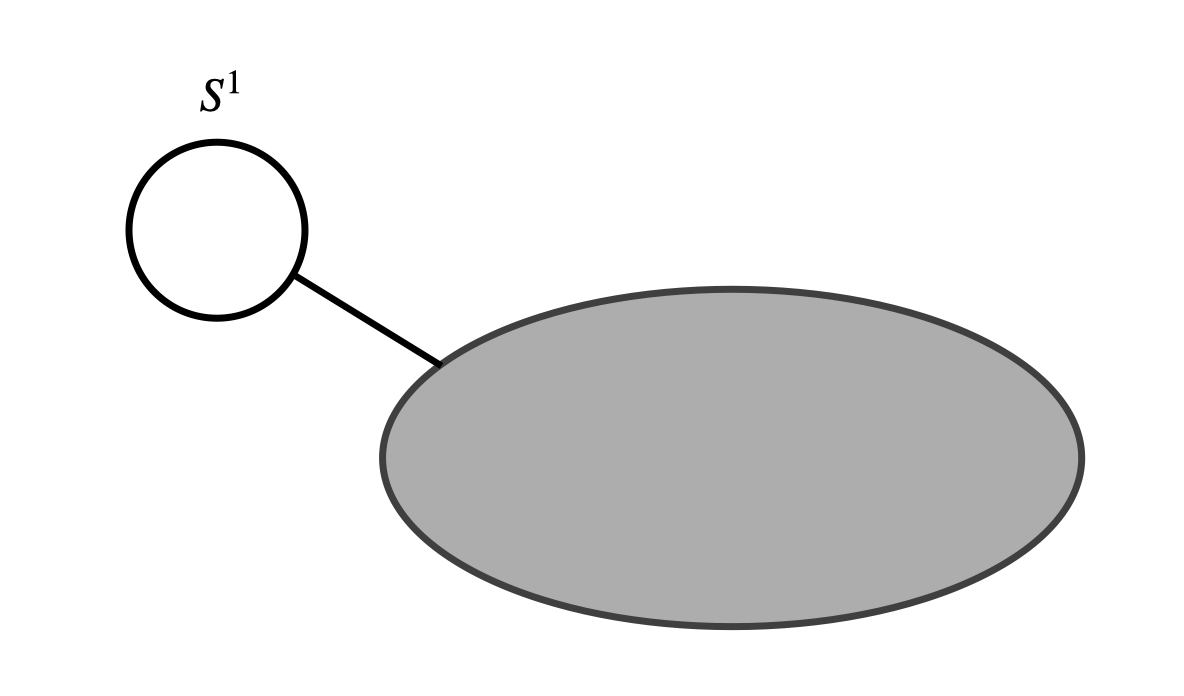}}} +
\vcenter{\hbox{\includegraphics[scale=0.2]{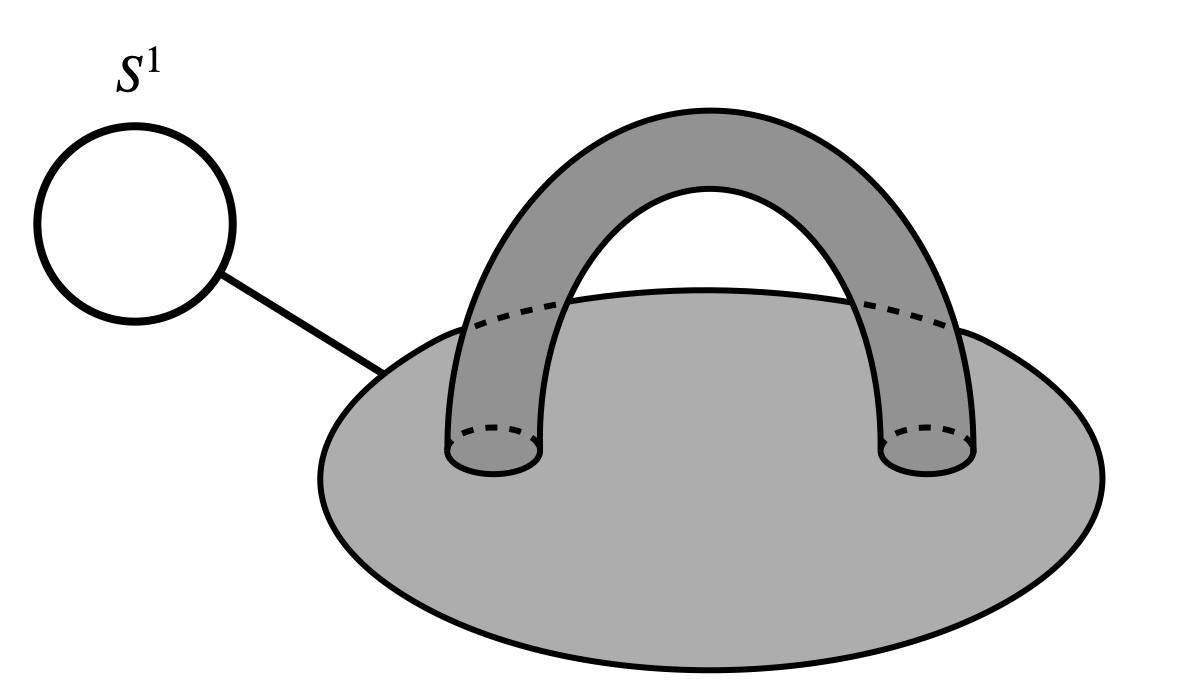}}} + \cdots
\end{align}
 The matrix model also predicts an answer for the gravity partition function on  multi-boundary geometries such as the 3-boundary torus wormhole shown below.   
\begin{align}\label{3toruswh}
     \braket{\prod_{i=1}^{3}\rho (\Delta_{i},s_{i}))}_{conn} = \vcenter{\hbox{   \includegraphics[scale=.25]{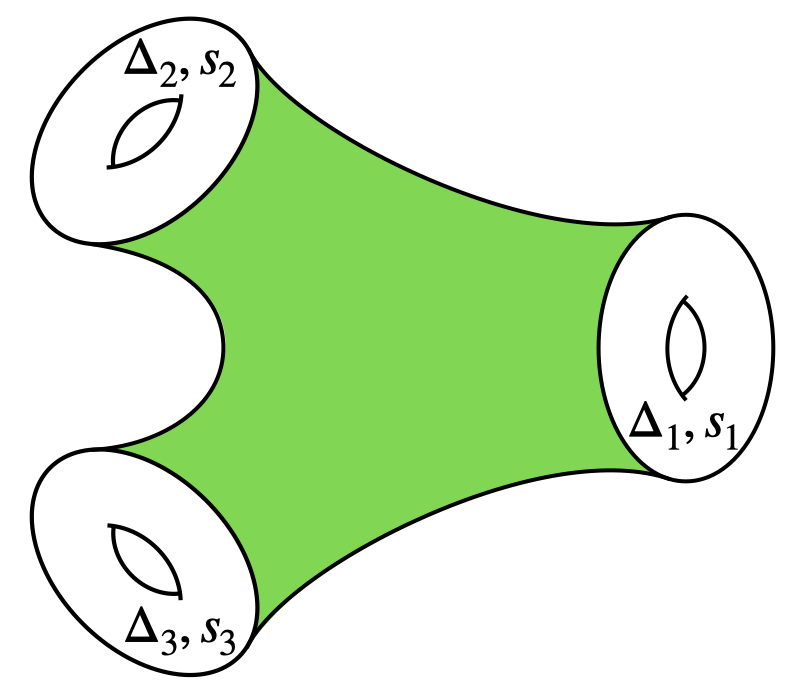}}} + \cdots
\end{align} 
Since the different spin sectors are independent, this connected correlator is only non zero when all the three spin variables are equal.  The remaining dependence on the conformal dimensions $\Delta_{i}$ can be obtained via topological recursion.   Indeed, we can make use of analogous computations in the SSS matrix model \cite{saad2019jt}.
  Specifically, we have
\begin{align}\label{rho3}
\braket{\prod_{i}\rho (\Delta_{i},s_{i}))}_{\text{conn}} = \begin{cases} \frac{8 \sqrt{2s}}{\pi^3 \sinh(2 \pi b \sqrt{s}) \sinh(\frac{2\pi \sqrt{s}}{b} )} \braket{\prod_{i}\rho (\Delta_{i})}_{SSS}  \quad \text{if  } s_{1}=s_{2}=s_{3} \equiv s \\ 
  0 \qquad \qquad \qquad  \text{   otherwise}
\end{cases},
\end{align} 
where $\braket{\prod_{i}\rho (\Delta_{i})}_{SSS}$ is the connected correlator computed via topological recursion as outlined in \cite{saad2019jt}.  In the 2D JT gravity langauge, $\braket{\prod_{i}\rho (\Delta_{i})}_{SSS}$ is obtained by gluing 3 cylinders\footnote{The cylinder amplitudes we want belong to the GOE, which differ from the GUE cylinder amplitudes in \cite{saad2019jt} by a factor of 2. } to a 3 holed sphere (with geodesic boundaries) to produce a pair of pants geometry.   The 3 holed sphere  depends on the spectral density near the edge of the cut, and we have included a spin dependent factor that captures the difference between our spectral density and the one in SSS.

The multiboundary torus wormhole is obtained by attaching a circle to the pair of pants, and this corresponds to introducing the spin dependence in \eqref{rho3}. 
The full gravity path integral as a function of the boundary moduli $(\tau_{i} ,\bar{\tau}_{i} )$ is computed by
Laplace and Fourier transforming the dimensions and spin variables respectively, and then attaching $\eta$ functions that capture the contributions of the descendants.  It would be interesting to check this against a gravity computation. 

Finally, more nontrivial 3-manifolds are produced when we introduce the constraint squared potential.   This is due to the vertex $V_{S}$ that  implement $SL(2,\mathbb{Z})$ transformations along bulk toroidal cuts.   Let us illustrate this in more detail for the case of a genus 1 correction to the disk: 
\begin{align}
\vcenter{\hbox{\includegraphics[scale=0.3]{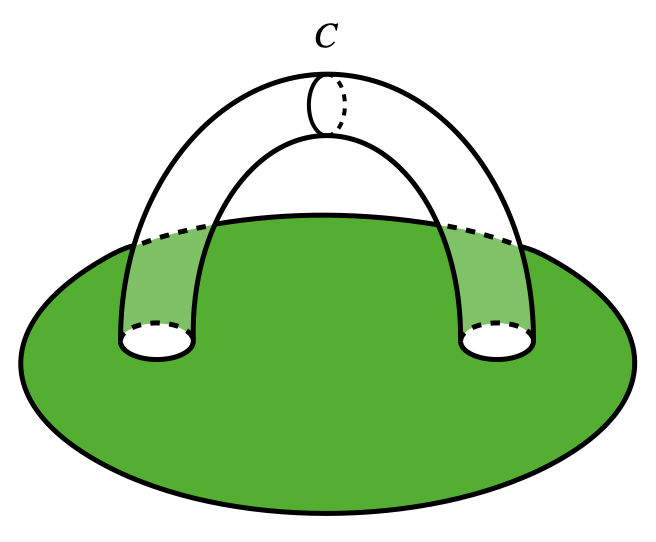}}} 
\end{align} 
In the 3D interpretation, the handle in this figure has the topology of  $T^2 \times I$, describing the time evolution of a torus.  
The matrix model expansion will insert a vertex $V_{S}$ into this propagation: 
\begin{align}
\vcenter{\hbox{\includegraphics[scale=0.3]{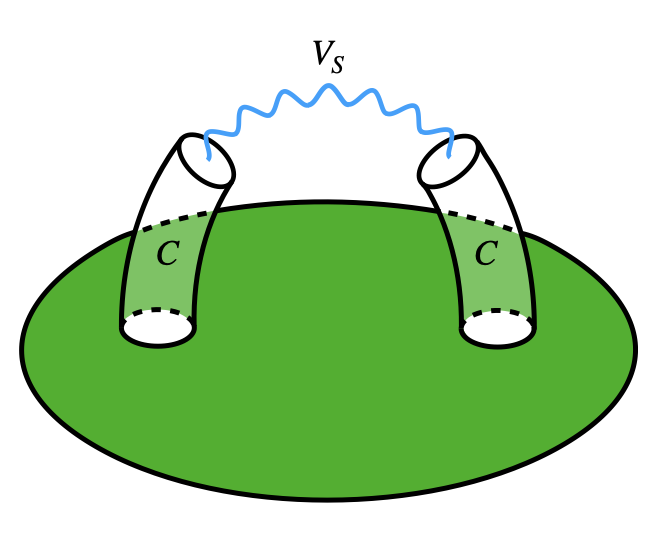}}}
\end{align}
In the 3D interpretation, $V_{S}$ produces two types of evolutions of the torus, shown in the RHS below: 
\begin{align} \label{CVC} \vcenter{\hbox{\includegraphics[scale=0.4]{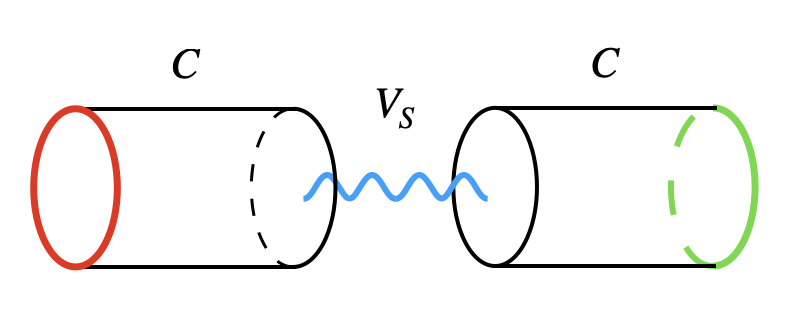} }} = \vcenter{\hbox{\includegraphics[scale=0.5]{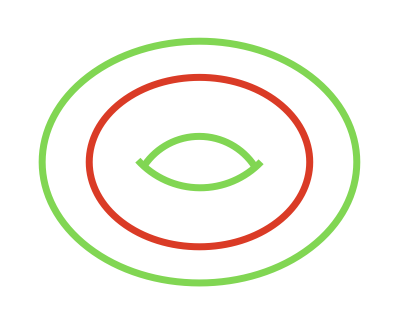} }} - \vcenter{\hbox{\includegraphics[scale=0.5]{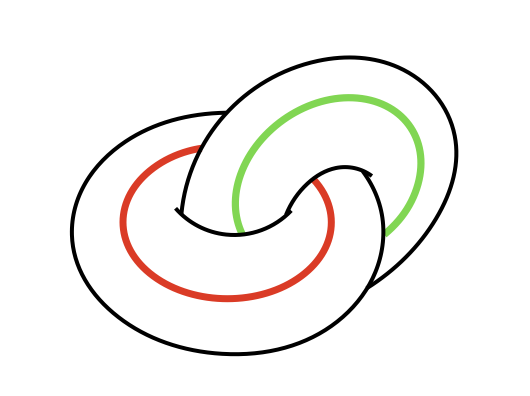} }}
\end{align}
 One is $C$, is just the original  $T^2 \times I$ propagator which we depicted as the expansion of a thin torus into a thicker one. The second term is $SC$, depicted by a $T^2$ evolution  (with $A$ cycle expanding) that is interrupted by an $S$ transformation: evolution beyond $S$ would proceed by contracting the $B$ cycle.  Thus the initial (red) and final (green) circle of the 2d cylinder on the LHS are linked in the 3d topology.   A general $SL(2,\mathbb{Z})$ transformation is produced when we combine $V_{S}$ with the sum over $T$ transformations needed to project on to integer spin. 

An important consequence of these $SL(2,\mathbb{Z})$ gluings in internal wormholes is the production of Seifert manifolds, which are circle bundles over a compact Riemann surface $\Sigma$, possibly with orbifold singularities.    This is an important class of manifolds because they are needed to cure a pathology due to the negative density of states produced by the sum over $SL(2,\mathbb{Z})$ BTZ black holes \cite{Maxfield_2021}.     To see how Seifert manifolds are produced, consider how a general matrix model manifold is constructed from the gluing of basic building blocks given by direct product of $S^1$ and a pair of pants geometry.   A nontrivial circle bundle over a smooth $\Sigma$ is obtained by gluing these building blocks together with $T$ transforms, which acts on the circle fibers.  To see that circle bundles over singular $\Sigma$ are also obtained, we start with such a manifold and excise the part of bundle sitting over small disks containing the  orbifold points.  The remaining part of the manifold can be decomposed into pairs of pants times $S^1$, glued together with $T$ transforms.   The excised region is a smooth 3-manifold, which must be just an $SL(2,\mathbb{Z})$ solid torus: we expect that the orbifold singularity corresponds to the shrinking of a particular cycle of the solid torus.  

\subsection{Gravity interpretation of the combined matrix-tensor integral}
Finally, there are nontrivial interactions between the matrix and tensor integral that are important for generating the full sum over 3-manifolds.   This is because the Feynman rules of the  tensor model dictates  that loops appearing in the triple lines diagrams are accompanied with a sum over the (approximate) CFT spectrum. This sum effectively inserts the density of states $\rho$ into the matrix integral.  When multiple Wilson loops are inserted, the matrix integral produces connected higher moments of $\rho$, which connects these loops with multi-boundary torus wormholes.  For example, the matrix model connect two internal Wilson loops with the full $T^2 \times I$ propagator (including the $SL(2,\mathbb{Z})$ sum) that captures the connected part of the second moment $\braket{\rho \rho}$.  In this way, the matrix integral can connect tensor model manifolds that would have been disconnected from the point of view of the tensor model integral alone. 

Finally, there are disconnected contributions to the moments of $\rho$, corresponding to the case when the matrix integral inserts a Cardy density on each Wilson loop inside a tensor model manifold.
 This implements toroidal surgery on the tensor model manifold, which is essential for obtaining an exact  match of our model with 3d gravity (see section \ref{SchwingerDyson} for a discussion of this point). In particular, surgery operations on Wilson lines provide a mechanism to construct 3-manifolds with arbitrary boundaries.  For this reason, we give an extended pedagogical introduction to surgery below, and explain its relevance to our model.

\subsubsection{Surgery on Wilson loops }
 Surgery on a manifold $M$ removes a tubular neighborhood of a knot, and then glues it back in with an $SL(2,\mathbb{Z})$-twist.  A simple example is an unknot C, shown below as a dotted line 
\begin{align} \label{solidtorusC}
\vcenter{\hbox{\includegraphics[scale=0.55]{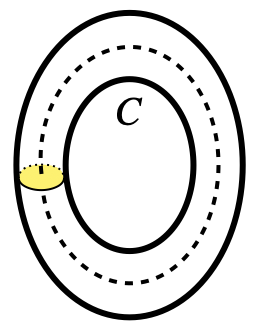} }} 
\end{align} 
(This figure shows a submanifold inside $M$, which is not drawn)
The tubular neighborhood has the topology of a solid torus $\mathcal{T}= D \times S^1$, so the excision creates a manifold $M \backslash \mathcal{T}$ with a toroidal boundary.  This is the torus shown in the figure above.  Gluing $\mathcal{T}$ back in with an $SL(2,\mathbb{Z})$ twist means that the boundary of  $\mathcal{T}$ and $M \backslash \mathcal{T} $ are glue together with an $SL(2,\mathbb{Z})$ transformation.  In particular, an $S$ twist would change the contractible A cycle of $\mathcal{T}\subset M$ into one that is non contractible inside $
\mathcal{T}$.   This is illustrated below
\begin{align}
\vcenter{\hbox{\includegraphics[scale=0.4]{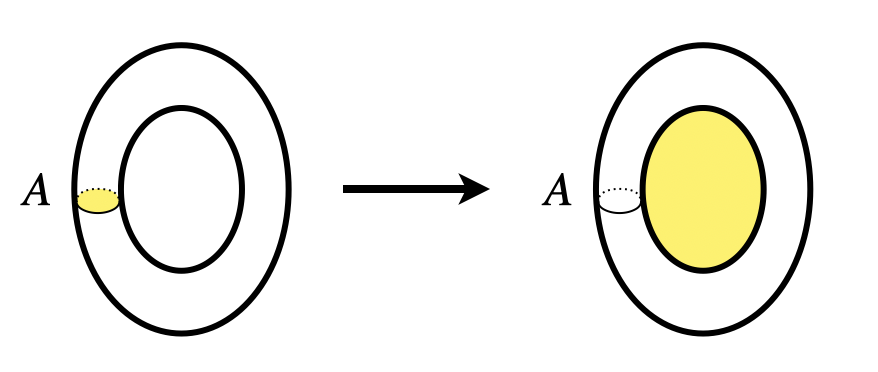} }}
\end{align}
Notice the same boundary $T^2$ contracts into a different solid torus before and after the surgery. 
If the modular parameter of the original $\mathcal{T}$ was $\tau$, the new solid torus has modular parameter $-1/\tau$.   In addition to $S$, we can also apply a $T$ transformation in the surgery. This corresponds to applying a Dehn twist, i.e. a $2\pi $  rotation on the A cycle of $\mathcal{T}$, before gluing it back in. 

Surgery can be used to eliminate $S^2$ handles on a manifold, which is a chunk of the manifold containing a non-contractible cycle. For example,  figure \ref{bulksurgery_dotted}  shows a manifold obtained from gluing together two 4-boundary wormholes.
\begin{align}\label{bulksurgery_dotted}
\vcenter{\hbox{\includegraphics[scale=0.4]{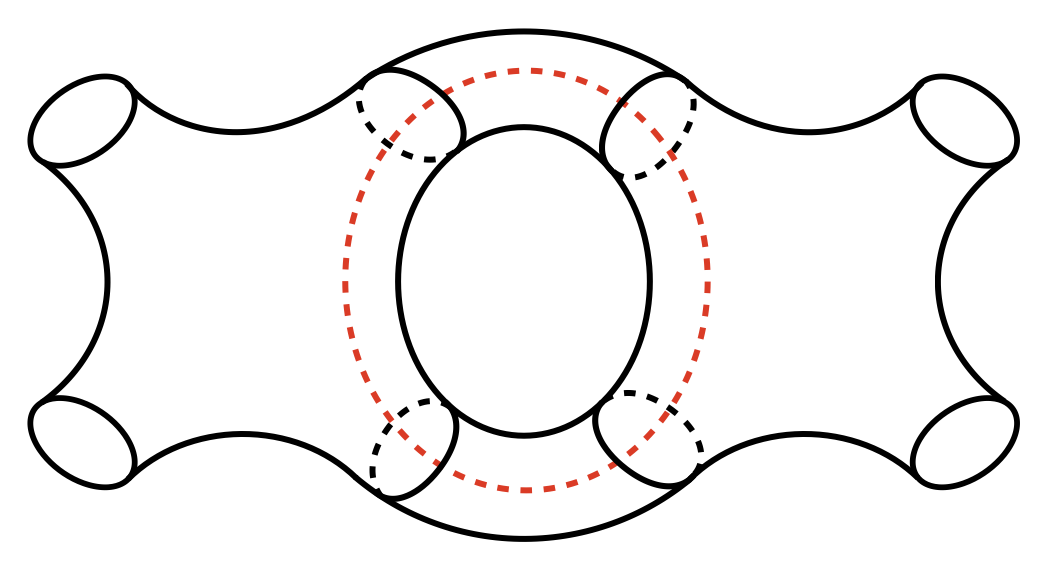} }} \end{align}
In this depiction of the manifoild, a cross section  (drawn as a circle) is really an $S^2$.  So the non contractible red loop in the figure lives inside an $S^{2} \times S^1$ submanifold.   Surgery on the red loop makes this cycle contractible.

\paragraph{TQFT representation}
There is a well known representation of surgery in TQFT \cite{Witten:1988hf}  which will provide a useful background for our discussion.  In a 3D TQFT, every 2D surface  is assigned a Hilbert space with a representation of the mapping class group, which are large diffeomorphisms of the surface.  For the torus, the large diffeos are exactly the $SL(2,\mathbb{Z})$ transformations used in surgery to glue along toroidal boundaries.  In the TQFT, a knot whose tubular neighborhood is removed in surgery is represented by the identity Wilson loop operator, and the $S$ transform is represented by the modular $S$ matrix.  Surgery along a Wilson loop determines a toroidal surface that separates a manifold $M$ into two halves.  We can then represent  the TQFT partition function $Z(M)$ as an overlap:
\begin{align}
    Z(M) = \braket{\Psi|\mathbbm{1} , \bar{\mathbbm{1}}}
\end{align}
where $\ket{\mathbbm{1} , \bar{\mathbbm{1}}} \in \mathcal{H}_{T^2}\otimes \mathcal{H}_{T^2}$ is the state on the torus Hilbert space corresponding to the empty solid torus $\mathcal{T}$, and $ \ket{\Psi}$ is a state representing $\mathcal{M} \backslash \mathcal{T} $.  Surgery creates a new manifold $M'$, whose partition function is 
\begin{align}\label{surgery1}
    Z(M') = \langle \Psi|S|\mathbbm{1} , \bar{\mathbbm{1}}\rangle 
\end{align}
There is a simple but useful way to re-interpret surgery in the TQFT langauge. We just write out how $S$ acts on $|\mathbbm{1} , \bar{\mathbbm{1}}\rangle$:
\begin{align}
 Z(M') = \braket{\Psi| \sum_{i} S_{\mathbbm{1} P_{i}} S_{\mathbbm{1} \bar{P}_{i}}|P_{i},\bar{P}_{i}} = \sum_{i} S_{\mathbbm{1} P_{i}} S_{\mathbbm{1} \bar{P}_{i}}
 \braket{\Psi|P_{i},\bar{P}_{i}}
\end{align}
The overlap $ \braket{\Psi|P_{i},\bar{P}_{i}}$ is the original manifold with a Wilson loop in the representation $(P_{i},\bar{P}_{i})$ inserted, so this represents a superposition of Wilson loops with weight $S_{\mathbbm{1} P_{i}} S_{\mathbbm{1} \bar{P}_{i}}$.  This is called an Omega loop \cite{Burnell_2010}, and we simplify the notation further by writing 
\begin{align}
    Z(M')= \braket{\Psi|\Omega},\qquad \ket{\Omega} \equiv \sum_{i} S_{\mathbbm{1} P_{i}} S_{\mathbbm{1} \bar{P}_{i}}\ket{P_{i} ,\bar{P}_{i}}
\end{align}
  Thus doing an S- surgery on the knot $C$ is equivalent to replacing it with the Omega loop.   This is shown below:
\begin{align}
\vcenter{\hbox{\includegraphics[scale=0.55]{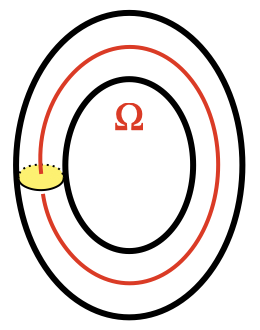} }}
\end{align}

\paragraph{Surgery on framed knots}
To include effects of $T$-transforms in
surgery, we must first frame each Wilson loop that makes up the Omega loop by thickening them into ribbons: we review this framing procedure in more detail in \ref{sec:ribbons}. Then the $T$ transform corresponds to the twisting of these ribbons, which adds phases in the superposition describing the  Omega loop.   For example,  doing $T^{n}S$ surgery on $M$ leads to the partition function
\begin{align}
    Z(M') &= \langle \Psi| T^n S|\mathbbm{1} , \bar{\mathbbm{1}}\rangle\nn
    &=  \sum_{P}  S_{\mathbbm{1} P_{i}} S_{\mathbbm{1} \bar{P}_{i}} \langle \Psi| T^n |P_{i},\bar{P}_{i} \rangle \nn
&= \sum_{P} S_{\mathbbm{1} P_{i}} S_{\mathbbm{1} \bar{P}_{i}} e^{ 2\pi n i  s_{i} }\langle \Psi  |P_{i}, \bar{P}_{i} \rangle
\end{align}
The phases provides a representation of the twists on each ribbon, determined by the spin.  In a theory with integer spins, these phases are trivial.  However, the topological expansion of 3d gravity will involve non integer spins, making these phases manifest.   
\paragraph{The ensemble representation of surgery }
As we have alluded to in section \ref{sec:mappingclass} in the text, in gravity the \emph{bulk} mapping class group is gauged.  This implies that the gravity Hilbert space on a bulk spatial slice is different than in TQFT.  Rather than implementing surgery via an $S$ matrix on $\mathcal{H}_{T^2} \otimes \mathcal{H}_{T^2}$, gravity implements surgery by attaching the disk density of the  matrix model integral.  The most trivial example of this is the computation of the average CFT partition function  (Note that the trace below is only over the primaries)
\begin{align} \label{trqL}
\braket{\Tr q^{L_{0}} \bar{q}^{\bar{L}_{0}} }= \sum_{i} \braket{q^{P_{i}^{2}+\frac{c-1}{24}} \bar{q}^{\bar{P}_{i}^{2}+\frac{c-1}{24}}}.
\end{align} In the 2D language, this observable is represented by a circle, and the leading order matrix model integral fills in this circle with a disk, corresponding to a dense covering of 't Hooft graphs.    In terms of formula, this replaces the sum over states with an integral over the Cardy density: 
\begin{align} \label{matrixS}
\braket{\Tr q^{L_{0}} \bar{q}^{\bar{L}_{0}}}=  \int d P \,\,
 S_{\mathbbm{1} P} S_{\mathbbm{1} \bar{P}} q^{P^{2}+\frac{c-1}{24}} \bar{q}^{\bar{P}^{2}+\frac{c-1}{24}}
\end{align} 
If we multiplied both sides with $\left|\frac{1}{\eta(q)}\right|^2$, this becomes a superposition of torus characters that is exactly the insertion of an Omega loop.  Thus we could view ``filling in the disk" as a surgery operation on a dual boundary torus.  This corresponds to the fact that the BTZ black hole related to thermal $\text{AdS}_3$ ($\text{TAdS}_3$) by a modular transformation. 
These three equivalent descriptions are shown below: 
\begin{align}
   \vcenter{\hbox{\includegraphics[scale=0.4]{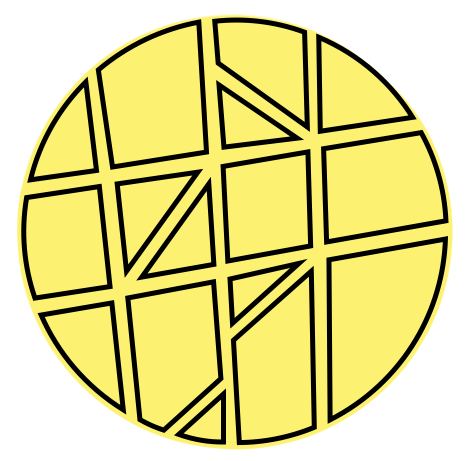} }}  = \qquad \vcenter{\hbox{\includegraphics[scale=0.55]{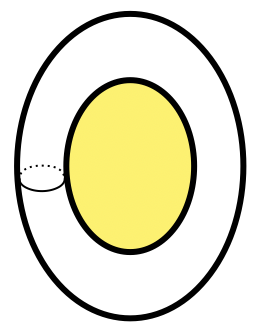} }} 
   =\qquad \vcenter{\hbox{\includegraphics[scale=0.55]{figures/fin_omegacycle.png} }} 
\end{align}
Next, consider internal loops that appear in the triple line diagram of the tensor model.   These loops are also observables in the matrix model. The leading order matrix integral once again attaches a Cardy density to these loops, which are identified as Wilson loops in the 3-manifold interpretation.  This effectively inserts an Omega loop that produces  a bulk surgery.  This is illustrated below.
\begin{align} \label{bulksurgery}
    \vcenter{\hbox{\includegraphics[scale=0.3]{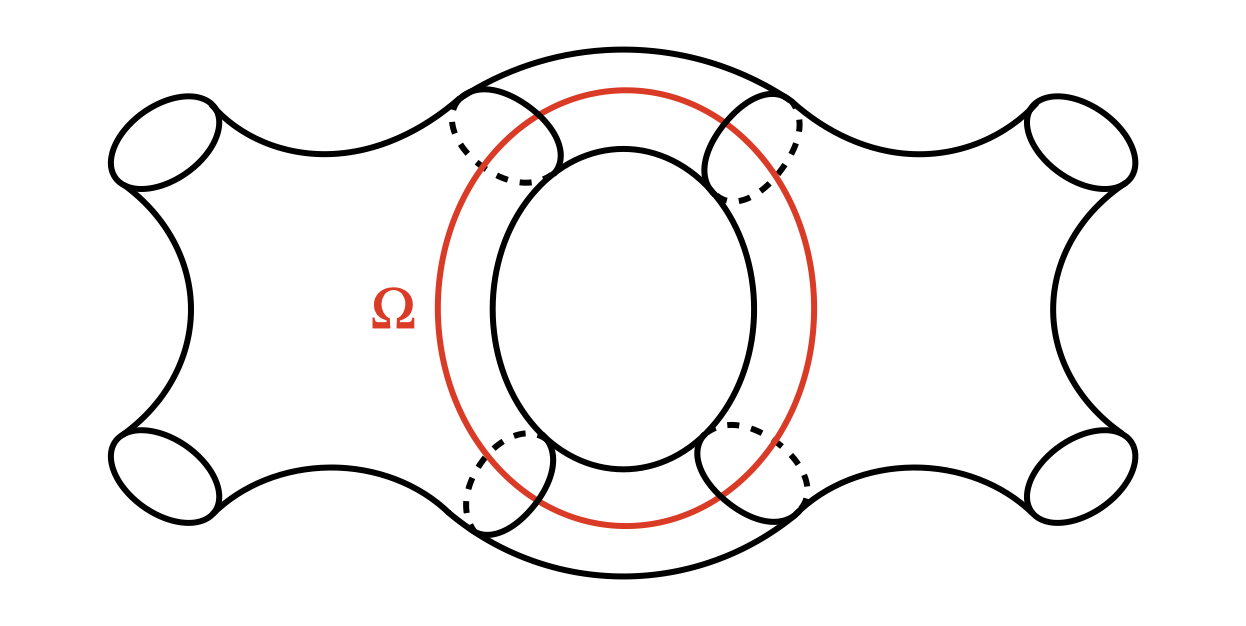} }} \rightarrow \vcenter{\hbox{\includegraphics[scale=0.3]{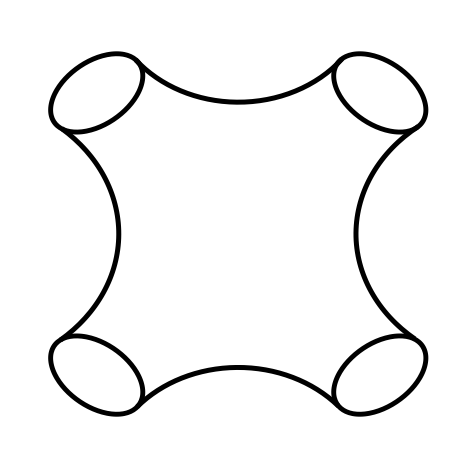} }}
\end{align}
To check that our surgery interpretation is correct, we must check that the amplitudes we assign to 3-manifolds do indeed satisfy the surgery relations.   Borrowing the  language of TQFT, we must show that our mapping between manifolds and amplitudes is ``functorial".   Since the gravity amplitudes are integrals of Virasoro crossing kernels, these surgery relations must correspond to integral identities of these kernels.  We check this in section \ref{sec:checks}.

\subsubsection{Surgery on Wilson lines and boundary manifolds }\label{sec:boundarysurgery}
 There is another type of surgery that can be viewed as a $\mathbb{Z}_{2}$ quotient of the surgery on knots described above.   Here we excise a tubular neighborhood of an open curve $C_{open}$ on a 3-manifold that that ends on two boundaries.  We can view the boundaries as the $Z_{2}$ fixed point that arise when we quotient the solid torus in \eqref{solidtorusC} by a reflection about the plane containing its contractible  A cycle.  We then glue in a solid torus that has been cut in half like a bagel. This manifoid $\mathcal{T}_{open}$ is the quotient of the dual torus by a reflection about the plane containing its non contractible B cycle. This is illustrated below. 
\begin{align}\label{opensurgery}
\vcenter{\hbox{\includegraphics[scale=0.4]{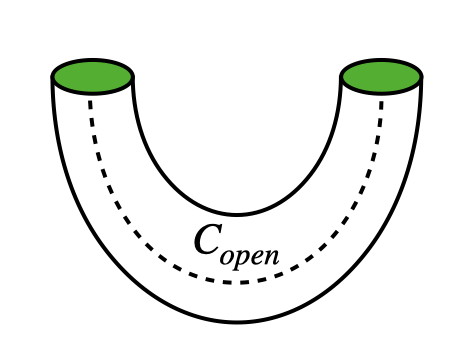} }} \xrightarrow[]{\text{surgery}} \vcenter{\hbox{\includegraphics[scale=0.4]{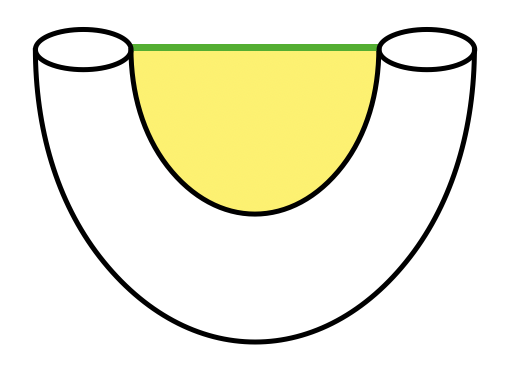} }} = \vcenter{\hbox{\includegraphics[scale=0.4]{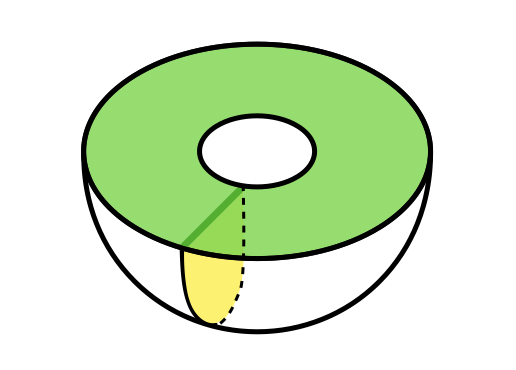} }} =\mathcal{T}_{open}
\end{align}
The green boundaries corresponds to the fixed point of the $Z_{2}$ quotient: note that surgery changes the green boundary's  topology from two disconnected disks to one connected annulus.  The latter is a boundary that has been created around the curve $C_{open}$. Below is another topological equivalent picture which illustrates this excision.
\begin{align}\label{opensurgery2}
\vcenter{\hbox{\includegraphics[scale=0.4]{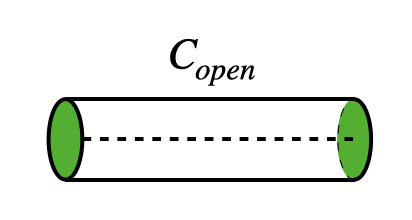}}} \xrightarrow[]{\text{surgery}} \vcenter{\hbox{\includegraphics[scale=0.4]{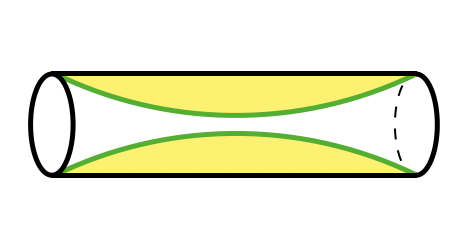} }} = \quad \underbrace{\vcenter{\hbox{\includegraphics[scale=0.4]{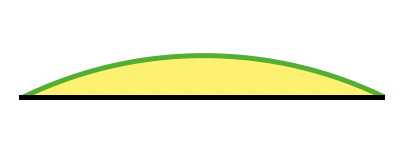}}}}_{\text{matrix model diagram}} \, \times \, S^1
\end{align}

\paragraph{TQFT interpretation}
As in the case of surgery on closed loops, surgery on an open segment has an  interpretion as the insertion of particular superposition of Wilson lines  into a TQFT- we might refer to this as the Omega line.  In this case the weight of the superposition also includes conformal blocks that depend on boundary moduli.   For example, consider the following TQFT identity  
\begin{align} \label{TQFTos}
\int d^{2}P_{p} S_{\mathbbm{1}P_{p}} S_{\mathbbm{1}\bar{P}_p} \left| \vcenter{\hbox{\includegraphics[scale=0.2]{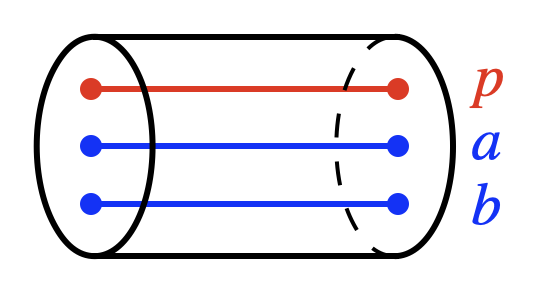}}}\right|^{2}\left| \vcenter{\hbox{\includegraphics[scale=0.2]{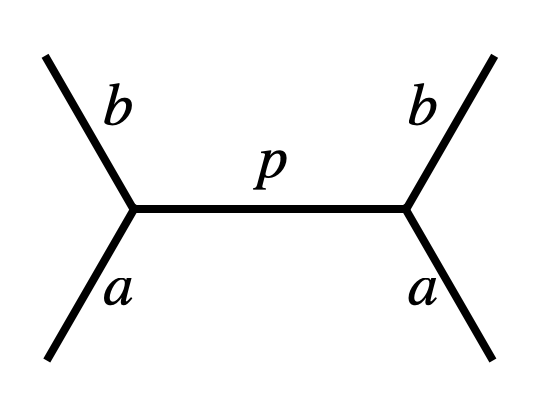}}}\right|^2  = \left| \vcenter{\hbox{\includegraphics[scale=0.2]{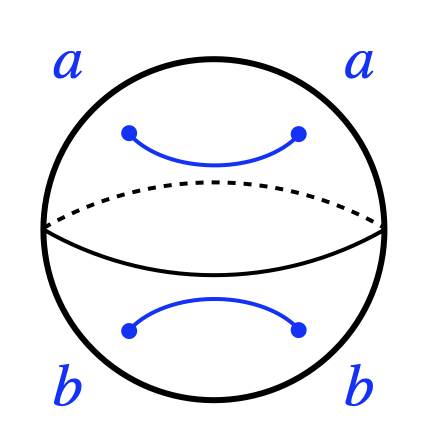}}}\right|^{2}(z)
\end{align}
 On the left hand side we have inserted a superposition of Wilson lines (in red) on the $C_{0}$ manifold, weighted by the Cardy density and the 4 point conformal block on the sphere: the latter has the cross ratio $z$ as the modulus. We interpret this superposition as an Omega line.   On the RHS we have the TQFT path integral on a solid 3-ball with two Wilson lines that end on the boundary.  To go between the topologies on the left and right hand side, we must remove a solid cylinder neighborhood of the $\Omega$ line (in red), which effectively connects the two boundary spheres.   But this is exactly what is accomplished by the surgery operation in \eqref{opensurgery}.   

The TQFT identity \eqref{TQFTos} can be understood as follows. The TQFT path integrals on both sides of this equation are related to Virasoro conformal blocks. On the LHS we have
\begin{align}\label{C0Z}
Z \left( \vcenter{\hbox{\includegraphics[scale=0.2]{figures/fin_cylinder1.png} }} \right) =|C_{0}(abp)|^2 
\end{align}
So substituting this gives
\begin{align}
\int d^{2}P_p S_{\mathbbm{1}P_p} S_{\mathbbm{1}\bar{P}_p} |C_{0}(abp)|^2   \left| \vcenter{\hbox{\includegraphics[scale=0.2]{figures/fin_4ptabp1.png}}}\right|^2= \left| \vcenter{\hbox{\includegraphics[scale=0.25]{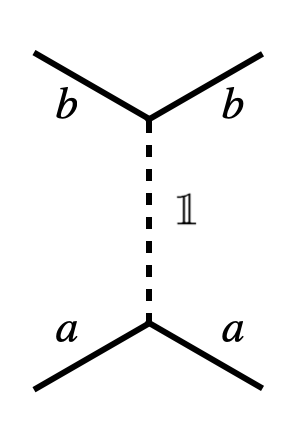}}}\right|^2,
\end{align}
where we have used the fact that $S_{\mathbbm{1}\bar{P}_p} C_{0}(abp) $ is the identity crossing kernel for the 4 point block.  This gives equation \eqref{TQFTos} upon substituting
\begin{align} 
Z \left( \vcenter{\hbox{\includegraphics[scale=0.2]{figures/fin_sphereab.png} }}\right) = \left| \vcenter{\hbox{\includegraphics[scale=0.25]{figures/fin_4ptid.png}}}\right|^2
\end{align} 

Notice that this type of surgery changes the boundary manifold from two 3-punctured spheres into a single 4 -punctured sphere. 
\paragraph{The ensemble representation}
To illustrate how the matrix-tensor model integral implements surgery on open curves, consider the average of the 4-point function on the sphere: 
\begin{align}
    \sum_{k} \braket{C_{abk}C_{abk}^* }_{0} \left| \vcenter{\hbox{\includegraphics[scale=0.25]{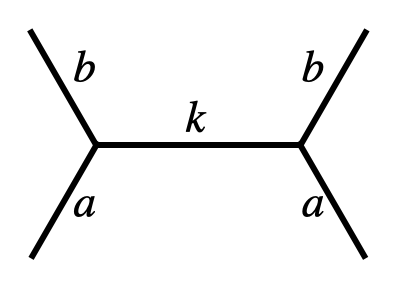} }}\right|^2= f(\hbar) \int d^{2}P_p S_{\mathbbm{1}P_p} S_{\mathbbm{1}\bar{P}_p} |C_{0}(abp)|^2 \left| \vcenter{\hbox{\includegraphics[scale=0.2]{figures/fin_4ptabp1.png}}}\right|^2 
\end{align}
Here $\braket{}_{0}$ refers to the leading order\footnote{This is leading order in the genus expansion for the matrix, corresponding to the half disk. For the tensor integral, this leading order averaging produces the expected manifold in \eqref{C0Z} up to a function $f$ of  $\hbar$.   We will explain how to deal with the $\hbar$ factors in section \ref{SchwingerDyson}. } matrix and tensor integral, which introduces the Cardy density and $C_{0}(abp)$.  Comparing  with the LHS of \eqref{TQFTos}, we see that the ensemble average effectively inserts an Omega line that implements surgery on the bulk $C_{0}$ manifold, turning it into a solid ball with two Wilson lines as in $\eqref{TQFTos}$.  
\paragraph{Boundary manifolds} Here we explain the relation between surgery on open curves and the insertion of boundary manifolds in the gravity theory.  
From the point of view of the boundary CFT, each insertion of $C_{ijk}$ produces a thrice punctured sphere (or equivalently a pair of pants), and contracting the indices of these OPE coefficients with the appropriate conformal blocks glues the punctures together to create an arbitrary 2-manifold. This ``plumbing" procedure creates a general CFT partition function $Z_{CFT}(\Sigma,q_{i})$ on a Riemann surface $\Sigma$ with modulus $q_{i}$, which is an observable that we insert into the matrix-tensor integral. Our formulation of surgery on Wilson lines gives an averaged description of CFT sewing in the bulk gravity theory.   In the gravity description, each
$C_{ijk}$ inserts a 3-punctured sphere boundary for some bulk 3-manifold such as the $C_{0}$ with Wilson lines inserted.   Averaging the plumbing construction produces surgery on the Wilson lines by excising tubular neighborhoods around them.  

To illustrate this interpretation, consider the following genus two CFT observable 
\begin{align}\label{genus2obs}
\sum_{ijk} \braket{ C_{ijk}C_{ijk }}  \mathcal{F}(h_i, h_j, h_k | q) \bar{\mathcal{F}} (\bar{h}_i, \bar{h}_j, \bar{h}_k | \bar{q})
\end{align} 
One diagram that contributes to this expectation value involves a $C_{0}$ manifold on which we perform surgery on 3 Wilson lines labelled by $i,j,k.$  This gives a handlebody bounded by a genus two surface
\begin{align}\label{genus2}
 \iiint & d^2 P_i d^2 P_j d^2 P_k |S_{\mathbbm{1} P_i} S_{\mathbbm{1} P_j} S_{\mathbbm{1} P_k}|^2 \vcenter{\hbox{\includegraphics[scale=.2]{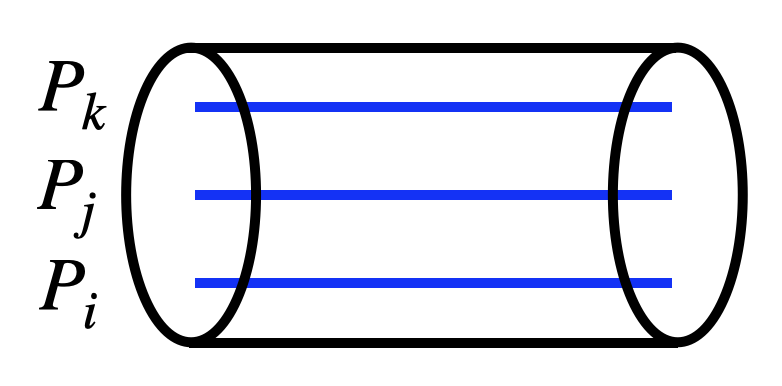}}} \mathcal{F}(P_i, P_j, P_k | q) \bar{\mathcal{F}} (\bar{P}_i, \bar{P}_j, \bar{P}_k | \bar{q})  \nn
&= \vcenter{\hbox{\includegraphics[scale=0.23]{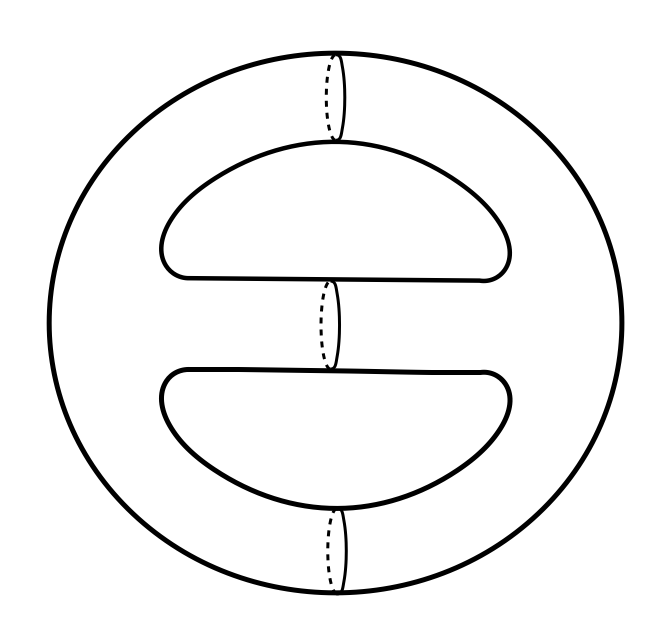}}} \, ,
\end{align}
where $\mathcal{F}(P_i, P_j, P_k | q)$ is the holomorphic conformal block of the theta graph. Note that in this picture handlebody is filling in the ``outside" of this surface. 

There are also contributions to the observable \eqref{genus2obs} from other topologies, which implement a more general kind of surgery.   In this more general scenario, instead of gluing in a half disk as in \eqref{opensurgery}, \eqref{opensurgery2}, one glues in a half disk with a hole representing a toroidal boundary. This is illustrated below
 
\begin{align}
    \vcenter{\hbox{\includegraphics[scale=0.4]{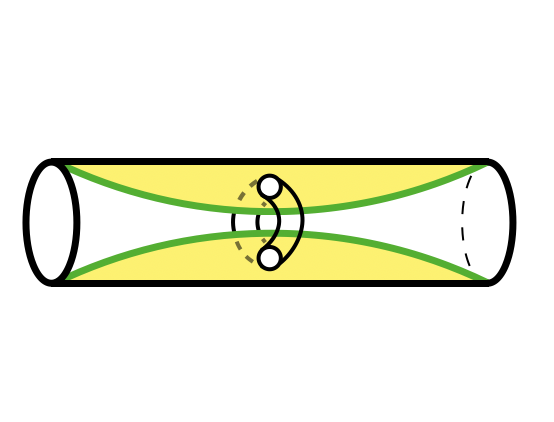} }} = \quad \underbrace{\vcenter{\hbox{\includegraphics[scale=0.4]{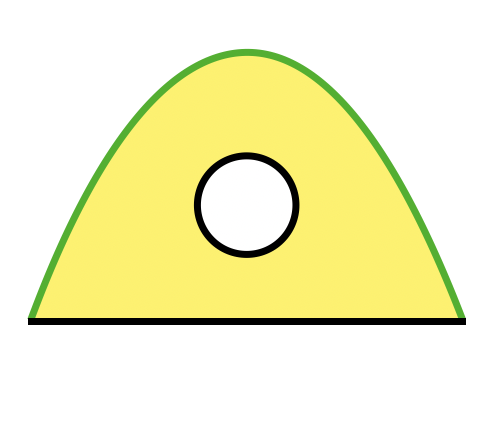}} }}_{\text{matrix model diagram}} \, \times \, S^1
\end{align}
On this toroidal boundary, we can then perform arbitrary $SL(2,\mathbb{Z})$ surgery by gluing in arbitrary power of the matrix model gadget \eqref{CVC}.    
\section{Checks of the 3d gravity interpretation}\label{sec:checks}
In the previous section, we gave a prescription for how the perturbative expansion of the CFT ensemble generates partition functions on 3-manifolds.  We now perform two types of consistency checks of the proposal.  The first type has to do with the internal consistency of our mapping between Feynman diagrams and 3-manifolds: for example, we check that surgery operations are properly represented, so that manifolds that are equivalent by surgery relations are assigned the same partition function.  We also give an argument for why the model produces all 3-manifolds. The second type of check compares our results with existing formulations of 3d gravity \cite{collier2023, Collier:2024mgv,Eberhardt:2022wlc}.  This is more subtle, because there is no universally accepted definition of pure $\text{AdS}_3$ quantum gravity on all manifolds.    

For hyperbolic manifolds, on which there are classical saddles, we can compare our computations with semi-classical $\text{AdS}_3$ gravity, which does have an accepted definition.   For example, up to an overall coefficient depending on $\hbar$, the gravity partition function on the $C_{0}$ manifold gives the Liouville 3-point function \cite{Collier_2020, Chandra_2022}:
\begin{align}
    Z_{\text{grav}}(C_{0}) \sim  |C_0(ijk)|^2 
\end{align}
 When $i,j,k$ label Wilson lines corresponding to massive geodesics,
\cite{Chandra_2022} showed that the large $c$ limit of $|C_0(ijk)|^2$  agrees with the on-shell action of classical gravity on the $C_{0}$ manifold.  For particles above the threshold, they found that classically, the $C_{0}$ manifold is replaced with a ball containing  2 Wilson lines that end on its boundary: this matches with the Wilson line surgery described in  \eqref{opensurgery2} which arises from averaging in the CFT ensemble.  This realizes the idea that semi-classical gravity performs a micro-canonical averaging for all states above the blackhole threshold.   Similar checks with classical gravity results can be done for 3-manifolds produced by the Gaussian part of our ensemble.   This includes higher genus, two-boundary wormholes, with Wilson lines crossing the wormhole. On the other hand, the approximate CFT ensemble also introduces non-Gaussianities due to the quartic terms in the tensor model potential.  These correspond to 4-boundary wormholes, whose classical on shell action has not yet been computed.

At the quantum level, we can obtain a check of our finite $c$ partition functions by comparing them to a definition of 3d gravity in terms of Virasoro TQFT \cite{collier2023}. This is a nonchiral TQFT that arises from a modification of  the integration contour for $PSL(2,\mathbb{R}) \times PSL(2,\mathbb{R} ) $ Chern Simons that accounts for features special to gravity such as the invertibility of the Vielbein  \cite{Mikhaylov_2018}.  It was originally formulated by Verlinde \cite{Verlinde:1989ua}, and  studied extensively by Kashaev and Anderson in terms of a state sum model in \cite{andersen2012tqft} \cite{andersen2013new}.   In \cite{Mertens:2022ujr,Wong:2022eiu}, this theory was formulated as an extended TQFT based on the representation category of the quantum group SL$_{q}(2,\mathbb{R})^+$, and applied to define bulk Hilbert space factorization and compute bulk entanglement entropies in AdS$_{3}$ gravity.  More recently, the ``modular functor" formulation of this TQFT \cite{teschner2007analog}  was further developed in \cite{collier2023, Collier:2024mgv} and applied to various computations, where it was reformulated as Virasoro TQFT.

In terms of the amplitudes $Z_{vir}$ of the Virasoro TQFT,  the ``Virasoro" 3d gravity partition functions takes the form 
\begin{align}
    Z^{\text{Virasoro}}_{\text{grav}}= \sum_{\text{topologies}}  |Z_{\text{Vir}} |^{2}
\end{align}
Below, we will explain why our gravity theory at fixed topology agrees with Virasoro TQFT for a large class of manifolds. The volume conjecture  \cite{collier2023,collier2023virasoro,kashaev1996hyperbolic} would then imply a match with semi-classical results in the large $c$ limit.   However, crucially, for non-hyperbolic manifolds such as  $T^2 \times I$, our gravity partition function does \emph{not} agree with Virasoro TQFT.   Instead, as we alluded to earlier,  our $T^2 \times I$ partition function matches with the gravity computation of Cotler and Jensen \cite{Cotler_2021}.  To give a self contained presentation, we will review aspects of VTQFT below and re-derive various formulas in a manner convenient for our discussion.

\subsection{Hyperbolic manifolds and Virasoro TQFT }
\paragraph{The modular functor for VTQFT}
For tensor model manifolds, the relation between our gravity theory and Virasoro TQFT can be understood most naturally via the definition of a TQFT as a modular functor. 
In this formulation, a 3D TQFT is defined by an assignment 
\begin{align}\label{ZH}
    Z:\Sigma_{g,n} \to \mathcal{H}_{\Sigma_{g,n}}
\end{align}
of a Hilbert space to each surface $\Sigma_{g,n}$ of genus g and n punctures, together with a representation of the mapping class group
\begin{align}
  \text{Map}(\Sigma_{g,n})\equiv \frac{\text{Diff} ( \Sigma_{g,n})}{\text{Diff}^{0}(\Sigma_{g,n})},
\end{align}
 which are diffeomorphisms of the surface modulo diffeomorphisms connected to the identity.  
Concretely, these mapping class group elements are crossing transformations on $\Sigma_{g,n}$.   For example, on the torus they are $SL(2,\mathbb{Z})$ transformations generated by Dehn twists and $S$-transform which exchanges the two cycles, and on the 4 punctured sphere these are changes in the slicing of the sphere as shown in \eqref{4ptcross}.   According to Moore-Seiberg, a representation of the mapping class group on any surface is generated by the F and R matrix describing fusion and braiding, together with the generator of $S$ and $T$ transformations on the punctured torus: 
\begin{align}\label{Crossmat}
\mathbb{F},\mathbb{R}, \mathbb{S}[i], \mathbb{T}
\end{align} 

The two ingredients \eqref{ZH},\eqref{Crossmat} are sufficient to  compute the TQFT amplitude on any  3-manifold with Wilson lines. This is because any 3-manifold $M$ admits a Heegaard splitting 
\begin{align}
    M= M_{+}\cup_{g}M_{-}
\end{align}
which decomposes $M$ into two halves\footnote{ For a closed manifolds $M$, $M_{\pm}$ are handlebodies.  When $M$ has boundaries, $M_{\pm}$ are in general compression bodies, which are just cobordisms between the gluing surface $\Sigma_{g,n}$ and an outer surface that is possibly disconnected.}  $M_{+},M_{-}$ that are glued together along a surface  $\Sigma_{g,n}$ via a mapping class group element $g \in  \text{Map}(\Sigma_{g,n})$.  In the TQFT, these halves are assigned to vectors on $\ket{M_{\pm}}\in \mathcal{H}_{\Sigma_{g,n}}$ and gluing the two halves together corresponds to the overlap
\begin{align}\label{MUM}
Z(M)= \braket{M_{+} | U(g)|M_{-}} ,
\end{align}
where $U(g)$ is a representation of $g$ generated by the crossing matrices \eqref{Crossmat}.

The discussion above is well known for TQFT associated to rational conformal field theories, which have a finite number of primaries:  $\mathcal{H}_{\Sigma_{g,n}}$  is then given by the finite dimensional Hilbert space of conformal blocks associated to the RCFT.   Remarkably, Teschner \cite{teschner2003liouville} showed that the space of Liouville conformal blocks endowed with the Verlinde inner product \eqref{Verlinde} provides an infinite dimensional generalization of a modular functor, in which the crossing matrices \eqref{Crossmat} are replaced by the Virasoro crossing kernels.   This defines the Virasoro TQFT.

\paragraph{VTQFT and the tensor model}
In the construction of the CFT ensemble in section \ref{ensembleDef}, we explicitly introduced elements of the modular functor associated to VTQFT.  For example, the crossing kernels $\mathbb{F},\mathbb{S}[i]$ appeared in our constrained square potential.  The braiding  kernel $\mathbb{R}$ is needed to formulate a consistent set of diagrammatic rules for the tensor model when the spins are no longer integers.  In particular, we will assign a ribbon junction to $C_{ijk}$ and impose the braiding rule:
\begin{align}
\mathbb{R}:\vcenter{\hbox{\includegraphics[scale=.25]{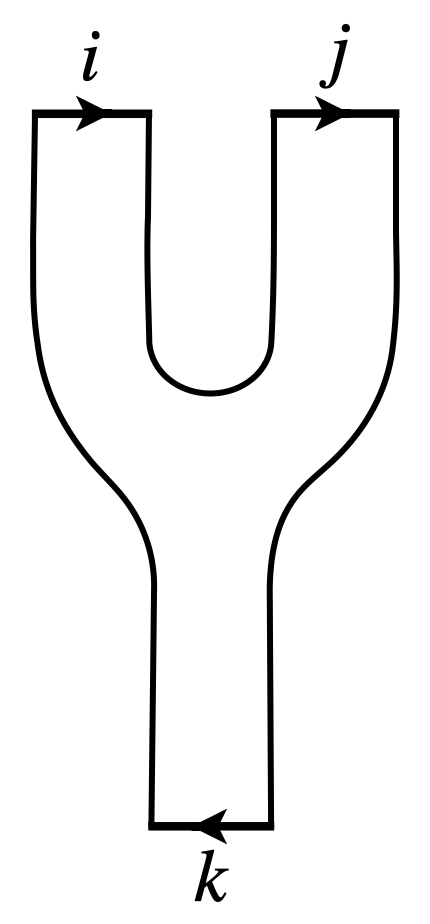}}}
   \to 
\vcenter{\hbox{\includegraphics[scale=.25]{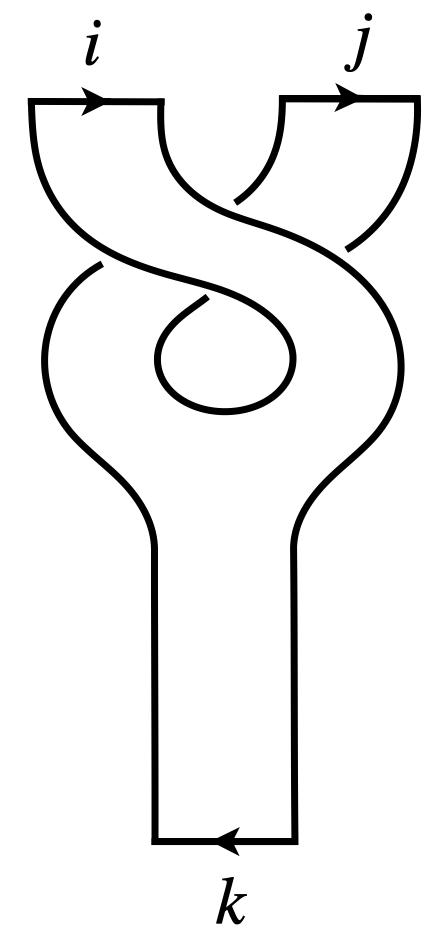}}}  = e^{i \pi (s_i + s_j - s_k)} \times \vcenter{\hbox{\includegraphics[scale=.25]{figures/fin_triple.png}}}
\end{align}
We will explain the associated ribbon calculus in detail in section \ref{sec:ribbons}.
Finally, the $T$ transform was introduced to the matrix model via the Dirac delta comb that imposed quantization of spin, and produces a $2\pi$ twisting of the ribbons. 

A direct connection between VTQFT and the tensor model can be seen in the mapping that produced gravity partitions from  the quartic vertices of the tensor model.  Indeed, according to the rules of the modular functor, the quartic vertices of the tensor model are VTQFT partition functions on 3-manifolds: 
\begin{align}\label{qvertices}
\left\langle \vcenter{\hbox{
\includegraphics[scale=.25]{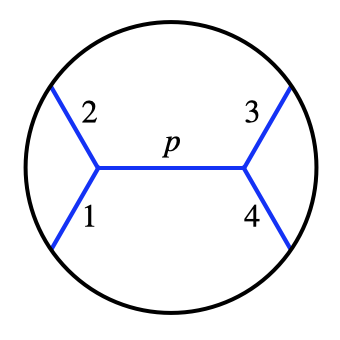}}} \right| \left. \vcenter{\hbox{\includegraphics[scale=.25]{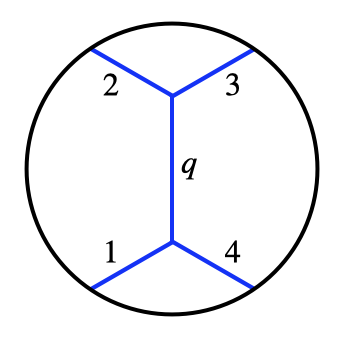}}}\right\rangle  &= Z_{\text{Vir}} \left(\vcenter{\hbox{\includegraphics[scale=.25]{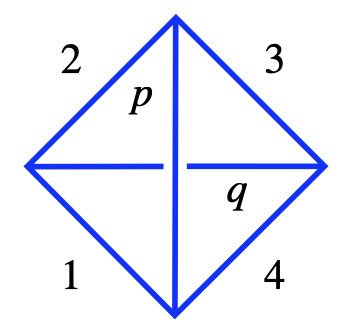}}}\right) \nn
\left\langle\vcenter{\hbox{\includegraphics[scale=.25]{figures/fin_4ptfunctionp_circle.png}}}\right| \left.\vcenter{\hbox{\includegraphics[scale=.25]{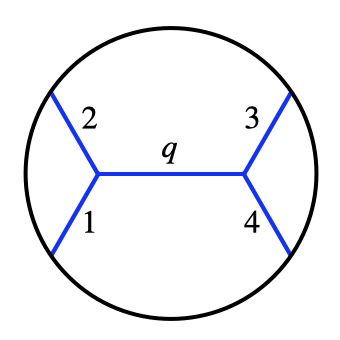}}}\right\rangle  &= Z_{\text{Vir}} \left(\vcenter{\hbox{\includegraphics[scale=.25]{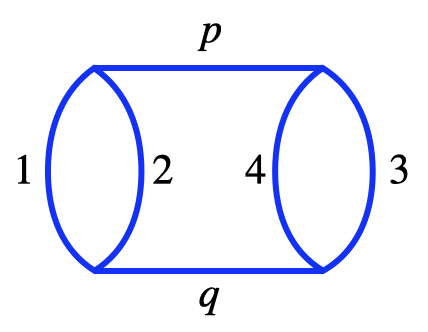}}}\right)
\end{align}

On the LHS, the circle represents an $S^2$ boundary of a solid ball containing a Wilson line network.  The overlap glues together these balls, and connects the Wilson lines to make a closed network inside $S^3$ in either the 6J or Pillow pattern.  This is a direct application of \eqref{MUM} with $U(g)$ being $\mathbb{F}$ and $\mathbbm{1}$ respectively.

To see the match between the tensor model and VTQFT on a general manifold, recall that our diagrammatic rules assigns  4-boundary wormholes to the quartic vertices, obtained by removing solid balls around the vertices of the tetrahedron and pillow graphs on the RHS of \eqref{qvertices}.  We then glue the resulting manifolds with triple line propagators.  Up to braiding phases, this can be interpreted in the VTQFT language as implementing a type of connect sum of 3-spheres containing the 6J or Pillow network.  

To explain this in more detail, consider the conventional connect sum $M_{1}\# M_{2}$
of two manifolds $M_{1}$, and $M_{2}$.  This is defined by removing a solid ball from $M_{1}$ and $M_{2}$, and then gluing the resulting manifolds along the $S^2$ boundaries. In the TQFT, removing the solid ball corresponds to doing the path integral over it, creating a state in the Hilbert space of the sphere.  Gluing is then just an application of the TQFT inner product.   In VTQFT, we can consider a special case of the connect sum where one integrates out a ball that contains a trivalent junction of Wilson lines- creating a state on $\mathcal{H}_{0,3}$- and then gluing along the resulting 3-punctured sphere.    The trivalent junction is necessary because the VTQFT Hilbert space only has normalizable states on a sphere with least 3 punctures.

To see how the tensor model implements this type of connect sum, we first note that the Hilbert space $\mathcal{H}_{0,3}$ is one dimensional. A basis element is given by the 3-point Virasoro block $\ket{\mathcal{F}_{0,3} (ijk)}$ on the sphere with a particular choice of normalization, which we identify as the TQFT path integral on a ball containing the 3-point junction (see \eqref{3ptnorm} below). 
For the unit normalization 
\begin{align} \label{triv}
    \ket{\mathcal{F}_{0,3}(ijk)}\equiv \frac{1}{z_{12}^{h_{i}+h_{j}-h_{k}} z_{23}^{h_{j}+h_{k}-h_{i}} z_{31}^{h_{i}+h_{k}-h_{j}}  }
\end{align}
the conformal block has the VTQFT norm
\begin{align}\label{3ptnorm}
    \braket{\mathcal{F}_{0,3} (ijk)| \mathcal{F}_{0,3} (ijk)}= \left\langle \vcenter{\hbox {\includegraphics[scale=.09]{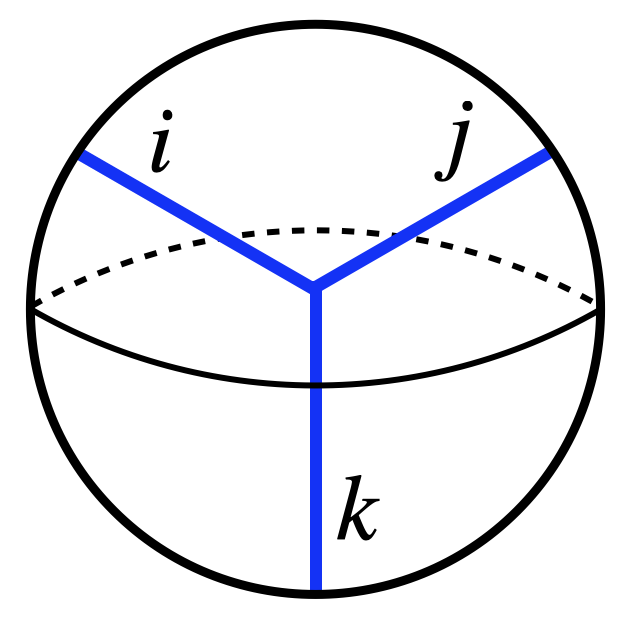}} } \right| \left. \vcenter{\hbox {\includegraphics[scale=.09]{figures/fin_3ptsphere.png}} } \right\rangle =\frac{1}{C_{0}(ijk)}
\end{align}

The VTQFT connect sum of two 3-spheres with tetrahedral networks then takes the following form:
\begin{align}\label{connect}
    Z_{\text{Vir}} \left( \vcenter{\hbox{
    \includegraphics[scale=.2]{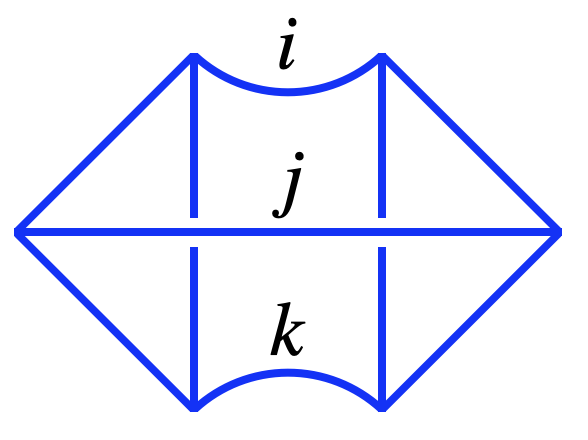}}} \right) &= 
    \frac{
    \left\langle \vcenter{\hbox {\includegraphics[scale=.2]{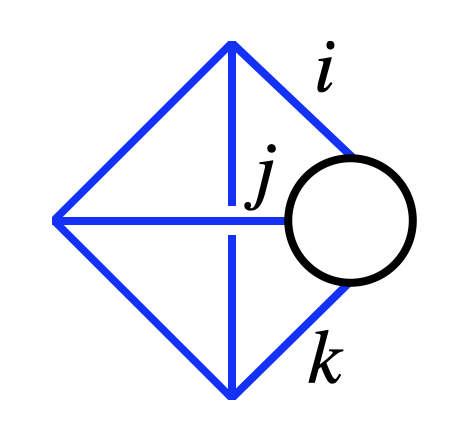}}} \right| \left. \vcenter{\hbox {\includegraphics[scale=.14]{figures/fin_3ptsphere.png}}} \right\rangle 
    \left\langle \vcenter{\hbox {\includegraphics[scale=.14]{figures/fin_3ptsphere.png}}} \right| \left. \vcenter{\hbox {\includegraphics[scale=.2]{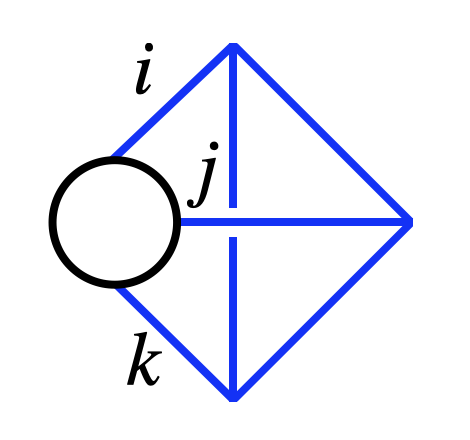}}} \right\rangle 
    }{
    \left\langle \vcenter{\hbox {\includegraphics[scale=.14]{figures/fin_3ptsphere.png}}} \right| \left. \vcenter{\hbox {\includegraphics[scale=.14]{figures/fin_3ptsphere.png}}} \right\rangle
    } \nn
    &= \frac{
    Z_{\text{Vir}} \left( \vcenter{\hbox {\includegraphics[scale=.2]{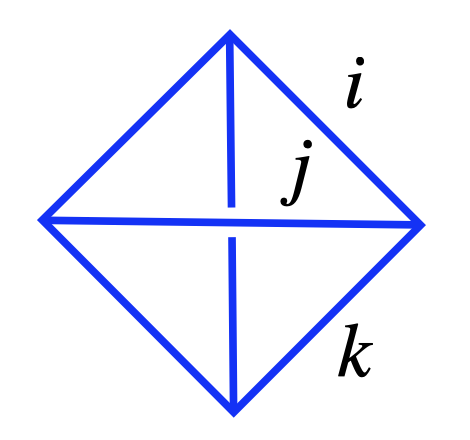}}} \right)
    Z_{\text{Vir}} \left( \vcenter{\hbox {\includegraphics[scale=.2]{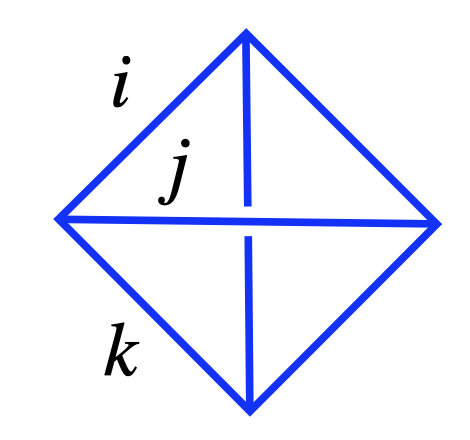}}} \right)
    }{
    \left\langle \vcenter{\hbox {\includegraphics[scale=.14]{figures/fin_3ptsphere.png}}} \right| \left. \vcenter{\hbox {\includegraphics[scale=.14]{figures/fin_3ptsphere.png}}} \right\rangle
    }
\end{align}
In the first line, we have merely  have inserted a resolution of identity  $ 
\mathbbm{1} = \frac{\ket{\mathcal{F}_{0,3} (ijk)} \bra{\mathcal{F}_{0,3} (ijk)}}{\braket{\mathcal{F}_{0,3} (ijk)| \mathcal{F}_{0,3} (ijk)}}
$ on $\mathcal{H}_{0,3}$ to implement the gluing.
\eqref{connect} is just the standard TQFT formula for the connect sum, where the the division by $\left\langle \vcenter{\hbox {\includegraphics[scale=.07]{figures/fin_3ptsphere.png}} } \right| \left. \vcenter{\hbox {\includegraphics[scale=.07]{figures/fin_3ptsphere.png}} } \right\rangle $gives a factor of  $C_{0}(ijk)$ according to  \eqref{3ptnorm}.  Up to a phase, this matches the tensor model Feynman rules for gluing two  6J vertices with a propagator. 

To understand the origin for the nontrivial phases in the propagator of the tensor model, note that we can introduce a crossing transformations into the connect sum gluing described above, just as we did in the Heegaard splitting. In this case the crossing transformation is just a braiding implemented by the $R$ matrix.  Depending on the relative orientation of the junctions being glued,  the Feynman rules of the tensor model prescribes a particular symmetric braiding of the three Wilson lines at the junctions.  We will say more about this braiding rule in section \ref{sec:ribbons}.  Notice that when we introduce the matrix integral, the sum over $T$ transforms will produce all possible $2\pi$ twists on the ribbons, which includes possible braiding operations we can perform in the connect sum.

\paragraph{The junction rule, operator normalization and checks of simple manifolds}
To compare VTQFT partition functions to our 3d gravity partition functions on manifolds with 3-punctured spherical boundaries, we need to specify boundary conditions in VTQFT.  To begin with, consider the following definition of the VTQFT partition function $Z_{\text{Vir}} \left( \vcenter{\hbox{\includegraphics[scale=.23]{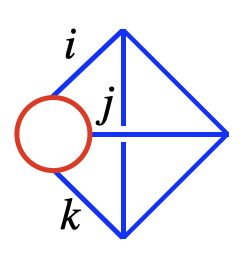}}} \right)$ on a manifold with an asymptotic  3-punctured sphere boundary, depicted in red:
\begin{align}
\left|\vcenter{\hbox{\includegraphics[scale=.15]{figures/fin_tetcut1.png}}}\right\rangle = Z_{\text{Vir}}\left(\vcenter{\hbox{\includegraphics[scale=.3]{figures/fin_tetcut1_red.png}}}\right) \left|\vcenter{\hbox{\includegraphics[scale=.1]{figures/fin_3ptsphere.png}}}\right\rangle
\end{align}
The left hand side is a state on the one dimensional Hilbert space $\mathcal{H}_{0,3}$, so it is a scalar coefficients times  $\left|\vcenter{\hbox{\includegraphics[scale=.1]{figures/fin_3ptsphere.png}}}\right\rangle$.    We have defined the red boundary so that it produces this coefficient when inserted into the VTQFT path integral. Equivalently, we 
can define this asymptotically $\text{AdS}_3$ boundary condition via the insertion of the VTQFT boundary state $ \left(\left\langle \vcenter{\hbox{\includegraphics[scale=.1]{figures/fin_3ptsphere.png}}}\right| \left. \vcenter{\hbox{\includegraphics[scale=.1]{figures/fin_3ptsphere.png}}}\right\rangle  \right)^{-1} \left\langle\vcenter{\hbox{\includegraphics[scale=.1]{figures/fin_3ptsphere.png}}}\right|$, since this picks out the desired coefficient:
 \begin{align}
  Z_{\text{Vir}} \left( \vcenter{\hbox{\includegraphics[scale=.3]{figures/fin_tetcut1_red.png}}} \right)& = \left(\left\langle \vcenter{\hbox{\includegraphics[scale=.1]{figures/fin_3ptsphere.png}}}\right| \left. \vcenter{\hbox{\includegraphics[scale=.1]{figures/fin_3ptsphere.png}}}\right\rangle  \right)^{-1}  \left\langle\vcenter{\hbox{\includegraphics[scale=.1]{figures/fin_3ptsphere.png}}}\right|\left.\vcenter{\hbox{\includegraphics[scale=.2]{figures/fin_tetcut1.png}}}\right\rangle\nn
  &=
 \left(\left\langle \vcenter{\hbox{\includegraphics[scale=.1]{figures/fin_3ptsphere.png}}}\right| \left. \vcenter{\hbox{\includegraphics[scale=.1]{figures/fin_3ptsphere.png}}}\right\rangle  \right)^{-1} Z_{\text{Vir}}\left(\vcenter{\hbox{\includegraphics[scale=.2]{figures/fin_tetblue1.png}}}\right)
 \end{align}
The normalization of the trivalent junction \eqref{3ptnorm} then implies introducing a red boundary around a junction corresponds to multiplying by $C_{0}(ijk)$.   This is the same as the junction rule in \cite{collier2023,collier2023virasoro}.

On the other hand, according to the Feynman rule -- 3-manifold mapping that defines our 3d gravity theory, the boundary condition at the 3 punctured sphere boundaries of the quartic vertices correspond to insertions of the boundary state 
$\left\langle\vcenter{\hbox{\includegraphics[scale=.1]{figures/fin_3ptsphere.png}}}\right|$, \emph{without} the $C_{0}$ factor\footnote{ For example, the 6J 4-boundary manifold \eqref{6Jman} is assigned to $Z_{\text{Vir}} \left(\vcenter{\hbox{\includegraphics[scale=.2]{figures/fin_bluetet.png}}}\right)$, which is partition function obtained by gluing in the trivalent junction without the $C_{0}$ factor.}. We denote these bulk 3 punctured sphere boundaries by a green circle so that: 
\begin{align}
    Z_{\text{Vir}} \left(\vcenter{\hbox{\includegraphics[scale=.3]{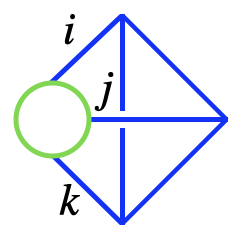}}}\right) \equiv \left\langle\vcenter{\hbox{\includegraphics[scale=.1]{figures/fin_3ptsphere.png}}}\right|\left.\vcenter{\hbox{\includegraphics[scale=.2]{figures/fin_tetcut1.png}}}\right\rangle = Z_{\text{Vir}}\left(\vcenter{\hbox{\includegraphics[scale=.2]{figures/fin_tetblue1.png}}}\right)
\end{align}

Applying this rule, we can obtain the VTQFT partition function on the $C_{0}$, 6J and Pillow manifold, which matches with the corresponding partition functions in our 3d gravity model. 
For example the the VTQFT partition function on the 6J manifold with $\text{AdS}_3$ boundary condition is 
\begin{align}
 Z_{\text{Vir}} \left(\vcenter{\hbox{\includegraphics[scale=0.2]{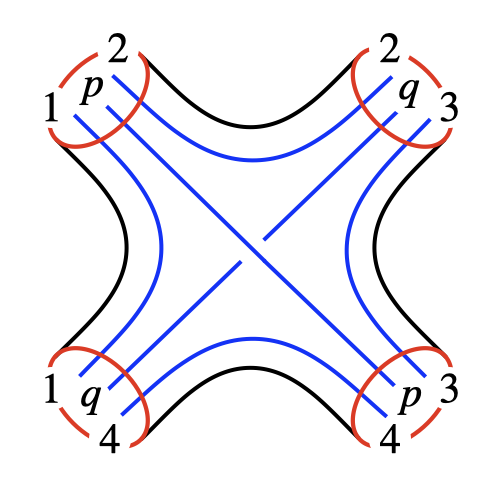}}}\right)&= C_{0}(12p)C_{0}(34p)C_{0}(14q)C_{0}(32q) \, Z_{\text{Vir}}\left(\vcenter{\hbox{\includegraphics[scale=.3]{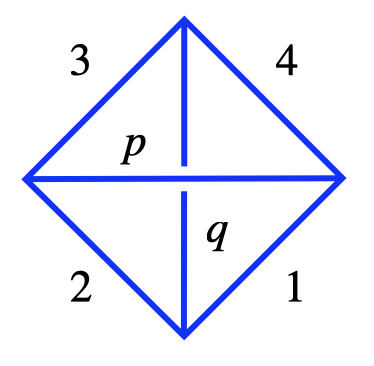}}}\right)\nn
 &=C_{0}(12p)C_{0}(34p)C_{0}(14q)C_{0}(32q) Z_{\text{Vir}}\left(\vcenter{\hbox{\includegraphics[scale=.2]{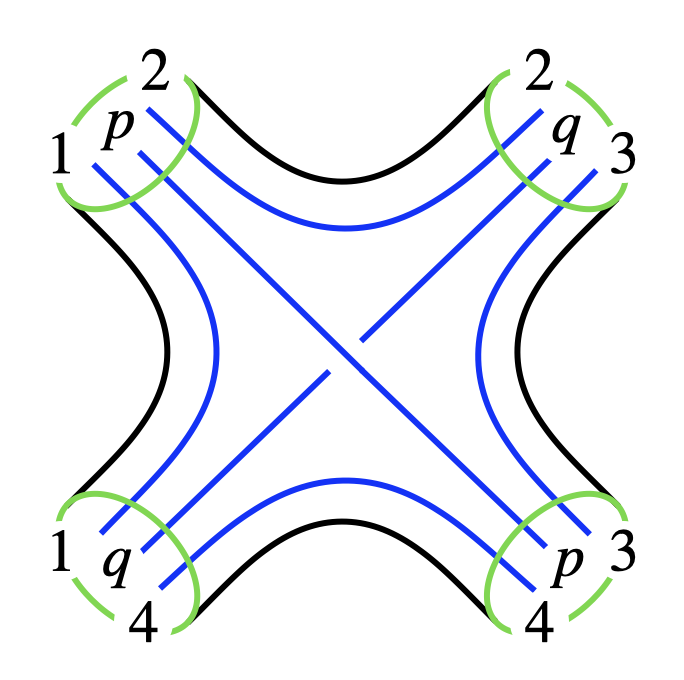}}}\right)
\end{align}
Interpreting the 4-boundary wormhole with green boundaries as our quartic vertex, this matches exactly the result of the CFT ensemble.

Finally,  notice that the definition of VTQFT partition functions given above depend on a choice of the state $\ket{\vcenter{\hbox{\includegraphics[scale=.09]{figures/fin_3ptsphere.png}}}}$  assigned to the ball with a trivalent junction \cite{collier2023, Collier:2024mgv}.   This choice affects the VTQFT partition function that contain this junction.  Due to the one dimensional nature of the Hilbert space, this is just a choice of normalization, which is the same as the choice of normalization of the 3-point block \eqref{triv}. In the tensor model, this corresponds to a choice of normalization of the OPE coefficients in the action, which can be viewed as a choice of measure in the tensor integral.    With the exception of the vacuum loop of $C_{ijk}$'s, this doesn't affect tensor model observables.  

\subsection{Matrix model manifolds and the bulk mapping class group }\label{sec:mappingclass}
\paragraph{Contrasting VTQFT and 3d gravity}
While the gravity partition function on manifolds associated to tensor model matches those of VTQFT, this is not true for matrix model  manifolds that have a nontrivial bulk mapping class group.  The most basic and important example of this is the non-hyperbolic  $T^2 \times I$ wormhole \eqref{T2I} - this is the 3d version of the double trumpet.  In our model, this is obtained from the Laplace transform of the kernel \eqref{annulus}, which computes the
 observable $\braket{\Tr (q_{1}^{L_{0}} \bar{q}_{1}^{\bar{L}_{0}}) \Tr (q_{2}^{L_{0} } \bar{q}_{2}^{\bar{L}_{0}})}_{0}$, with $q_{i}=e^{2 \pi i \tau_{i}}$.
This gives  
\begin{align}\label{T2Igrav}
Z^{\text{gravity}}\left(
 \tau_{1} \vcenter{\hbox{ \includegraphics[scale=.25]{figures/fin_2torus_wormhole.png}}} \tau_{2}\right)  = 2 \frac{\sqrt{\text{Im}(\tau_{1})\text{Im}(\tau_{2})}}{2 \pi^2 | \tau_{1} +\tau_{2}|^2}  
\end{align}

As alluded to previously, aside from the factor of 2,  \eqref{T2Igrav} is equal to the wormhole partition function as computed by Cotler-Jensen in \cite{Cotler_2021}, before imposing their sum over relative $SL(2,\mathbb{Z})$ images.  From the gravity point of view, the factor of 2 comes from geometries in which there is relative flip  $z\to - z$ in the torus coordinates between the two ends of the wormhole.   Notice that this keeps the modular parameter $\tau$ of the torus invariant, so it is not included in the sum over  $PSL(2,\mathbb{Z})$ images  in \cite{Cotler_2021}.  (It is incorporated by a sum over $SL(2,\mathbb{Z})$).  We interpret geometries with this  coordinate flip as an orientable uplift of a 2D non-orientable geometry corresponding to a cylinder with a relative parity transformation between the two ends.  This geometry is included in the GOE ensemble for $\Delta_{s}$.  The 3D uplift is orientable because we flip an extra coordinate in the 3rd dimension.  This is a direct generalization of the example given in \cite{yan2023torus}, in which a 2D cross cap is lifted to an orientable 3-manifold.   In our model, the full relative $SL(2,\mathbb{Z})$  sum is implemented via the perturbative expansion of the matrix model described in  section \ref{sec:gravity}.

On the other hand the VTQFT  $T^2 \times I$ partition function is a continuation of the (regularized) Liouville partition function on the torus in which one sets  $(\tau, \bar{\tau}) = (\tau_{1}, -\bar{\tau}_{2})$:  
\begin{align}\label{VTQFTDT}
    Z^{\text{VTQFT}} \left(\tau_{1}\vcenter{\hbox{ \includegraphics[scale=.25]{figures/fin_2torus_wormhole.png}}} \tau_{2}\right) =\frac{1}{|\tau_{1} +\tau_{2}|} 
\end{align}
Note that this is holomorphically factorized, but the gravity partition function is not. 

This discrepancy arises because 3d gravity imposes a gauging of the \emph{bulk}\footnote{Not to be confused with the 2D mapping class group on a fixed time slice, which was discussed in the context of the VTQFT Hilbert space earlier.} mapping class group on $T^2 \times I$, while VTQFT does not.   This issue becomes manifest when we construct the double trumpet by gluing two single trumpets along a bulk torus.  In 3D, what we call the trumpet is a torus wormhole between an asymptotic $T^2$ boundary and a $T^2$ boundary in the bulk.  Equivalently, the trumpet geometry can be viewed as a solid torus with a Wilson loop inserted in the interior, labelled by Liouville momenta $(P,\bar{P})$ \cite{Mertens_2023}.  In both 3d gravity and VTQFT, this gives a Virasoro character: 
\begin{align}
Z_{\text{trumpet}}(\tau,\bar{\tau}, P,\bar{P} )= \chi_{P}(-1/\tau)\bar{\chi}_{\bar{P}}(-1/\bar{\tau})
\end{align}
    To glue two trumpets along the bulk torus, we identify the trumpet partition functions as a complete set of states on the bulk $T^{2}$ Hilbert space, and insert a resolution of identity.   For VTQFT, the bulk Hilbert space $\mathcal{H}_{T^{2}} \times \mathcal{H}_{T^2}$ is just the space of conformal blocks endowed with the Verlinde inner product.   Indeed the characters can be interpreted as the overlap
\begin{align}
    \chi_{P}(\tau) = \braket{\tau|P} 
\end{align}
between the holomorphic basis $\ket{\tau}$ and the basis $\ket{P}$ on $\mathcal{H}_{T^2} \otimes \mathcal{H}_{T^2}$ \cite{Terashima:2011qi} .

According to the Verlinde inner product, $\ket{P}$ forms a delta-function normalized orthornomal basis, so the resolution of identity in this basis is trivial.  Using the notation $\ket{P,\bar{P}}= \ket{P}\otimes \ket{\bar{P}}, \ket{\tau_{1},\tau_{2}}= \ket{\tau_{1}} \otimes \ket{\tau_{2}}$, the gluing of two single trumpets gives
\begin{align}\label{identity}
    Z^{\text{VTQFT}}_{T^{2}\times I}(\tau_{1},\tau_{2}) = \int dP \int d\bar{P}\braket{ \tau_{2},\tau_{1}| P, \bar{P} } \braket{P,\bar{P} | \tau_{1},\tau_{2}}  
\end{align}
This corresponds to gluing the Virasoro characters with the $dP$ measure.  The formula \eqref{identity} makes explicit the fact that VTQFT assigns the identity operator to wormhole geometries of the form $\Sigma_{g,n} \times I$: in the holomorphic basis, this  translates into the continuation of the Liouville  partition function on $\Sigma_{g,n}$. 
The partition function \eqref{identity} is actually divergent due to the continuum of states: subtracting this divergence leads to formula \eqref{VTQFTDT}

The statistics of the random matrix ensemble associated to VTQFT can be read off from the double trumpet partition function in the momentum basis (this is what we previously referred to as the annulus): 
\begin{align}
Z^{\text{VTQFT}}_{T^{2}\times I}(P_{1},\bar{P}_{1};P_{2},\bar{P}_{2}) = \braket{P_{1} \bar{P}_{1} |\mathbbm{1}| P_{2} \bar{P}_{2}}  =\delta^{2}(P_{1}-P_{2} )
\end{align}
Since the inverse of this kernel gives the Vandemonde determinant, this implies that there is a delta function eigenvalue repulsion in the VTQFT ensemble, in contrast to the long range logarithmic repulsion of the gravity model.

While the gravity partition function on $T^2 \times I$  also corresponds to the identity operator on the gravitational torus Hilbert space, this Hilbert space differs from  $\mathcal{H}_{T^2} \otimes \mathcal{H}_{T^2}$  because the gauging of the mapping class group makes the Virasoro characters non-orthogonal.    The mapping class group for the torus wormhole is $\mathbb{Z}\times \mathbb{Z}$, corresponding to the $2\pi$ Dehn twists which can be applied to either cycles of a bulk toroidal cut:
\begin{align}
\vcenter{\hbox{\includegraphics[scale=.1]{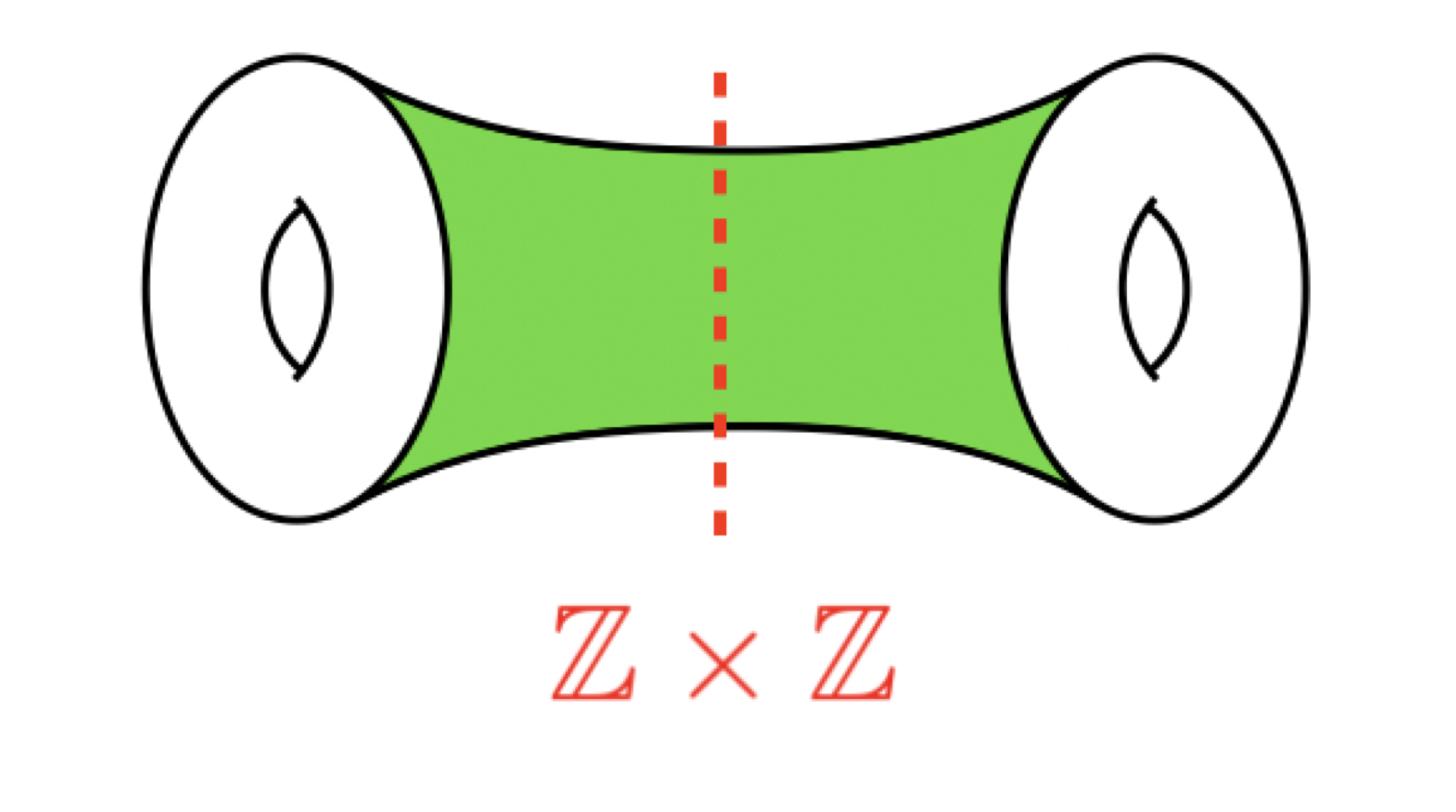}}}
\end{align}

The $\mathbb{Z}\times \mathbb{Z}$ gauging must modify the overlaps of the Virasoro characters, in order for the gluing of the single trumpets to produce the gravity $T^2 \times I$ partition function. The new overlaps can be read off from \eqref{annulus},  which we interpret as the components of the resolution of identity in a non-orthorgonal basis. 
In general, given the metric $g_{ij}=\braket{i|j}$ of overlaps, the resolution of identity takes the form $$\mathbbm{1} = \sum_{i,j}  g^{ij} \ket{i}\bra{j}.$$ Thus, the inverse Vandermonde kernel \eqref{annulus} corresponds to the inverse metric $g^{ij}$ on the gravitational torus Hilbert space.    The gravity inner product is therefore given by the Vandermonde \eqref{matrix_measure}.  In terms of Liouville momenta this is 
\begin{align}
    \braket{P_{1},\bar{P}_{1} |P_{2},\bar{P}_{2}}_{\text{grav}} &=  P_{1} \bar{P}_{1} P_{2} \bar{P}_{2} \delta( P_{1}^{2}- \bar{P}_{1}^{2} -(P_{2}^{2}- \bar{P}_{2}^{2}) ) \log \left| P_{1}^{2}+ \bar{P}_{1}^{2} - (P_{2}^{2}+ \bar{P}_{2}^{2})  \right| 
    \end{align}
\subsection{Representation of surgery}
In section \ref{sec:gravity}, we explained how ensemble averaging over Wilson loops and Wilson lines corresponds to surgery on 3-manifolds. This introduced a set of diagrammatic relations in the Feynman rules  of the ensemble.  For surgery corresponding to averaging over Wilson loops, these relations are generated by the operation in figure  \eqref{bulksurgery}. 
To show that our surgery interpretation is correct, we must show that the amplitudes assigned to each diagram, which are explicit functions of the  weights and the central charge, also satisfy these relations.  In other words, we want to show that our gravity amplitudes provides a representation of surgery.   Since these the amplitudes are integrals of the Virasoro crossing kernels,  surgery must be represented as integral identities of these kernels.  

A complete set of these identities are given by the Moore-Seiberg relations \cite{Moore:1988qv}.  These are generated by  the orthogonality of the 6J symbol\footnote{This is equivalent to the statement that the F-matrix is idempotent, since applying crossing twice reverts to the original channel.}, the hexagon and pentagon identities.   Here we explain the equivalence between surgery relations and the 6J orthogonality and the Hexagon ID.

\paragraph{Orthogonality}
\begin{figure} 
    \centering
    \includegraphics[scale=0.5]{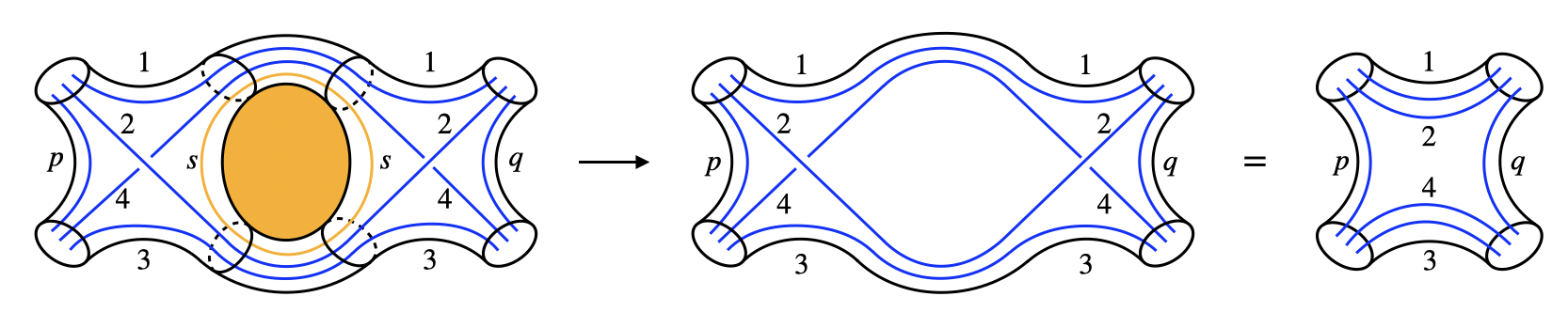}
    \caption{Gluing two 4-boundary wormholes gives the pillow manifold (without external propagator insertions). This is because the insertion of an Omega loop around an $S^2$ handle effectively fills in the associated non contractible cycle, thereby removing the handle.}\label{fig:ortho_mfld}
\end{figure}
The orthogonality relation of the $\mathbb{F}$ crossing kernel is given by:
\begin{align}\label{orthog}
\int \frac{d P_s}{2} \, \mathbb{F}_{p s} 
    \begin{bmatrix}
    3 & 4 \\ 2 & 1
    \end{bmatrix}
    \mathbb{F}_{s q} 
    \begin{bmatrix}
    3 & 2 \\ 4 & 1    
    \end{bmatrix}
 = \delta(P_p-P_q)
\end{align}
Here we illustrate how this identity provides a representation of the surgery relation in figure  \ref{fig:ortho_mfld}, which relates the gluing of two 6J manifolds to a pillow manifold via surgery.  It will be useful to express the amplitude assigned to the 6J manifold in terms of the $\mathbb{F}$ crossing kernel:
\begin{align}
    \vcenter{\hbox{\includegraphics[scale=0.3]{figures/fin_4ptvertex_1.png}}} = \frac{\left| \mathbb{F}_{q p} \begin{bmatrix} 3 & 4 \\ 2 & 1 \end{bmatrix}\right|^2}{\left|\rho_0(q) C_0(12q) C_0(34q)\right|^2}
\end{align} (here and below, we will simplify the diagrams a bit by only drawing the lines, leaving the multi-boundary geometry implicit.)
Note that since this combination of functions possesses tetrahedral symmetry, we may draw the diagram in a more convenient way to help us with the diagrammatics of the calculations:
\begin{align}
    \vcenter{\hbox{\includegraphics[scale=0.3]{figures/fin_4ptvertex_1.png}}} = \vcenter{\hbox{\includegraphics[scale=0.44]{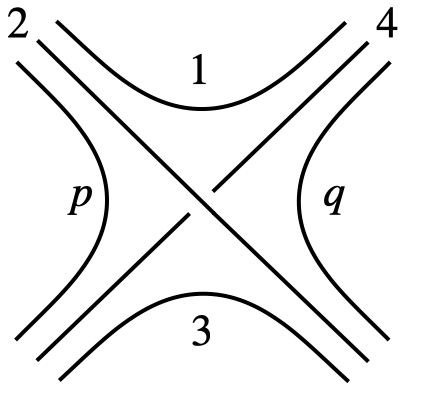}}}
\end{align}
From this representation of the vertex, we see that the gluing of 6J manifolds in figure \ref{fig:ortho_mfld} corresponds to the following integral of $\mathbb{F}$ crossing kernels:  
\begin{align}
& \vcenter{\hbox{\includegraphics[scale=0.25]{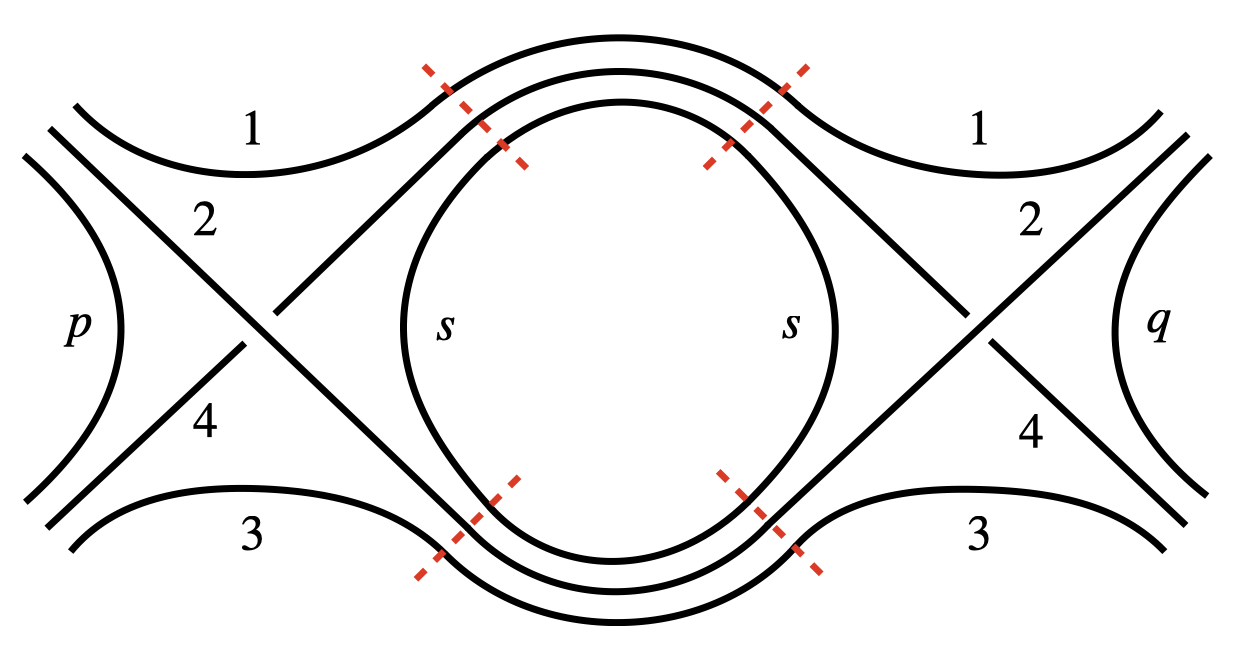}}}  \nn
    &=  \int d^2P_s \, \left|\rho_0(s)C_0(12s)C_0(34s)\right|^2  \frac{\left| \mathbb{F}_{p s} \begin{bmatrix} 3 & 4 \\ 2 & 1 \end{bmatrix} \mathbb{F}_{s q} \begin{bmatrix} 3 & 2 \\ 4 & 1 \end{bmatrix}\right|^2}{\left|\rho_0(s)\rho_0(q)C_0(12s)C_0(34s)C_0(14q)C_0(23q)\right|^2} \nn
    &= \frac{1}{\left| \rho_0(q) C_0(14q) C_0(23q) \right|^2} \int d^2 P_s \, \left| \mathbb{F}_{p s} \begin{bmatrix} 3 & 4 \\ 2 & 1 \end{bmatrix} \mathbb{F}_{s q} \begin{bmatrix} 3 & 2 \\ 4 & 1 \end{bmatrix} \right|^2 
    \end{align}
In the first equality, in addition to gluing the 6J manifolds with propagators, we have averaged over the index contraction between the two $\mathbb{F}$ matrices using the matrix integral.   This produces an integral over the weight $P_s$ with Cardy density of states $\rho_{0}(s)$.
Applying the orthogonality relation then gives:
\begin{align}\label{linesortho}
    \vcenter{\hbox{\includegraphics[scale=0.25]{figures/fin_orthogfuns.png}}} &=  \frac{1}{\left|\rho_0(q) C_0(14q) C_0(23q)\right|^2} \delta^{(2)}(P_q-P_p) \nn
    &= \vcenter{\hbox{\includegraphics[scale=0.35]{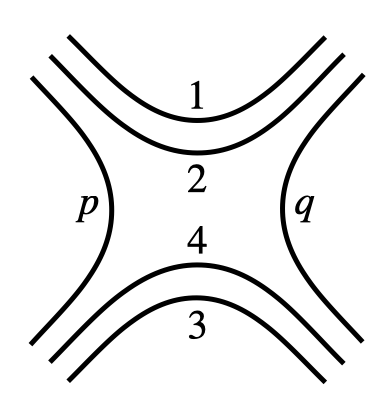}}}
\end{align}
The produces the desired surgery relation.

\subsection{Triple line 6J self contraction }
We now consider Feynman diagram in which the surgery relation corresponds to the hexagon identity. 
Consider the correlator
\begin{align}
\braket{C_{ijk} C_{jik} } 
\end{align}
One diagram contributing to this involves one self contraction of the 6J vertex which connects two legs,  and with the remaining two legs contracted with the external operators. 
We make the contractions explicit below
\begin{align}
\braket{C_{ijk} C_{jik} }_{\text{self contract}}=
\sum_{i_{1},i_{2},i_{3},i_{4}}\sum_{p,q} \wick{ \c1{C_{ijk}}   ( \c3{C_{i_1 i_2 p}} \c3{ C_{ i_3 i_4 p}}  \c1{C_{q i_1 i_{4}}}  \c2{C_{q i_3 i_2 }}  ) \c2{C_{jik} }}  .
\end{align} 

According to \eqref{Prop},these contractions set
\begin{align}
    i_{1}=i_{3}=k,\qquad i_{2}=i_{4}=j, \qquad q=i,
\end{align}
which gives the diagram

\begin{align}
\vcenter{\hbox{\includegraphics[scale=.16]{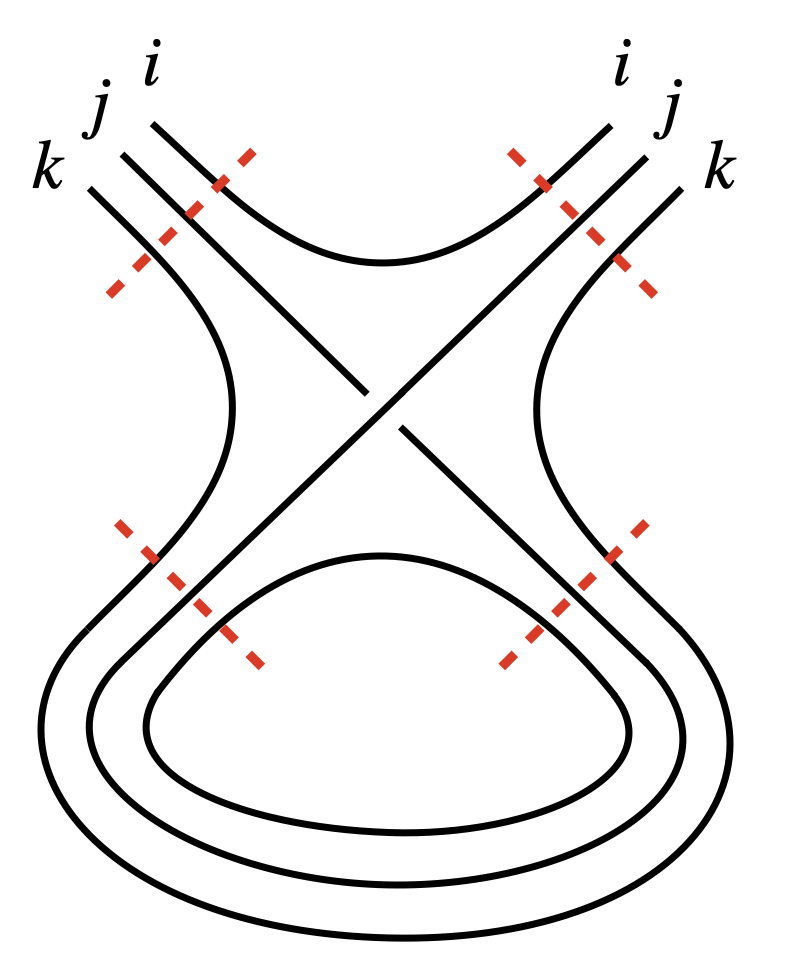} }}  = \int d^{2}P_p  \color{red}|\rho_{0}(p) |^{2} |C_{0}(jkp)|^{2} e^{- i\pi ( s_{j}+s_{k}+s_{p})} & \color{blue}   |C_{0}(jki)|^{4}  e^{-i\pi( s_{i}+s_{j}+s_{k})} \nn
& \times \color{black}  \frac{\left|\begin{Bmatrix}
    i & j & k \\
 p & j & k \end{Bmatrix}\right|^{2} }{|C_{0} (jkp)|^{4} |C_{0}(jki)|^{4}}, \label{6jself1}
\end{align} where we used the red dotted lines to emphasized where we have attached a propagator.  The blue factors in the integrand comes from the external propagator and the red ones from the self contraction. 

Our surgery relation implies that we can remove the loop in the self contraction along with the $S^2$ handle created by it.    This should give a diagrammatic relation of the form 
\begin{align}
    \vcenter{\hbox{\includegraphics[scale=.16]{figures/fin_selfcontr.png} }}= \text{phase} \, \vcenter{\hbox{\includegraphics[scale=.16]{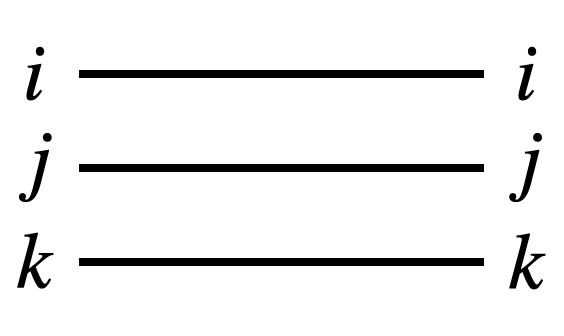} }}
\end{align}
To check that this relation holds for the corresponding amplitudes, we rewrite \eqref{6jself1} in terms of the $\mathbb{F}$ kernel as follows: 
\begin{align}
   \left| \begin{Bmatrix}
   i & j & k \\
 p & j & k \end{Bmatrix} \right|^2&=\rho(i)^{-2} |C_{0}(jkp)|^{4} \left|\mathbb{F}_{pi} \begin{bmatrix} k& j \\ j & k \end{bmatrix}  \right|^{2} \nn 
    \rho_{0}(p) C_{0} (jkp ) &= \mathbb{F}_{1p} \begin{bmatrix} j& k \\ j & k \end{bmatrix}
\end{align}
This gives 
\begin{align}
\vcenter{\hbox{\includegraphics[scale=.2]{figures/fin_selfcontr.png}} } =e^{ -i \pi ( s_{i}) -2\pi i(s_{j}+s_{k})} \int d^{2}p e^{-i\pi (h_{p} -\bar{h}_{p})} \left |\rho(i)^{-1} \mathbb{F}_{1p} \begin{bmatrix} j& k \\ j & k \end{bmatrix}  \mathbb{F}_{pi}  \begin{bmatrix} k& j \\ j & k \end{bmatrix}  \right|^{2}
\end{align}
 Now we apply a special case of the Hexagon identity ($h_{1}$ refers to the dimension of the identity which we set to zero)\footnote{This is the special case for which there are only 2 distinct external dimensions $h_{i},h_{j}$. The general formula is $\int dp \,e^{\pi i( \sum_{a}  h_{a}) - h_{1} -h_{p}-h_{i} )} \mathbb{F}_{1p} \begin{bmatrix} j& k \\ j & k \end{bmatrix}  \mathbb{F}_{pi}  \begin{bmatrix} k& j \\ j & k \end{bmatrix} =\mathbb{F}_{1i} \begin{bmatrix} j& k \\ j & k \end{bmatrix}  $.}:
 \begin{align}\label{hexagonspecial}
        \int dp \,e^{\pi i( 2 h_{j}+2 h_{k} - h_{1} -h_{p}-h_{i} )} \mathbb{F}_{1p} \begin{bmatrix} j& k \\ j & k \end{bmatrix}  \mathbb{F}_{pi}  \begin{bmatrix} k& j \\ j & k \end{bmatrix} &=\mathbb{F}_{1i} \begin{bmatrix} j& k \\ j & k \end{bmatrix} \nn 
         \int d\bar{p} \,e^{-\pi i( 2 \bar{h}_{j}+2 \bar{h}_{k} - \bar{h}_{1} -\bar{h}_{p}-\bar{h}_{i} )} \mathbb{F}_{1p} \begin{bmatrix} j& k \\ j & k \end{bmatrix}  \mathbb{F}_{pi}  \begin{bmatrix} k& j \\ j & k \end{bmatrix} &=\mathbb{F}_{1i} \begin{bmatrix} j& k \\ j & k \end{bmatrix} 
\end{align}
So the diagram evaluates to
\begin{align}\label{6jselffin}
    \vcenter{\hbox{\includegraphics[scale=.2]{figures/fin_selfcontr.png}}} =e^{-4 \pi i(s_{j}+s_{k} )}|C_0(jki)|^2=e^{-4 \pi i(s_{j}+s_{k} )}     \vcenter{\hbox{\includegraphics[scale=.2]{figures/fin_3lines.png}}}
\end{align}
as expected from surgery.
\subsection{Ribbon graphs} \label{sec:ribbons}
The triple line diagrams we used to represent our contraction rules do not make manifest phases that appear in the correlators.  For example, the diagram in \eqref{6jselffin} doesn't contain enough information to determine the phase on the RHS.   To capture these phases diagrammatically, we need to frame the Wilson lines by thickening them into ribbons.    In order to incorporate all phase information into the twisting of these ribbons, it will be useful to consider a different normalization and  phase convention for the OPE coefficients. We define 
\begin{align}\label{BoldCijk}
\mathbf{C}_{ijk} \equiv  \frac{C_{ijk}}{C_{0}(ijk)} e^{i \frac{\pi}{2}(s_{i}+s_{j}+s_{k}) 
 }\end{align}
and consider its correlators.    The propagators are now unit normalized, but with a different phase:
\begin{align}
     \wick{ \c1{\mathbf{C}_{ijk}}  \c1{\mathbf{C}_{ijk}}}&= 1 \nn
      \wick{ \c1{\mathbf{C}_{ijk}}  \c1{\mathbf{C}_{jik} }}&= e^{i\pi (s_{i}+s_{j}+s_{k})}
\end{align}
But the quartic vertices also change by a phase involving the sum of all spins.
\begin{align}\label{hatquart}
    \vcenter{\hbox{\includegraphics[scale=.3]{figures/fin_4ptvertex_1.png}}} &= \frac{1}{\hbar} e^{-i\pi (\sum_{i} s_{i})}
    \left|\begin{Bmatrix}
        q & 4 & 1 \\
        p & 2 & 3
    \end{Bmatrix}\right|^2 \\
    \vcenter{\hbox{\includegraphics[scale=.52]{figures/fin_pillowvertex.png}}} &= \frac{1}{\hbar} \frac{e^{-i\pi (\sum_{i} s_{i})} \delta^{(2)}(P_p - P_q)}{\rho_0(p)}
\end{align}

\paragraph{Framing and junctions}
A framing of a line $C$  is defined by specifying a normal vector field along $C$.  We can think of the tip of this vector as specifying a second curve - the frame- shown in green below. 
\begin{align}
\vcenter{\hbox{\includegraphics[scale=0.2]{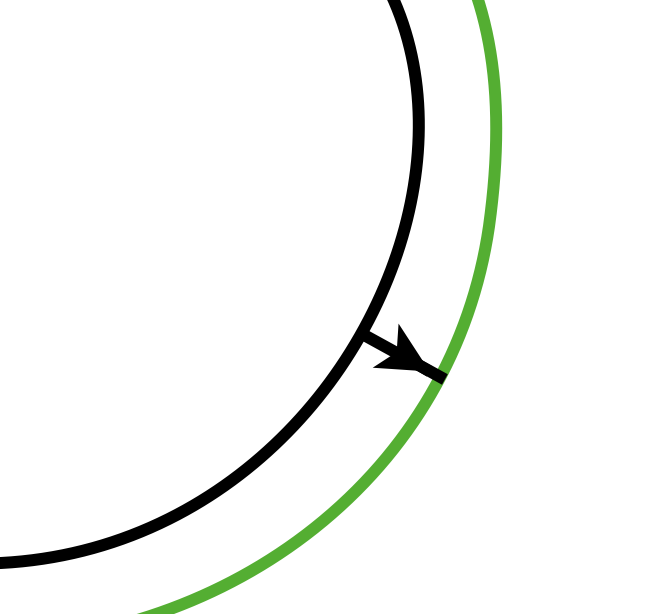}}} 
\end{align}
A framed knot can then be interpreted as a ribbon, whose twisting determines the self-linking number of the knot.
\begin{align}
\vcenter{\hbox{\includegraphics[scale=0.3]{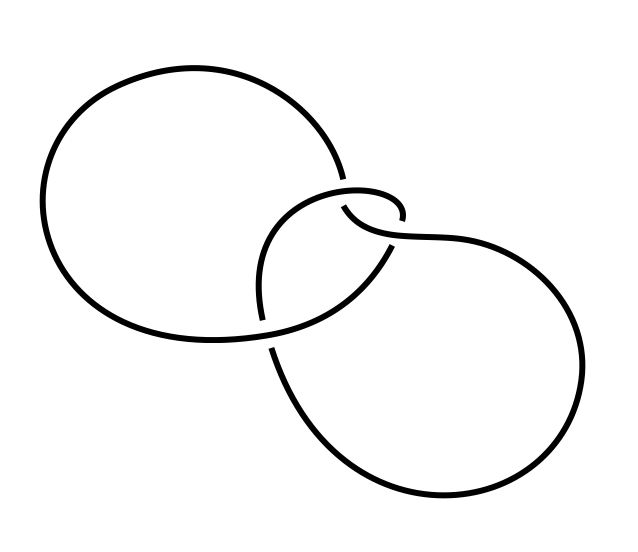}}} \to \vcenter{\hbox{\includegraphics[scale=0.3]{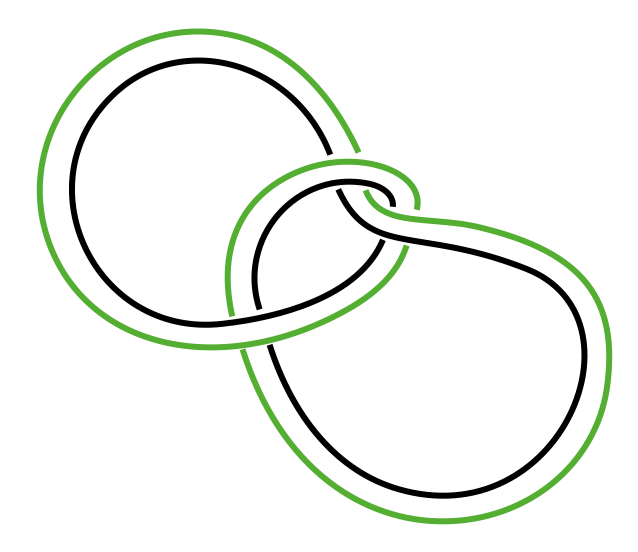}}} 
\end{align}
In the figure above, we chose a framing with linking number zero.  More generally, we can introduce nontrivial twists, which leads to phases determined by the spin:
\begin{align}
\vcenter{\hbox{\includegraphics[scale=0.3]{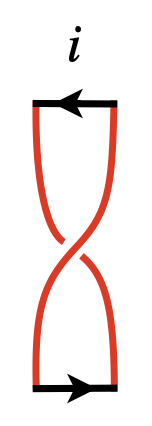}}} = e^{i \pi s_i} \vcenter{\hbox{\includegraphics[scale=0.3]{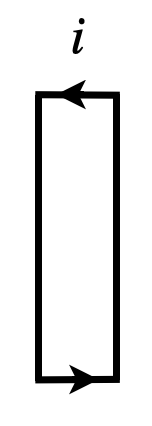}}} \, , \qquad \vcenter{\hbox{\includegraphics[scale=0.3]{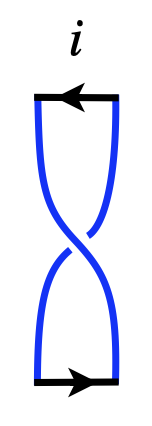}}} = e^{-i\pi s_i} \vcenter{\hbox{\includegraphics[scale=0.3]{figures/fin_blackribbon.png}}}
\end{align}

The Wilson lines in our Feynman diagram end on three-punctured spheres, which are glued together by propagators according to the rules \eqref{Prop}.   To capture these gluing rules, we introduce a junction to  keep track of the cyclic order of the 3 punctures produced by the framed lines.  Thus each OPE coefficient is mapped to a ribbon junction:
\begin{align} \label{Crib}
    C_{ijk} \simeq \vcenter{\hbox{\includegraphics[scale=0.25]{figures/fin_triple.png}}} , \quad C_{jik} \simeq \vcenter{\hbox{\includegraphics[scale=0.27]{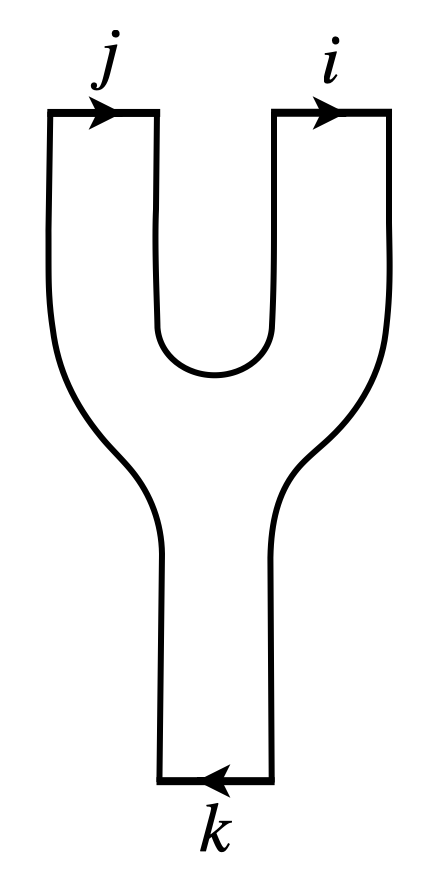}}} 
\end{align}
  The orientation of the three intervals attached to a junction satisfies a clockwise rule that also determines the orientation of the disk.   This determines the cyclic order of the indices in the OPE coefficient.    Braiding of the ribbons can be replaced with a combination of twists according to the rule:
 \begin{align}\label{braidtwist}
    \vcenter{\hbox{\includegraphics[scale=.2]{figures/fin_braid.png}}} = \vcenter{\hbox{\includegraphics[scale=.2]{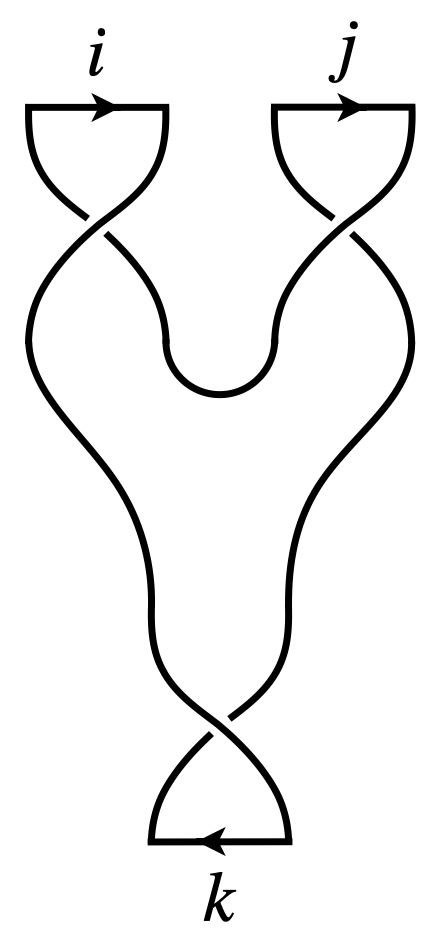}}} = e^{i \pi (s_i + s_j - s_k)} \times \vcenter{\hbox{\includegraphics[scale=.2]{figures/fin_triple.png}}}
\end{align}
 
After framing and inserting the junctions,  the 6J vertex becomes
\begin{align}
\vcenter{\hbox{\includegraphics[scale=.25
]{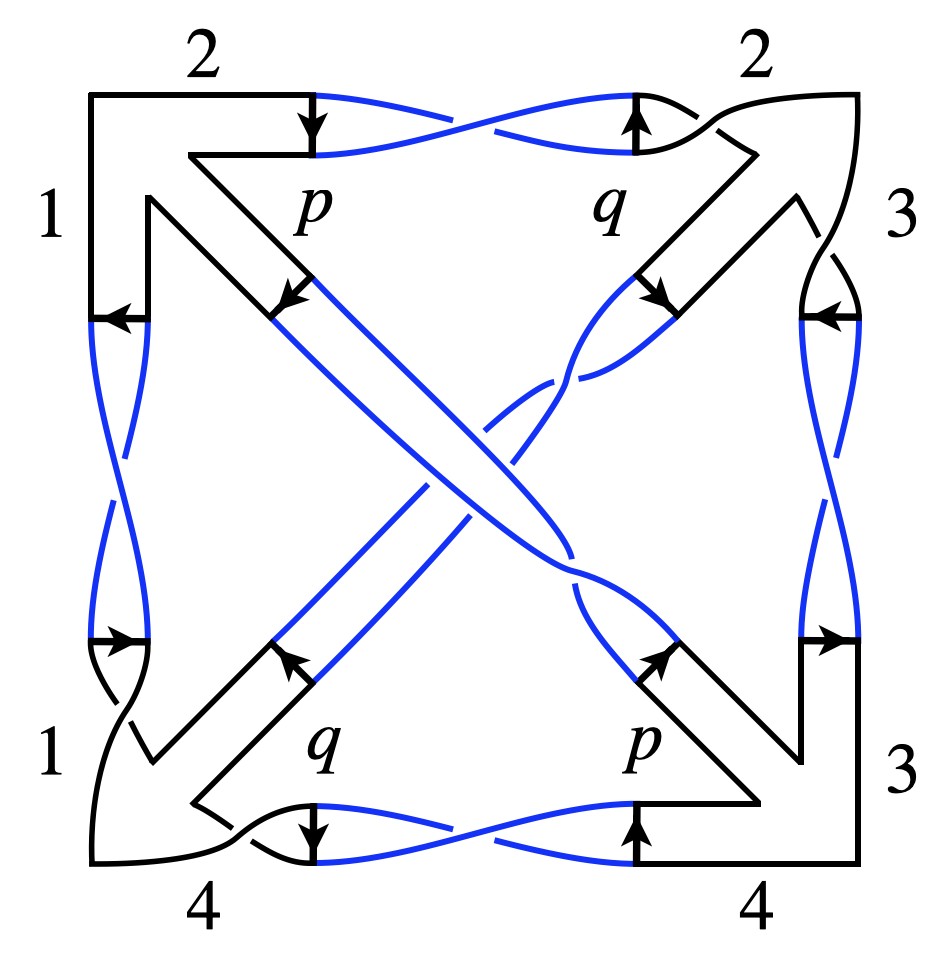}}}=\frac{1}{\hbar} e^{-i\pi (\sum_{i} s_{i})}\left|\begin{Bmatrix}
        q & 4 & 1 \\
        p & 2 & 3
    \end{Bmatrix}\right|^2
\end{align}
Prior to inserting the blue twists, this was a tetrahedral ribbon corresponding to a sphere with 4 holes, which gives the trivial framing corresponding to  to the $6$-J symbol.  The introduction of the left handed twists indicated by the blue ribbons gives rise to the phases.    Similarly, the framed pillow vertex is
\begin{align}
\vcenter{\hbox{\includegraphics[scale=.25]{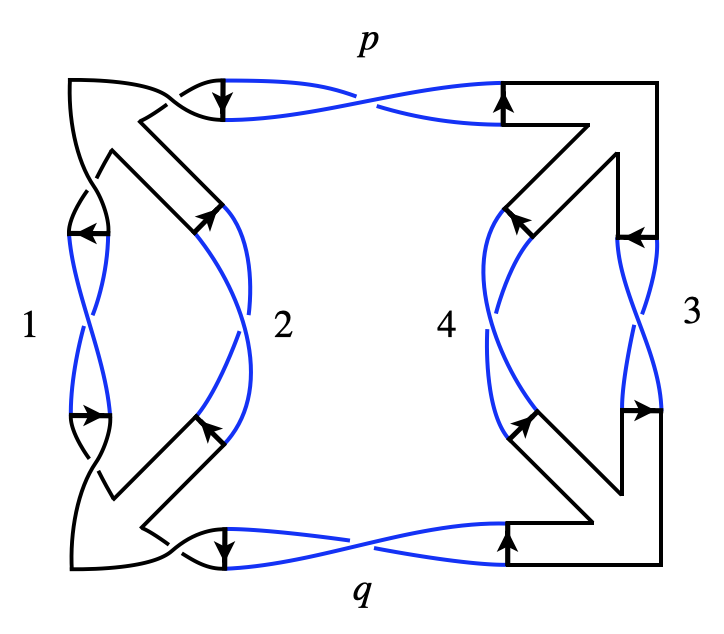}}} = \vcenter{\hbox{\includegraphics[scale=.25]{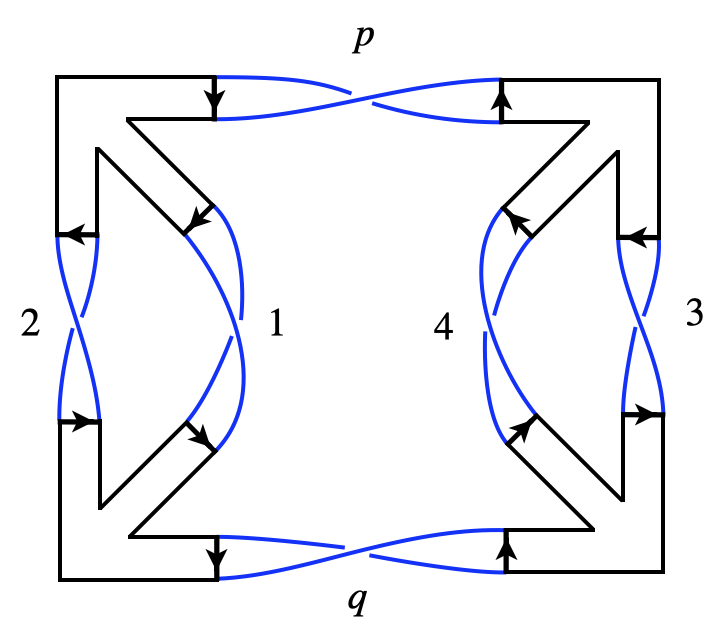}}}=\frac{1}{\hbar} \frac{e^{-i\pi (\sum_{i} s_{i})} \delta^{(2)}(P_p - P_q)}{|\rho_0(p)|^2 }
\end{align}
The kinetic term in the potential is a twisted theta graph obtained by setting one of the 6J ribbons to identity 
\begin{align}
\vcenter{\hbox{\includegraphics[scale=.25]{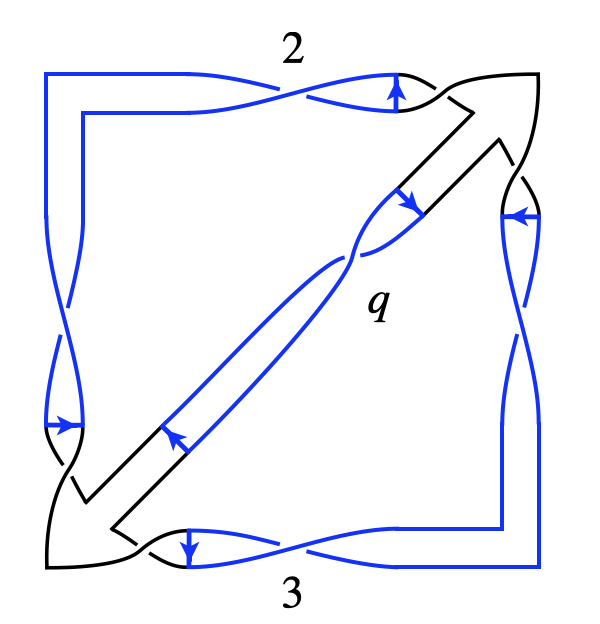}}} = \vcenter{\hbox{\includegraphics[scale=.3]{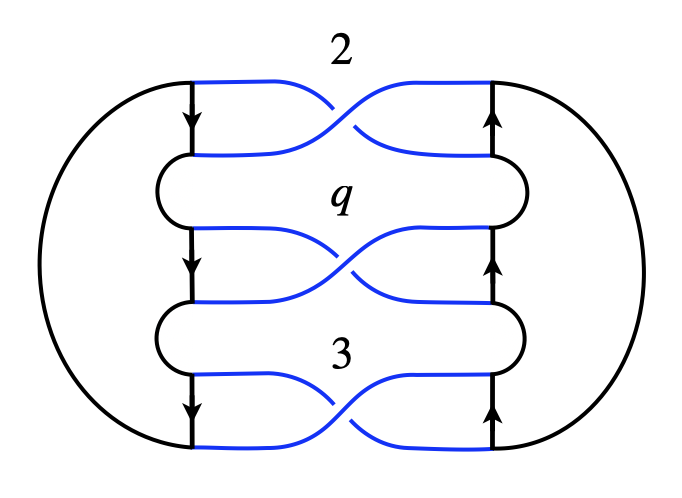}}} = \vcenter{\hbox{\includegraphics[scale=.2]{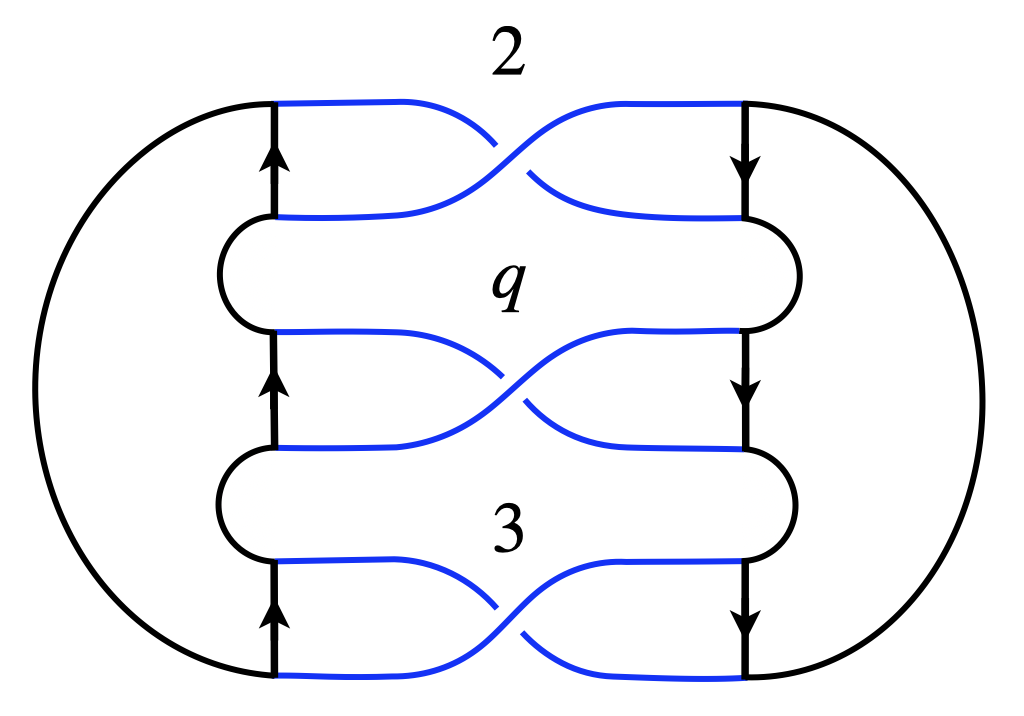}}} = e^{-\pi i (s_{i}+s_{j}+s_{k})}
\end{align}

To specify the tensor model gluing rules,  it will be useful to put the junctions in the same canonical position defined in \eqref{Crib}.  For example, in the case of the 6J ribbon this gives the topologically equivalent projection:
\begin{align} \label{quad}
\vcenter{\hbox{\includegraphics[scale=.4]{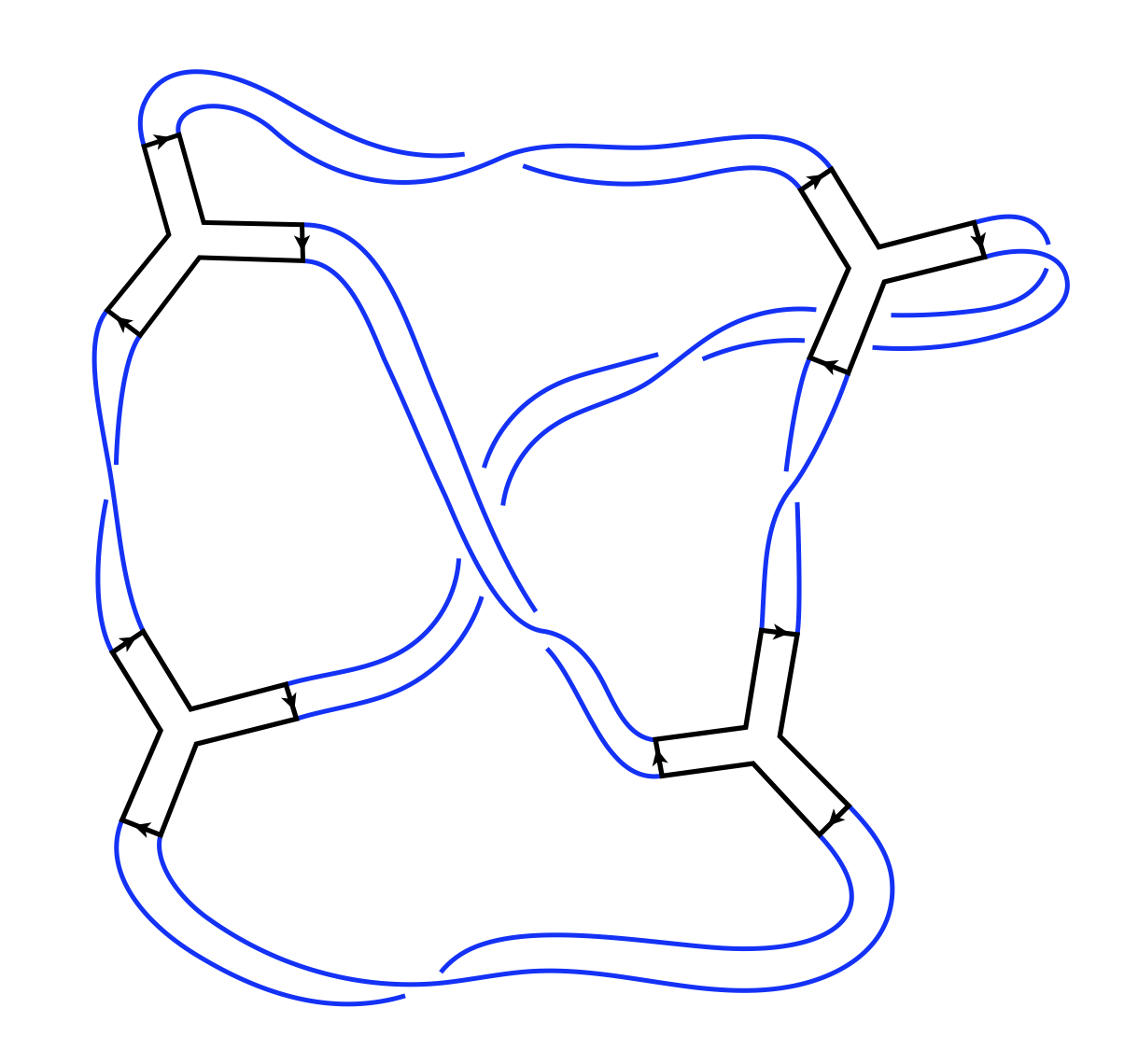}}}
\end{align}
Note that the internal junctions are placeholders that allow us to determine how to draw the framed propagator that connects different diagrams, i.e. there are no bulk junctions in 3d gravity.  Once we have determined how the ribbons graphs should be glued, we will remove these internal junctions. 
\paragraph{Gluing rules}
Now we define rules for gluing together the 6J or and Pillow ribbon graphs.   This  corresponds to insertion of propagators in the tensor model, and the connect sum around a junction in the  VTQFT language.   In formulating the gluing rules below, it will be useful to remember that we are just gluing the framed Wilson line represented by the oriented intervals: we are not gluing the face of the junctions together.  When connecting the ribbons, we should be careful to match the orientation of intervals: otherwise, we would be gluing the line to its framing, which is not allowed.

The tensor model defines contraction rules for the OPE coefficients, which corresponds to a specific  way to glue the ribbons together when they are attached to two junctions with the same cyclic order.  This is given by the following rule: 
\begin{align}\label{CijkCijk}
   \wick{\c1{\mathbf{C}_{ijk}} \c1{\mathbf{C}_{ijk}}} =  \vcenter{\hbox{\includegraphics[scale=0.2]{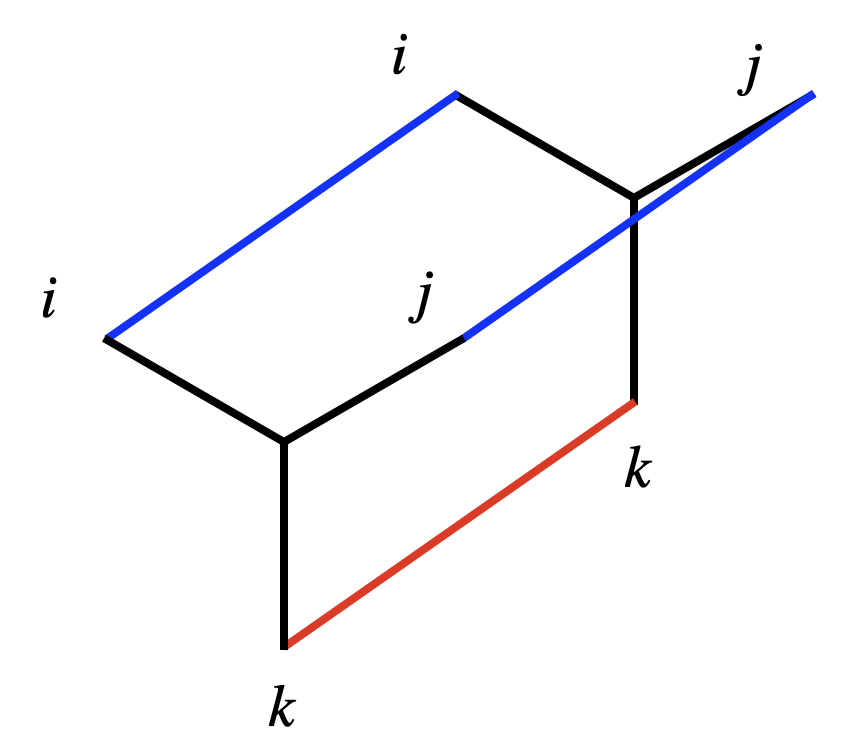}}} = \vcenter{\hbox{\includegraphics[scale=0.25]{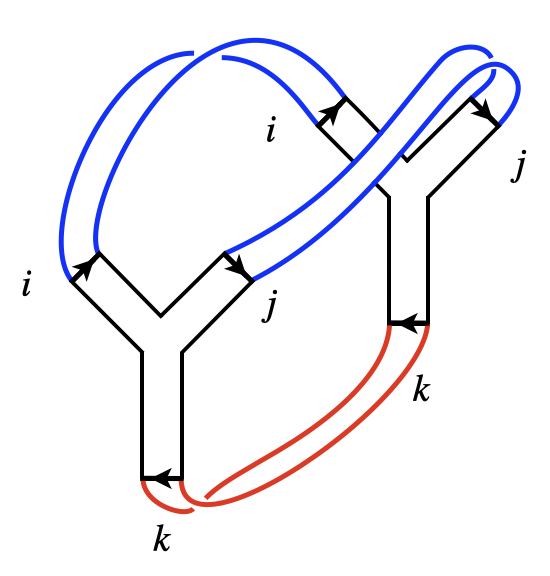}}} = \vcenter{\hbox{\includegraphics[scale=0.25]{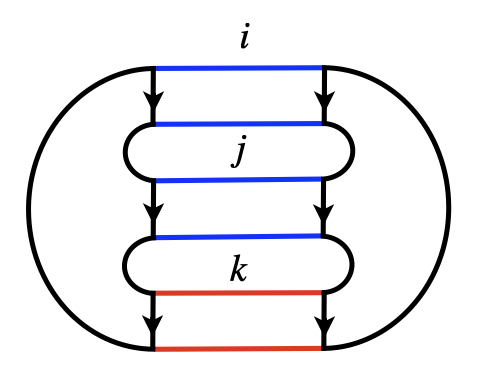}}} = 1
\end{align}
To get to the final, untwisted theta graph\footnote{ Notice that in the representation as a theta graph, we have flipped the junction on the right on its back.}, we applied a braiding of the i and j, which can be converted into twists via the rule \eqref{braidtwist}.

An equivalent 2d projection of this gluing is given by 
\begin{align} \label{caneven}
    \vcenter{\hbox{   \includegraphics[scale=.3]{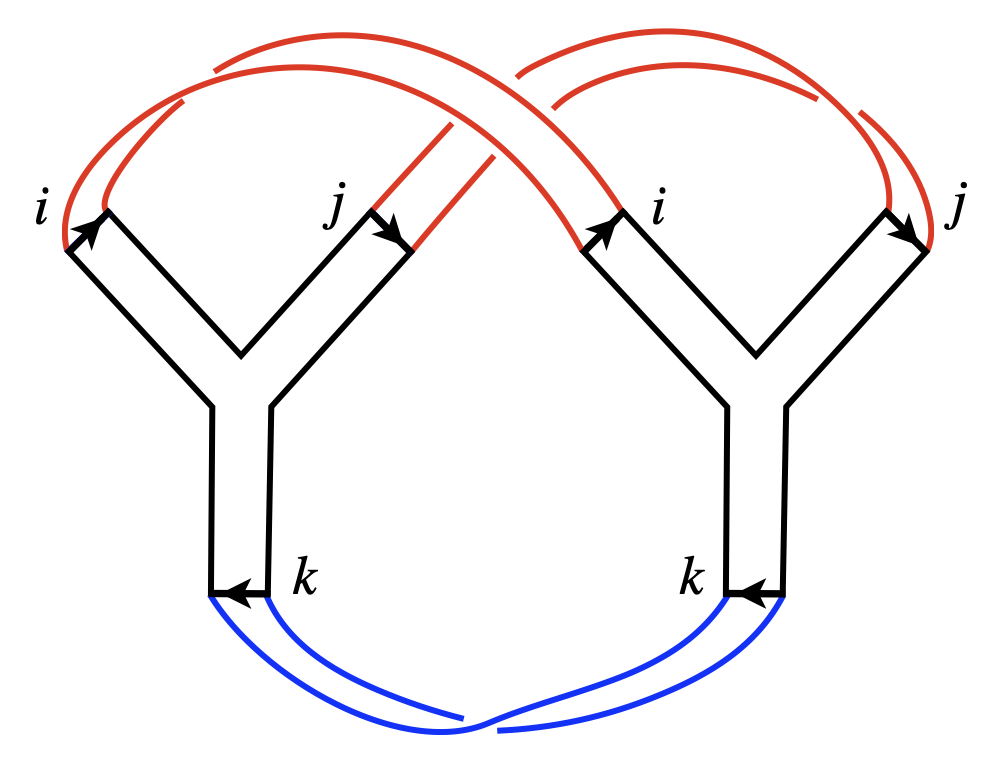}} } = 1
\end{align}
Here $i$ and $j$ are braided in the opposite sense from the analogous diagram in \eqref{CijkCijk}, but the ribbons also have the opposite twist so they are  topologically the same.  
Notice that up to a uniform 2 $\pi$ twists on each ribbon, \eqref{CijkCijk} is the unique choice of gluing that is symmetric in the 3 weights.

The tensor model propagator also joins two junctions that have weights in opposite cyclic order.  The corresponding gluing rule is 
\begin{align}\label{CijkCjik}
    \wick{\c1{\mathbf{C}_{ijk}} \c1{\mathbf{C}_{jik}}} = \vcenter{\hbox{ \includegraphics[scale=.2]{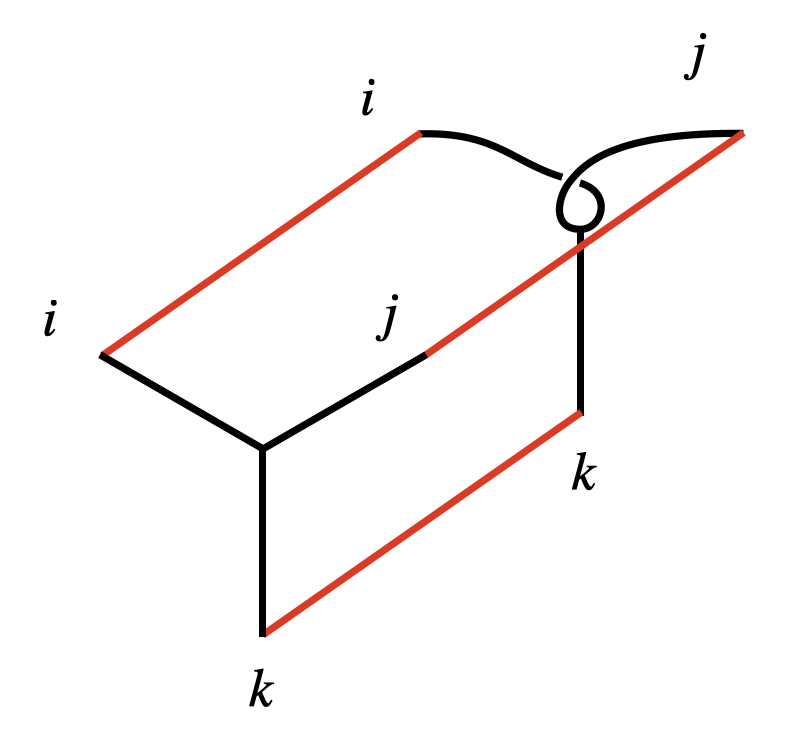}}} = \vcenter{\hbox{ \includegraphics[scale=.25]{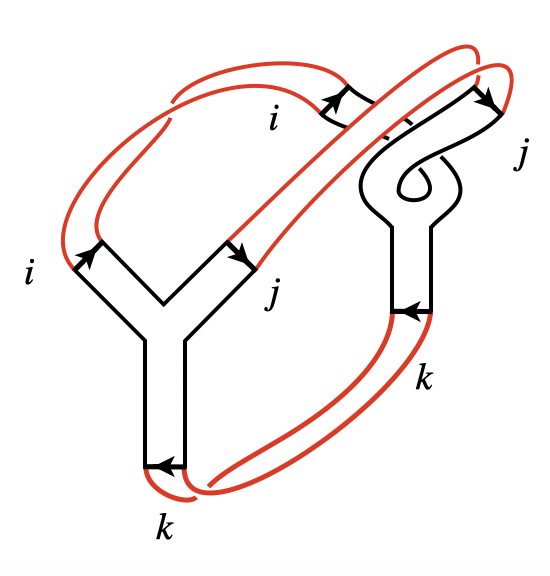}}} = \vcenter{\hbox{ \includegraphics[scale=.25]{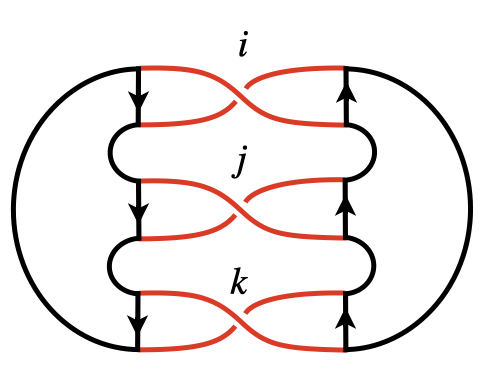}}}  = e^{i\pi (s_{i}+s_{j}+s_{k}) }
\end{align}
The theta graph representation shows that this is again a symmetric choice.  Sometimes (as in check of the 6J orthogonality below), it is useful to consider an equivalent depiction of \eqref{CijkCjik} where we make the junctions face each other

\begin{align}\label{oddface}
    \wick{\c{\mathbf{C}_{ijk}} \c{ \mathbf{C}_{jik}}} = \vcenter{\hbox{\includegraphics[scale=0.2]{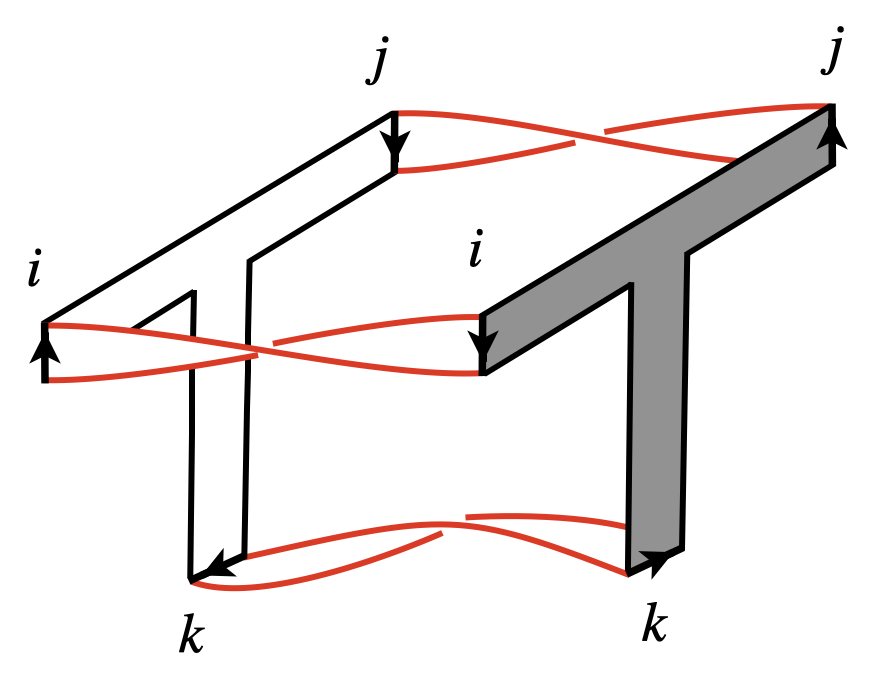}}}
\end{align}

The upshot is that we have two types of contractions, represented by ribbon graphs that have a relative phase: 
\begin{align}\label{odd}
    \vcenter{\hbox{ \includegraphics[scale=.25]{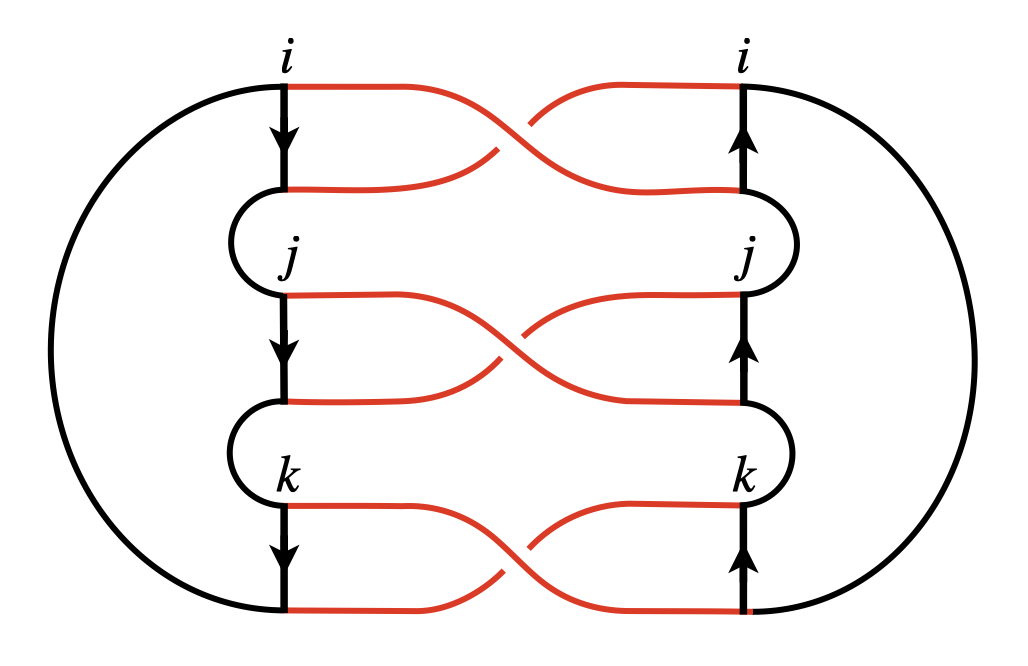}  }} = e^{i\pi (s_i+s_j+s_k)} \vcenter{\hbox{ \includegraphics[scale=.3]{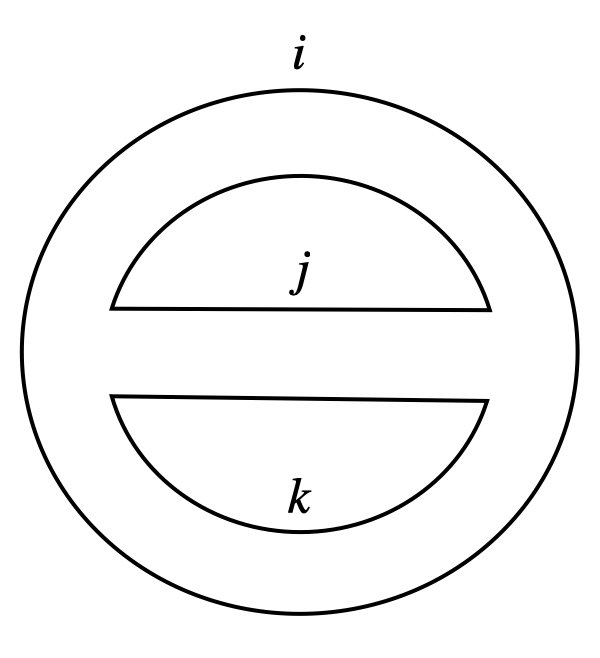}  }} 
\end{align}
As a simple example, these ribbon rules imply that to leading order, the two point function of $\mathbf{C}_{ijk}$  is given by
\begin{align}\label{CC*}
    \braket{\mathbf{C}_{ijk} \mathbf{C}_{lmn}} =  & (\delta_{il} \delta_{jm} \delta_{kn} + \delta_{im} \delta_{jn} \delta_{kl} + \delta_{in} \delta_{jl} \delta_{km}) \nn
    &+e^{i \pi (s_i+s_j+s_k)}\Big( \delta_{im} \delta_{jl} \delta_{kn} + \delta_{in} \delta_{jm} \delta_{kl} + \delta_{il} \delta_{jn} \delta_{km} \Big)
\end{align}

\subsubsection{Ribbon graphs for 6J  orthogonality }
To illustrate how the tensor model ribbon rules work, it is useful to repeat the 6J orthogonality relation with ribbon graphs.   First consider two 6J ribbon graphs, rotated with respect to each other: 
\begin{align}\label{orthogribbons}
\vcenter{\hbox{\includegraphics[scale=0.25]{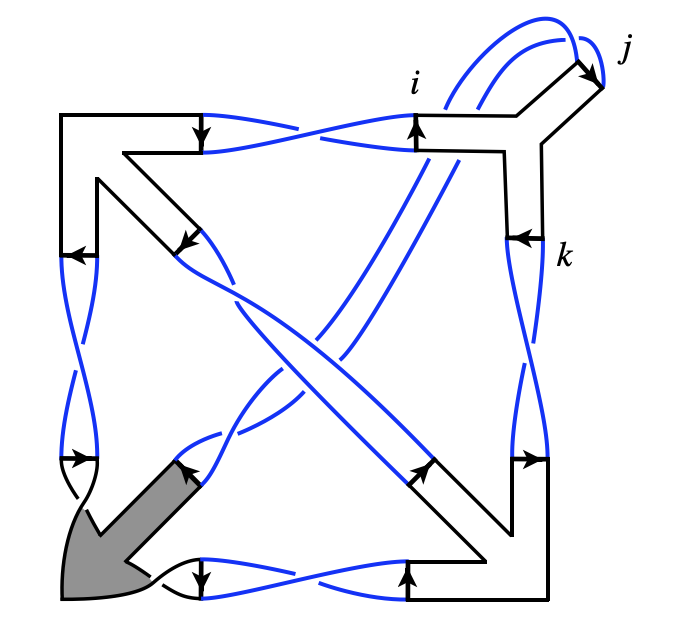}}} \qquad \vcenter{\hbox{\includegraphics[scale=0.25]{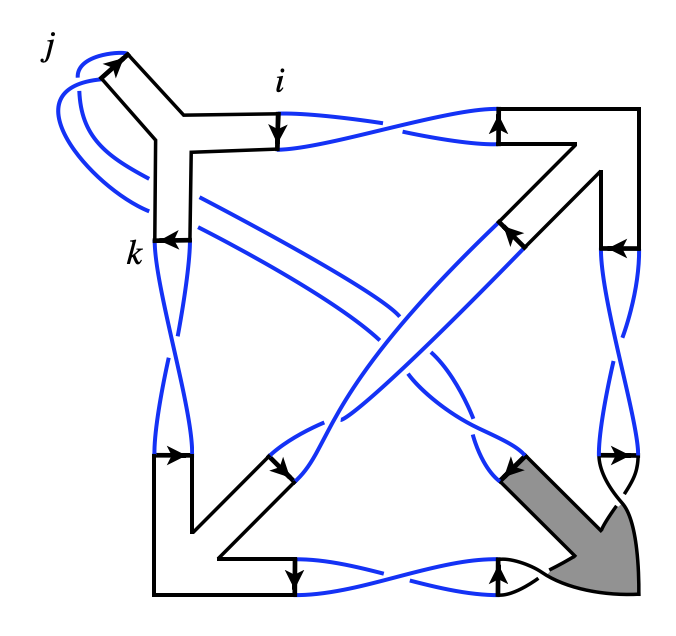}}}
\end{align}

\begin{figure}
    \centering
    \includegraphics[scale=0.2]{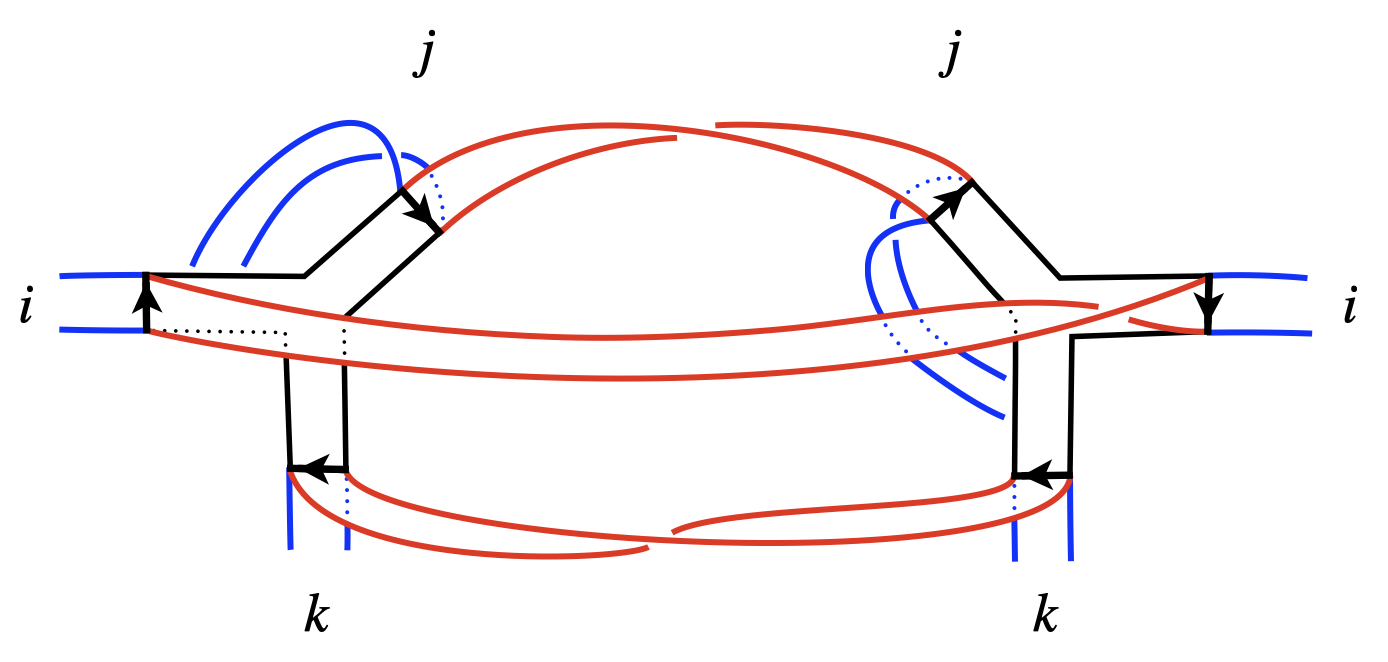} \, \includegraphics[scale=0.2]{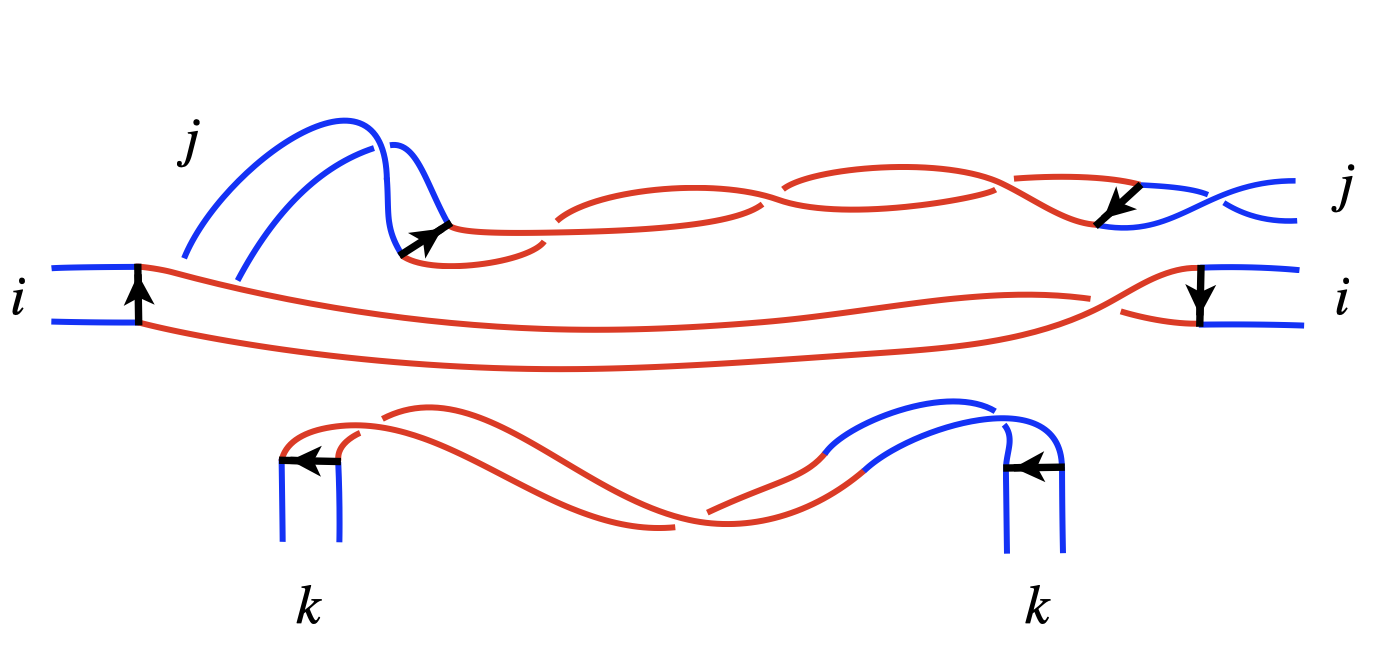} \, \includegraphics[scale=0.2]{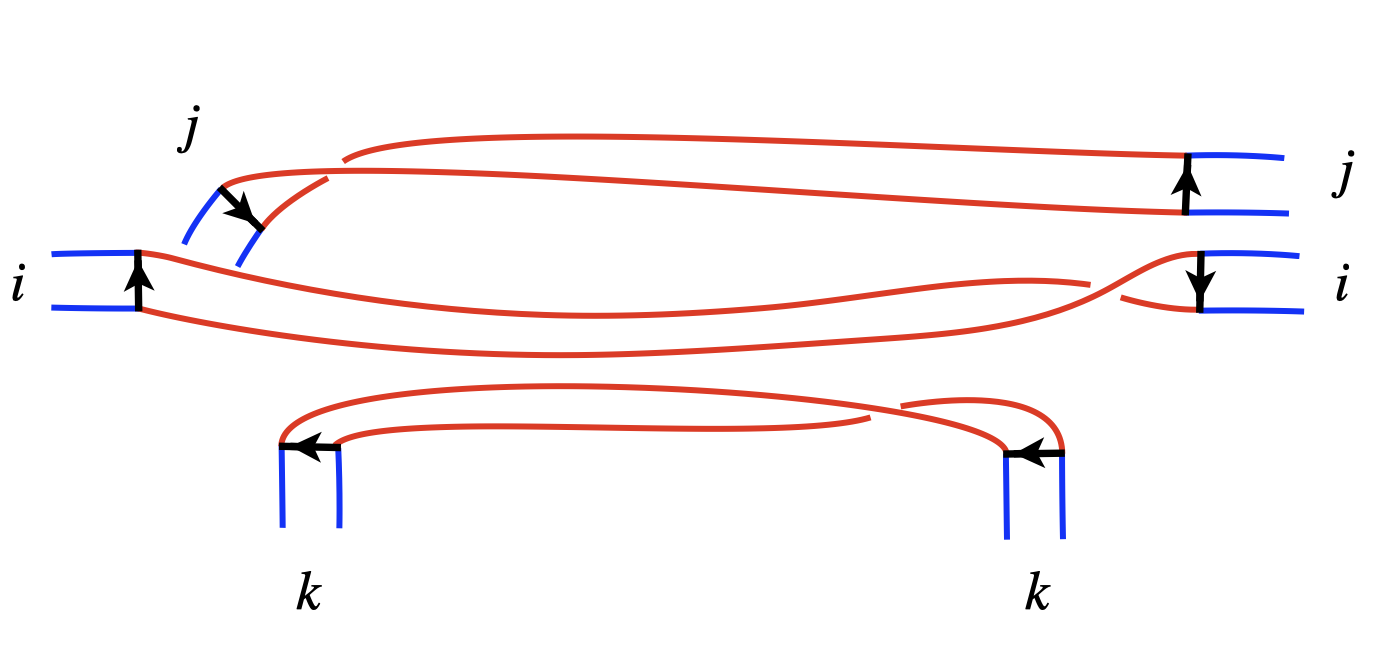}
    \caption{Explicit demonstration of how to resolve ribbons at junctions into a 2d picture.}
    \label{fig:resolve}
\end{figure}

We enfolded the junctions that will be glued next to each other, so they are in canonical position.  Note that they are in opposite cyclic order, so we use the gluing rule in \eqref{CijkCjik}.  In applying \eqref{CijkCjik} to \eqref{orthogribbons},  we need to be careful to resolve the over and under crossings that arise from projecting onto a plane. We demonstrate how to do this in figure \ref{fig:resolve}.
Alternatively, we can observe that the junctions can be made to face each other as in \eqref{oddface}:  so we could just remove the junctions continue the ribbons with right handed twists. Either way one finds that gluing to 6J ribbon graphs gives:
\begin{align}
    \vcenter{\hbox{\includegraphics[scale=.2]{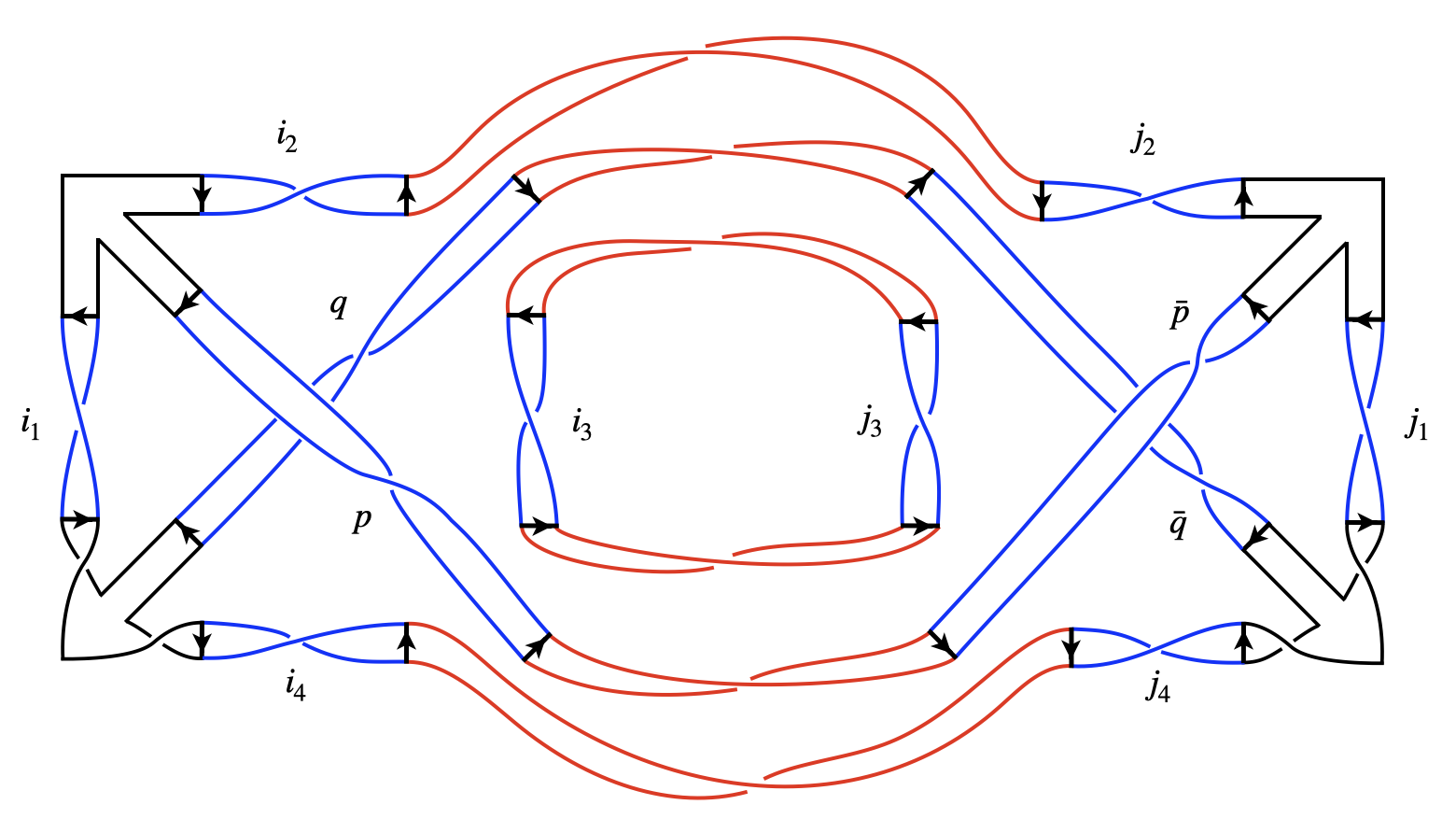}}} &= \vcenter{\hbox{\includegraphics[scale=.2]{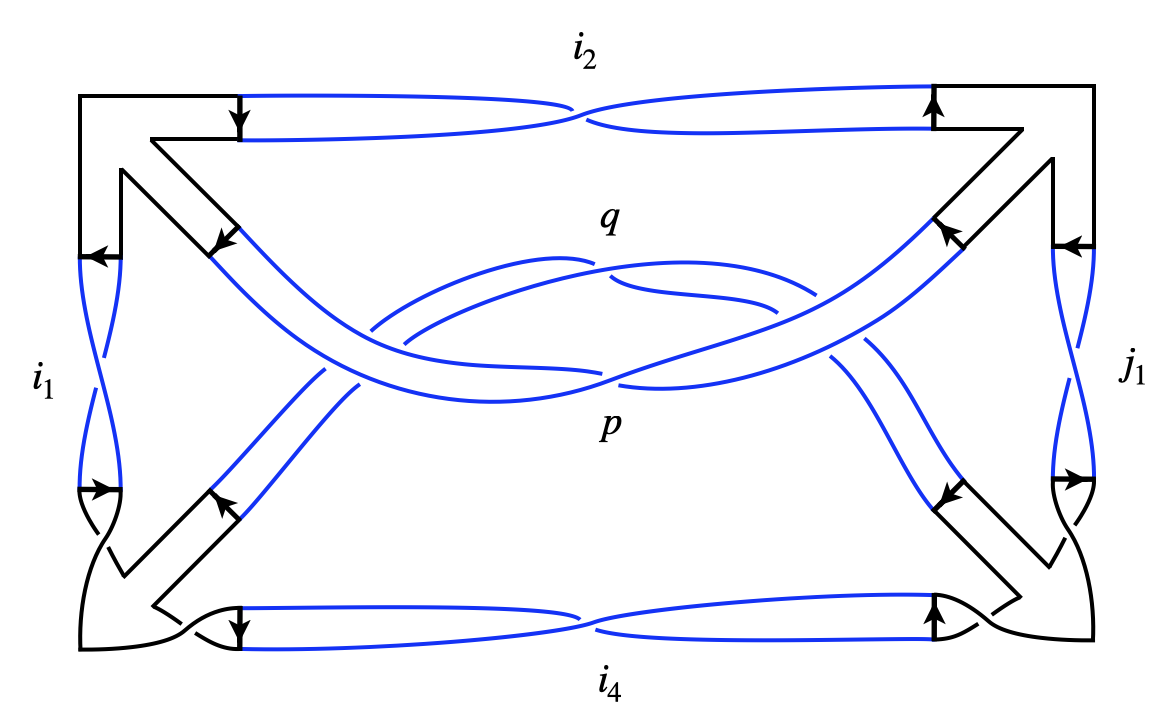}}} \\
    &= \vcenter{\hbox{\includegraphics[scale=.2]{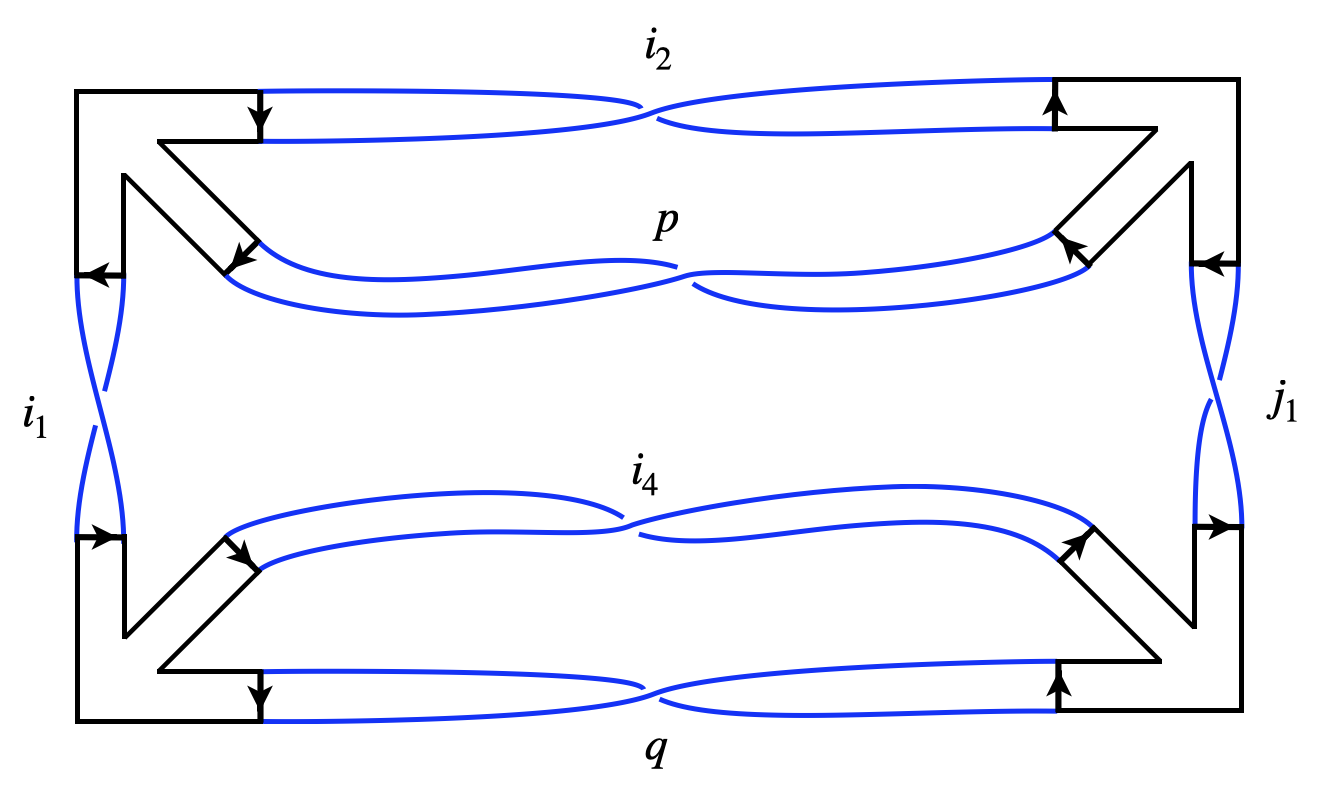}}}
\end{align}
where the final figure is the ribbon diagram for the pillow.
This reproduces the 6J orthogonality relation \eqref{linesortho}.

\subsection{Hexagon identity} 
To further illustrate the consistency of the ribbon calculus, we consider the gluing of two 6J ribbons with the \emph{same} orientation. 
We show that the surgery operation on the resulting diagram reproduces the Hexagon ID identity.

In terms of the F and R matrix, the Hexagon identity can be expressed schematically as
\begin{align}
    \mathbb{F} \mathbb{R} \mathbb{F}= \mathbb{R} \mathbb{F} \mathbb{R}  
\end{align}
Explicitly it is given by
\begin{align}\label{HexID}
\int_{0}^{\infty} dP_q \,\, \mathbb{F}_{pq} \begin{bmatrix} j& k \\ i & l \end{bmatrix}  e^{-i \pi (h_{l}+h_{q} -h_{i} )} \mathbb{F}_{qr} \begin{bmatrix} l& j \\ k & i \end{bmatrix}=e^{-i \pi (h_{k}+h_{l}-h_{p})}\mathbb{F}_{pr} \begin{bmatrix} j& l \\ i & k \end{bmatrix} e^{-i \pi (h_{j}+h_{l}-h_{r} )}
\end{align} 

In our gravity model, this appears from the gluing of two 6J ribbon graphs in a particular orientation.   To see how this arises, consider a rewriting of \eqref{HexID} in which we multiply both sides by a phase $e^{-2\pi i h_{i}}$.  Combining this with the analogous hexagon identity for the opposite chirality gives
\begin{align}\label{HexIDmod}
\int_{0}^{\infty} dP_q \,\, \left|\mathbb{F}_{pq} \begin{bmatrix} j& k \\ i & l \end{bmatrix}\right
|^2 & e^{-i \pi (s_{l}+s_{q} +s_{i} )} \left|\mathbb{F}_{qr} \begin{bmatrix} l& j \\ k & i \end{bmatrix} \right|^2 \nn
&= e^{-2\pi i s_{i}}e^{-i \pi (s_{k}+s_{l}-s_{p})} \left|\mathbb{F}_{pr} \begin{bmatrix} j& l \\ i & k \end{bmatrix}  \right|^2 e^{-i \pi (s_{j}+s_{l}-s_{r} )}
\end{align} 
Note the symmetric phase on the left hand side of this equation, which is the type of phase that arises from the tensor model propagator.  Writing this in terms of the tetrahedrally symmetric 6J symbols gives 
\begin{align}
   \int_{0}^{\infty} d^2 P_q |\rho(q) |^2  \left| \begin{Bmatrix}
    p & k & l \\
    q & i & j \end{Bmatrix} \right|^2  &e^{-i \pi (s_{l}+s_{q} +s_{i} )} 
     \left| \begin{Bmatrix}
     q & j & k \\
    r & i & l \end{Bmatrix} \right|^2
     \nonumber \\ &=e^{-2\pi i s_{i}}e^{-i \pi (s_{k}+s_{l}-s_{p})}  \left| \begin{Bmatrix}
     p & l & k \\
    r & i & j \end{Bmatrix} \right|^2 e^{-i \pi (s_{j}+s_{l}-s_{r} )}
\end{align}

Now consider gluing together 6J ribbon graphs as shown below 
\begin{align}
     \int_{0}^{\infty} d^2 P_q |\rho(q)|^2  \vcenter{\hbox{ \includegraphics[scale=.25]{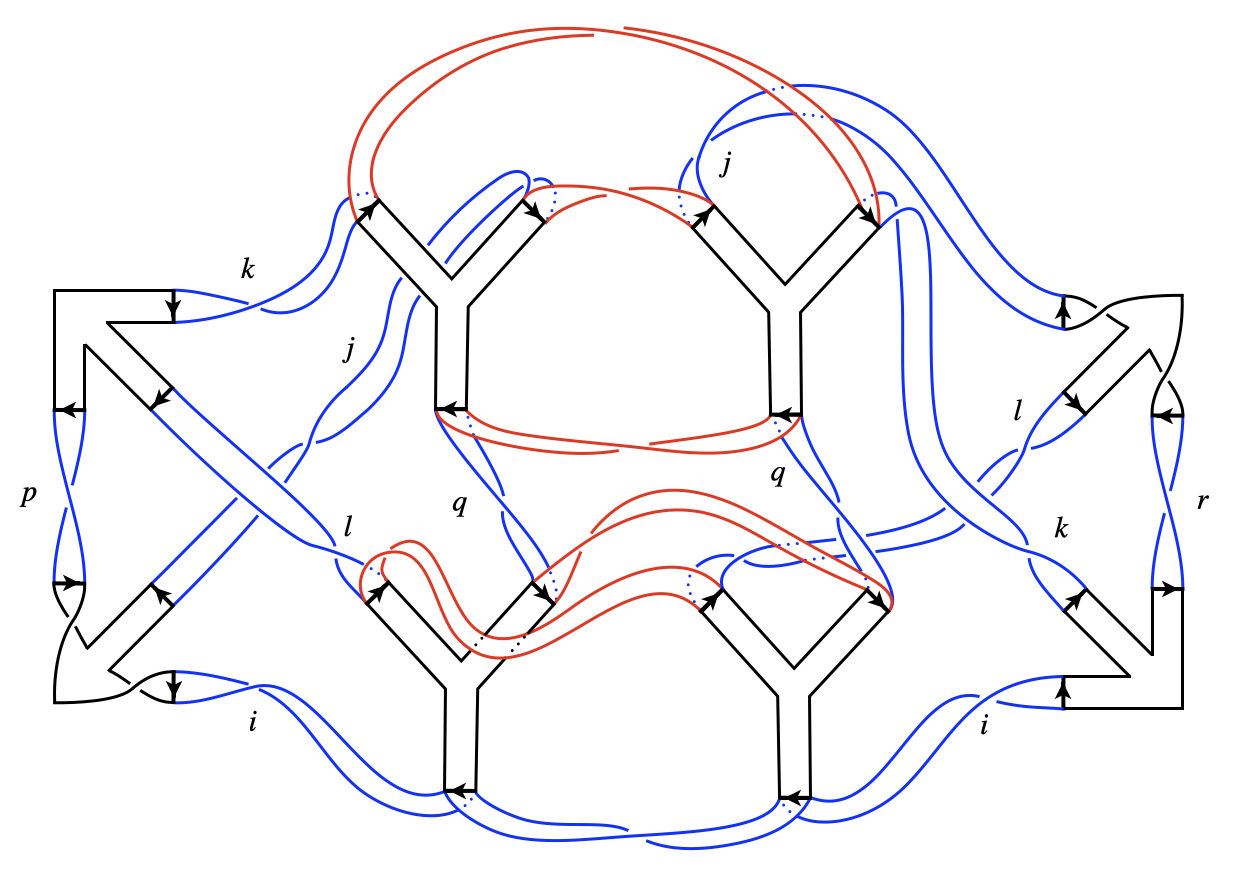}}}
\end{align}
Here we have included the junctions we are gluing to illustrate that the ribbons connecting them are consistent with the definitions  \eqref{CijkCijk}, \eqref{CijkCjik}.   Removing the glued  junctions gives 
\begin{align}
     \int_{0}^{\infty} d^2 P_q |\rho(q)|^2  
     \vcenter{\hbox {\includegraphics[scale=.25]{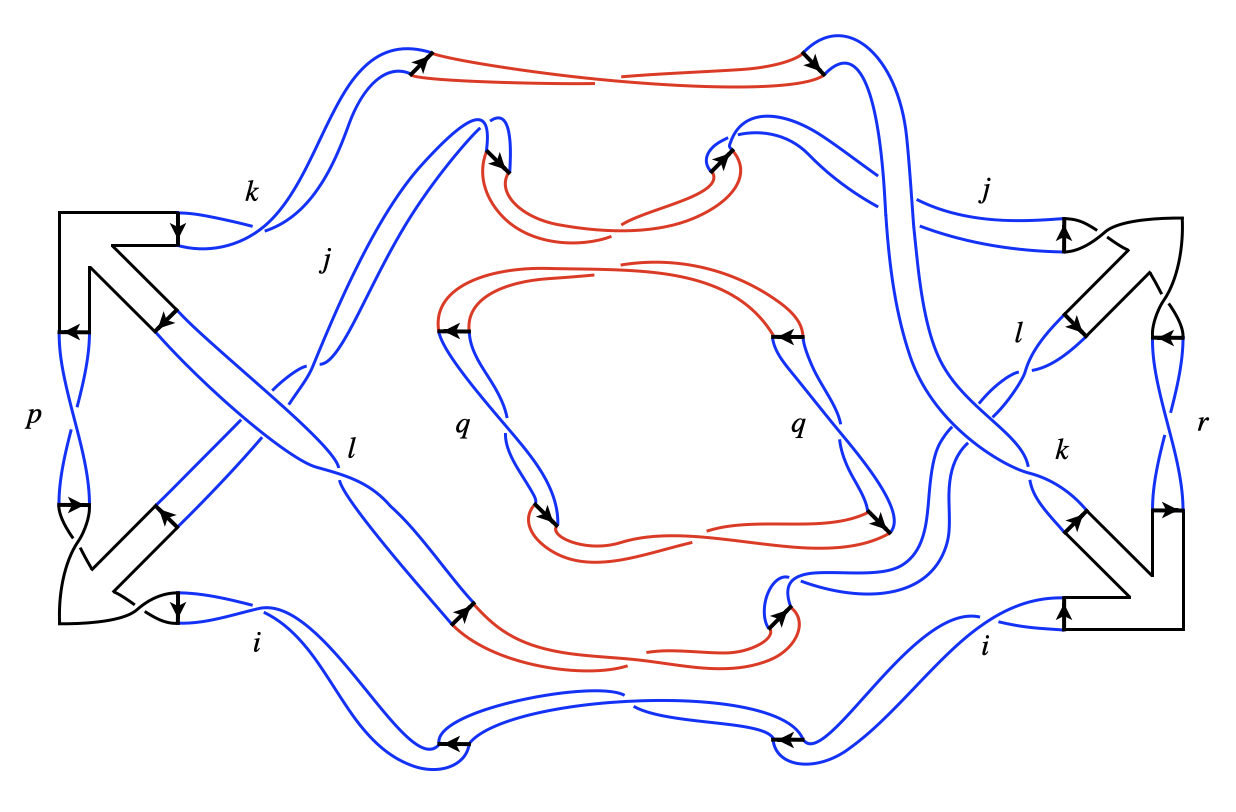}}}
\end{align}
The loop weighted by $q$ is untwisted, so this gives an Omega loop which implements surgery.   This gives  
\begin{align}
     \vcenter{\hbox {\includegraphics[scale=.2]{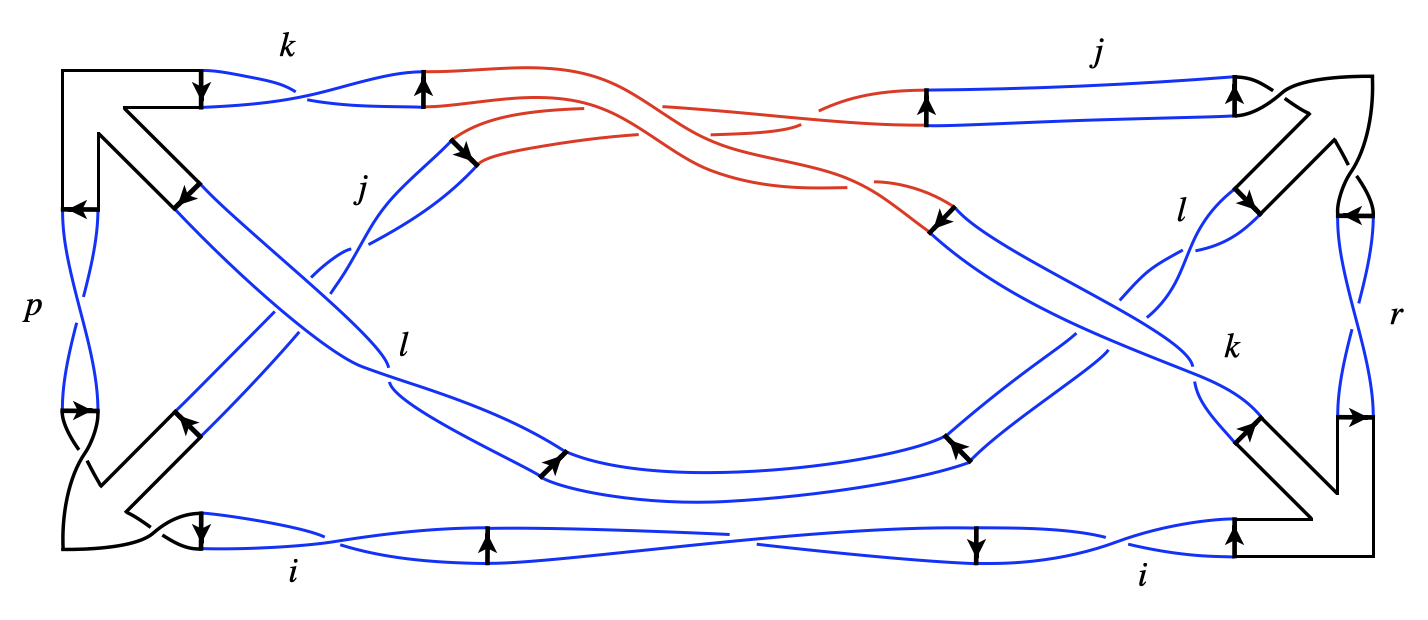}}}=    e^{-2\pi i s_{i}}  
     \vcenter{\hbox {\includegraphics[scale=.2]{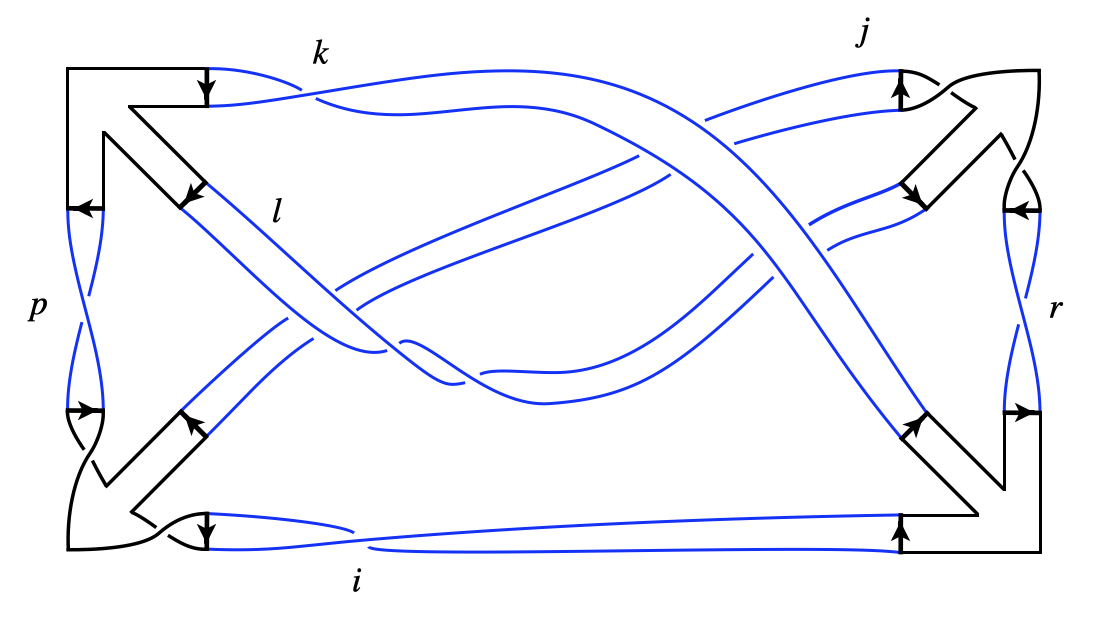}}}
\end{align}
On the RHS, we  canceled the red and blue twists on the ribbon, and we replaced a double twist on the ith ribbon with the phase $e^{-2
\pi s_{i}}$.  Finally, to arrive at the hexagon ID, we perform a simultaneously braiding of the junctions on the upper left and right corner: 
\begin{align}
     e^{-2\pi i s_{i}} 
     \vcenter{\hbox {\includegraphics[scale=.2]{figures/fin_hex2.png}}} &=e^{-2\pi i s_{i}}
     \vcenter{\hbox {\includegraphics[scale=.2]{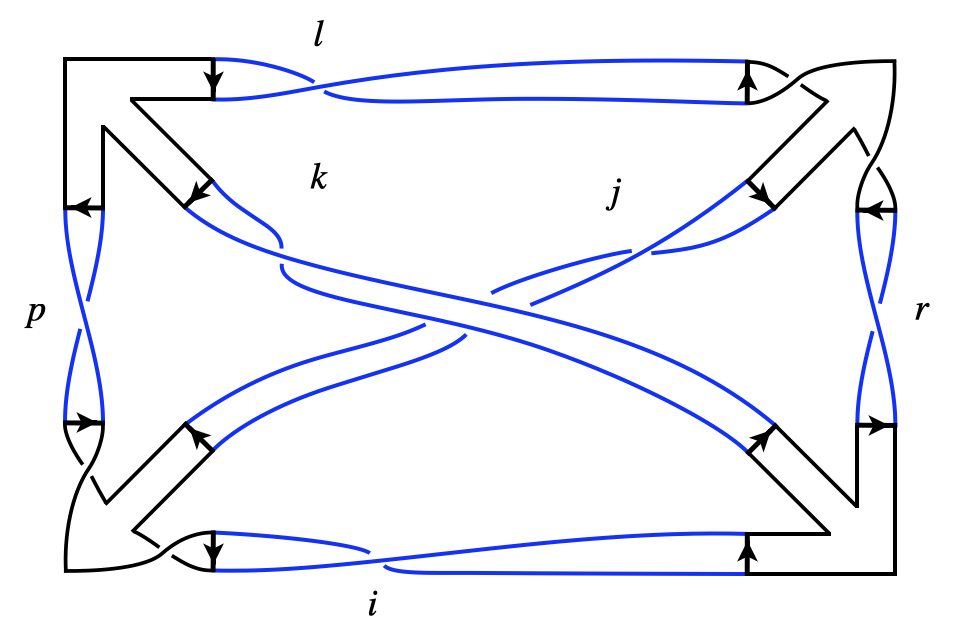}}}e^{-i \pi (s_{k}+s_{l}-s_{p})} e^{-i \pi (s_{j}+s_{l}-s_{r} )}  
\end{align}
Thus surgery produces the diagrammatic identity: 

\begin{align}
     \int_{0}^{\infty} d^2 P_q |\rho(q)|^2  &
     \vcenter{\hbox {\includegraphics[scale=.25]{figures/fin_hex_nojunctions.png}}} \nn
     &=  \vcenter{\hbox {\includegraphics[scale=.2]{figures/fin_hex3.png}}} \, e^{-i \pi (s_{k}+s_{l}-s_{p})} e^{-i \pi (s_{j}+s_{l}-s_{r} )} 
\end{align}
In terms of 6J symbols, this translates to: 
\begin{align}
e^{- i \pi (s_{p}+s_{k}+s_{j})} &
   \int_{0}^{\infty} d^2 P_q |\rho(q) |^2  \left| \begin{Bmatrix}
    p & k & l \\
    q & i & j \end{Bmatrix} \right|^2  e^{-i \pi (s_{l}+s_{q} +s_{i} )} 
     \left| \begin{Bmatrix}
     q & j & k \\
    r & i & l \end{Bmatrix} \right|^2
e^{-i \pi ( s_{i}+s_{r} +s_{k} )}
     \nonumber \\ &=e^{-2\pi i s_{i}}e^{-i \pi (s_{k}+s_{l}-s_{p})}  \left| \begin{Bmatrix}
     p & l & k \\
    r & i & j \end{Bmatrix} \right|^2 e^{-i \pi (s_{j}+s_{l}-s_{r} )} e^{- i \pi ( s_{p}+s_{l}+s_{k}+s_{r}+s_{i}+s_{j})}
\end{align}
which is equivalent to the hexagon identity \eqref{HexID}
\subsection{ Ribbon calculation of the 6J self contraction}
Let us now revisit the the 6J self contraction diagram for 
\begin{align}
\braket{\mathbf{C}_{ijk} \mathbf{C}_{jik} }   
\end{align}
using the ribbon graphs, and check that the correct phase is reproduced.   We start by focusing on the self contraction (leaving aside the gluing of the external propagator for the moment).  Applying the rule \eqref{CijkCijk}  to the bottom two junctions gives:
\begin{align}
\vcenter{\hbox{\includegraphics[scale=0.35]{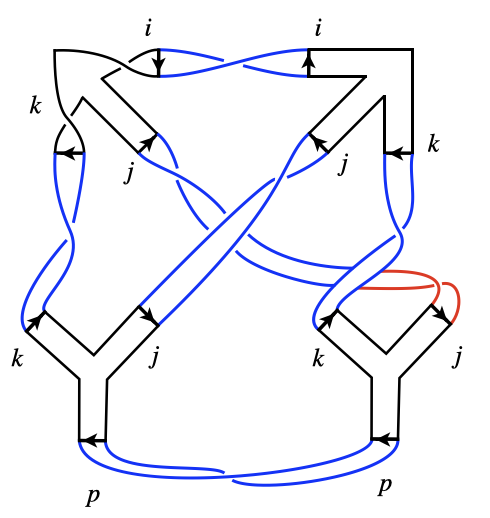}}} &= \vcenter{\hbox{\includegraphics[scale=0.35]{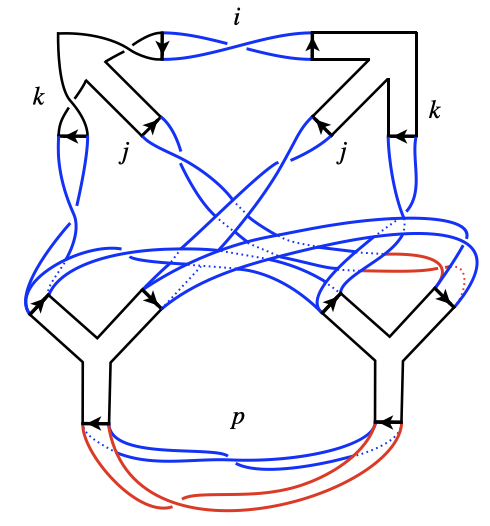}}} = \vcenter{\hbox{\includegraphics[scale=0.35]{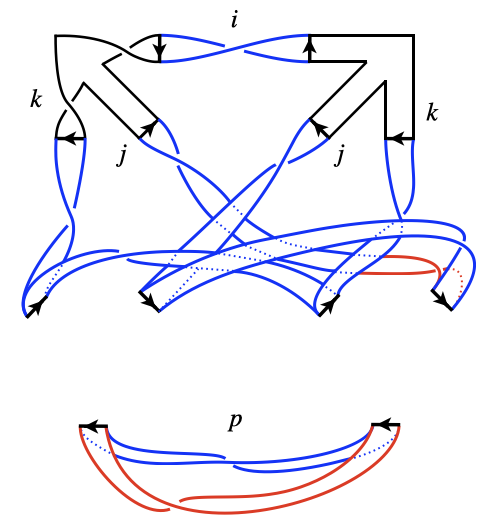}}} \nn
    &= \vcenter{\hbox{\includegraphics[scale=0.35]{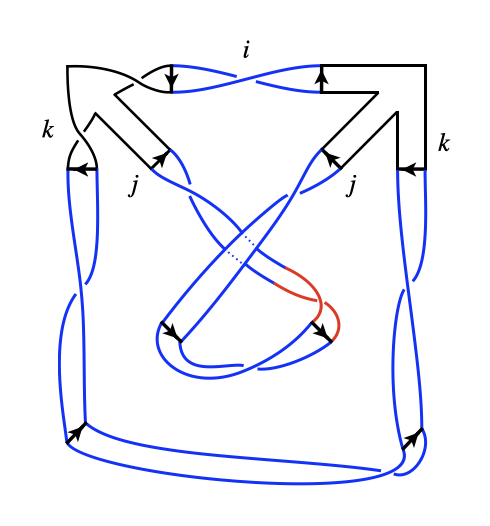}}} = \vcenter{\hbox{\includegraphics[scale=0.35]{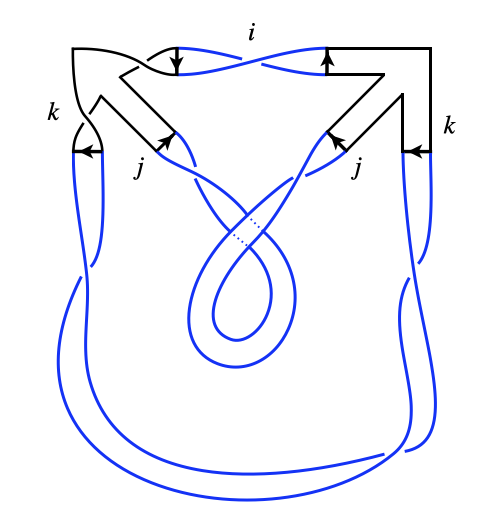}}} 
\end{align}
\begin{align}\label{greenselfcontract}
\vcenter{\hbox{\includegraphics[scale=0.3]{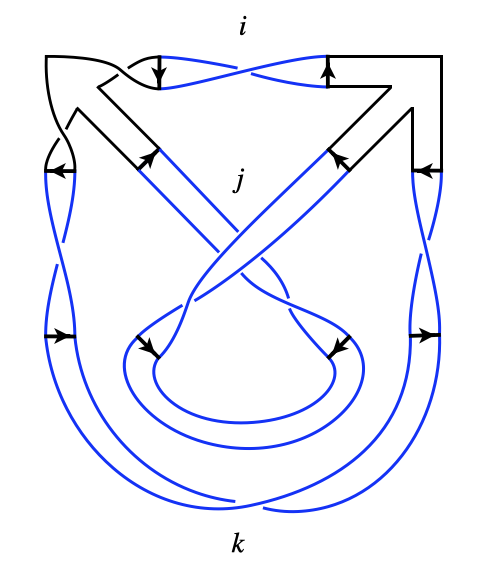}}} = e^{-4\pi i (s_j + s_k)} \vcenter{\hbox{\includegraphics[scale=0.3]{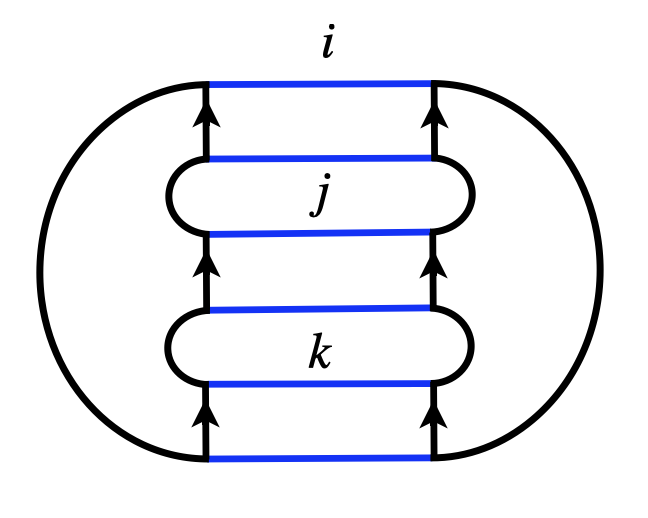}}}
\end{align}

Gluing in the external propagator using the rule \eqref{oddface} and \eqref{CijkCijk} just produces right handed twists on each leg of the upper left junction. This gives 
\begin{align}
\braket{\mathbf{C}_{ijk} \mathbf{C}_{jik}}_{\text{Self-contract}}
=\vcenter{\hbox{\includegraphics[scale=.25]{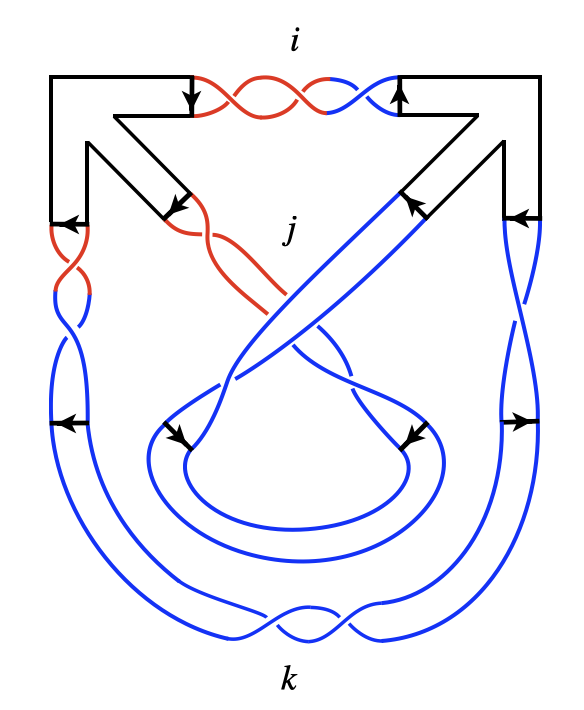}}} =e^{ -3 \pi i(s_{j}+s_{k} ) +i \pi s_{i} }|C_0(jki)|^2,
\end{align}
which is consistent with \eqref{6jselffin}, once we account for the phase difference between $C_{ijk}$ and $\mathbf{C}_{ijk}$ as given in \eqref{BoldCijk} 

\paragraph{A special case of the Hexagon ID}
We pointed out previously that the self contraction diagram above corresponds to a special case of the Hexagon equation, where one line is set to the identity.  In the framed version of the Hexagon equation, setting a ribbon to identity  is tricky: since the ribbon ends on a junction, one has to define the framing for the ribbons left behind.  However, we can check that the self contraction corresponds to setting a ribbon to identity in the following way:
\begin{align}
& \vcenter{\hbox{\includegraphics[scale=0.25]{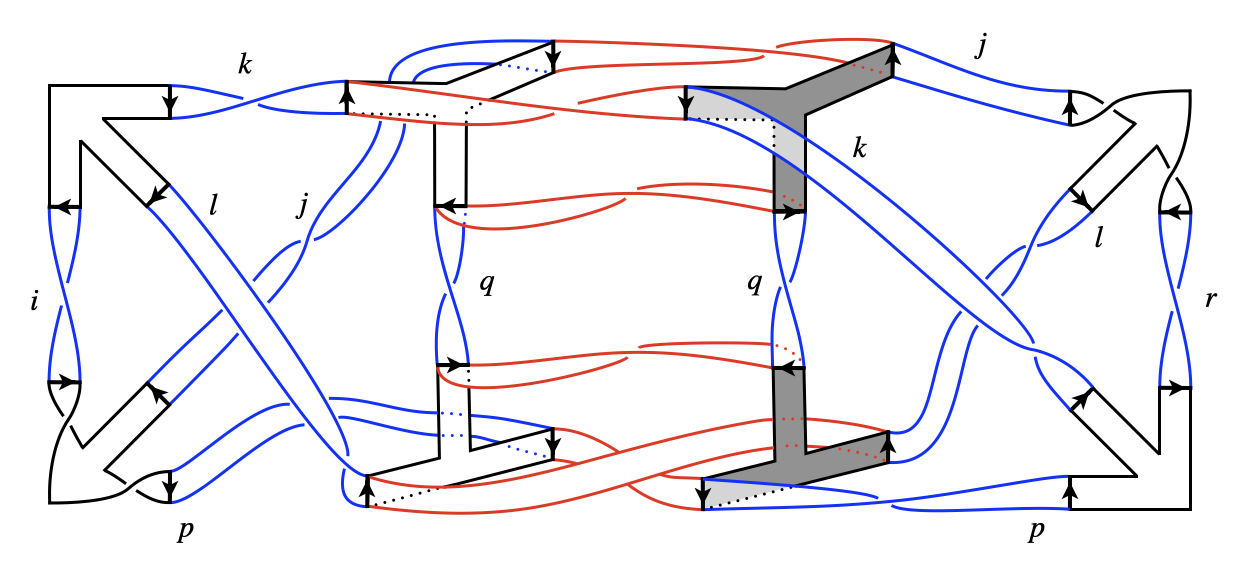}}} \xrightarrow[r = \mathbbm{1}]{} \vcenter{\hbox{\includegraphics[scale=0.25]{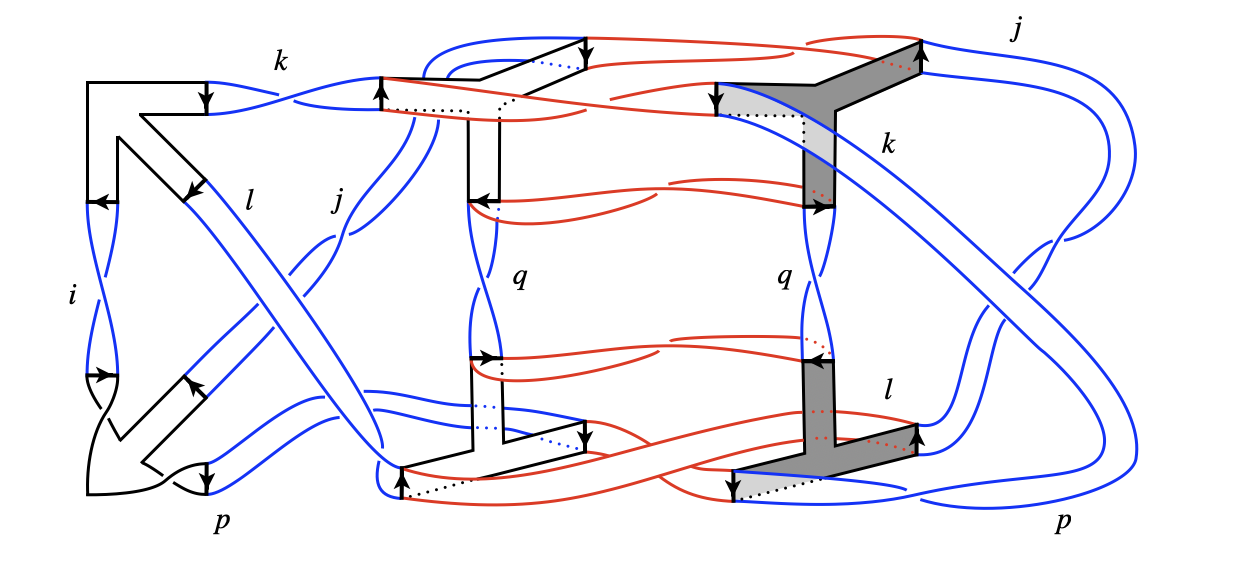}}} \\
& = \vcenter{\hbox{\includegraphics[scale=0.25]{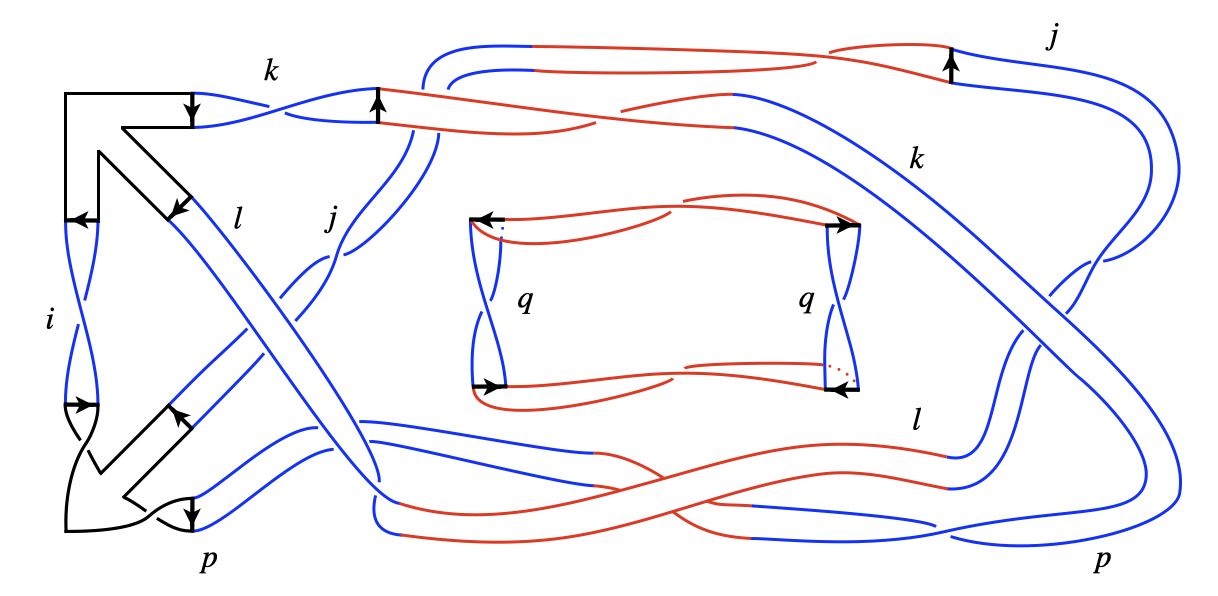}}} = \vcenter{\hbox{\includegraphics[scale=0.25]{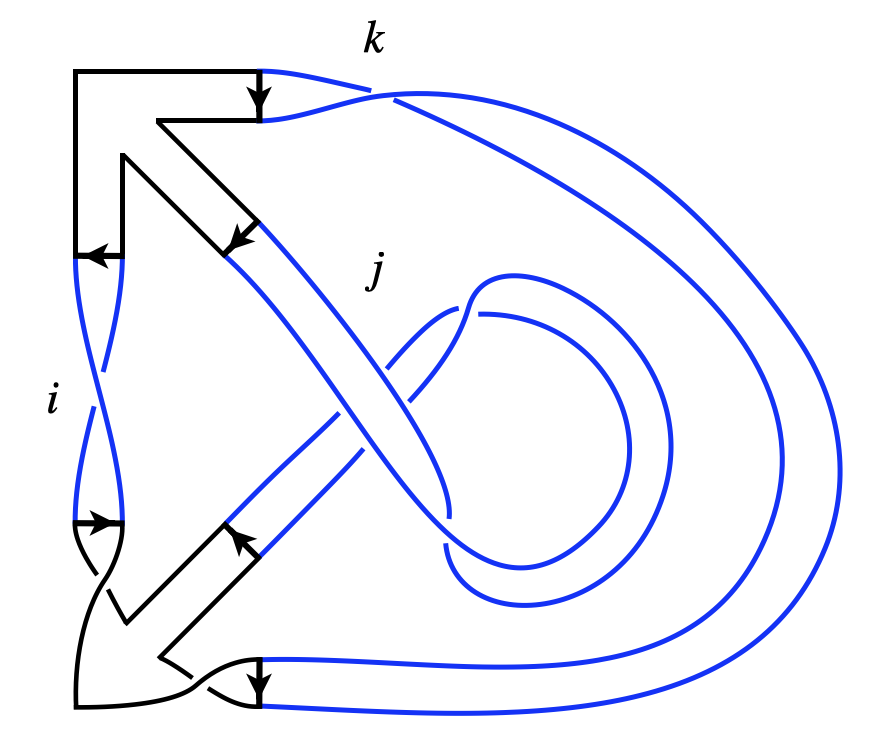}}}
\end{align}

\subsection{Generating all 3-manifolds} 

To show that the CFT ensemble is equivalent to the full 3d gravity path integral, we must show that its perturbative expansion produces a sum over all 3-manifolds.    
In this subsection,  we explain how the  sum over Wilson lines associated to the triple line diagrams, when combined with the surgery operations implemented by the matrix model, is sufficient to generate all compact, oriented 3-manifolds. 
We start by considering closed manifolds, and then generalize to manifolds with boundaries.

\paragraph{Closed Manifolds}
Closed Manifolds arise from the vacuum diagrams of the CFT ensemble.  According to Lickorish's theorem, 
all closed, orientable,  connected 3-manifolds can be obtained from surgery on links in $S^3$ \cite{Lickorish_62}.  Therefore, it suffices to show that the CFT ensemble generates all links\footnote{A link is set of disconnected knots that are linked together in a topologically nontrivial way. } inside $S^3$.   The key observation is that an arbitrary link can be  produced by gluing together 6J manifolds in an alternating pattern such as the one shown below:
\begin{align}
    \vcenter{\hbox{\includegraphics[scale=.3]{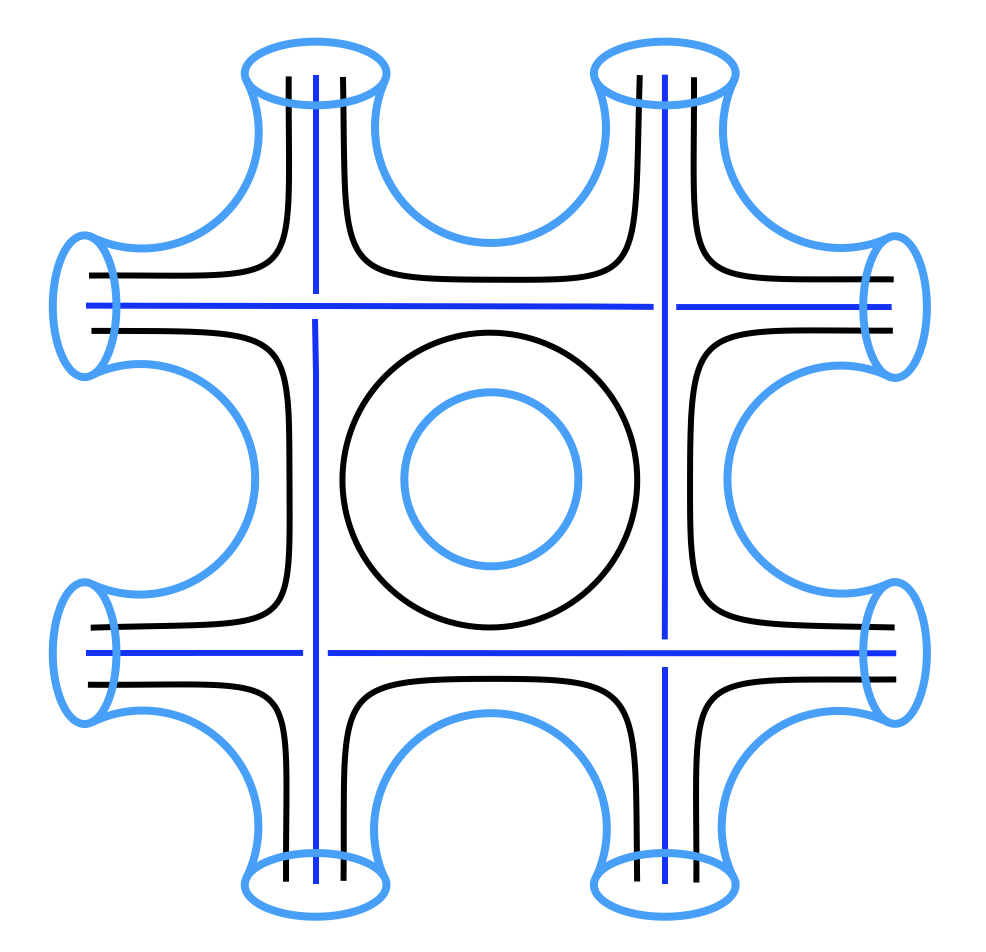}}}
\end{align}
Notice that a linking pattern of Wilson lines is produced by rotating the 6J manifolds by 90 degrees relative to each other, which produces a pattern of over and under crossings.   After creating the desired linking pattern, we close off the manifold by attaching $C_{0}$ manifolds to the thrice punctured sphere boundaries as shown below.
\begin{align}\label{6jclosed}
\vcenter{\hbox{\includegraphics[scale=.2]{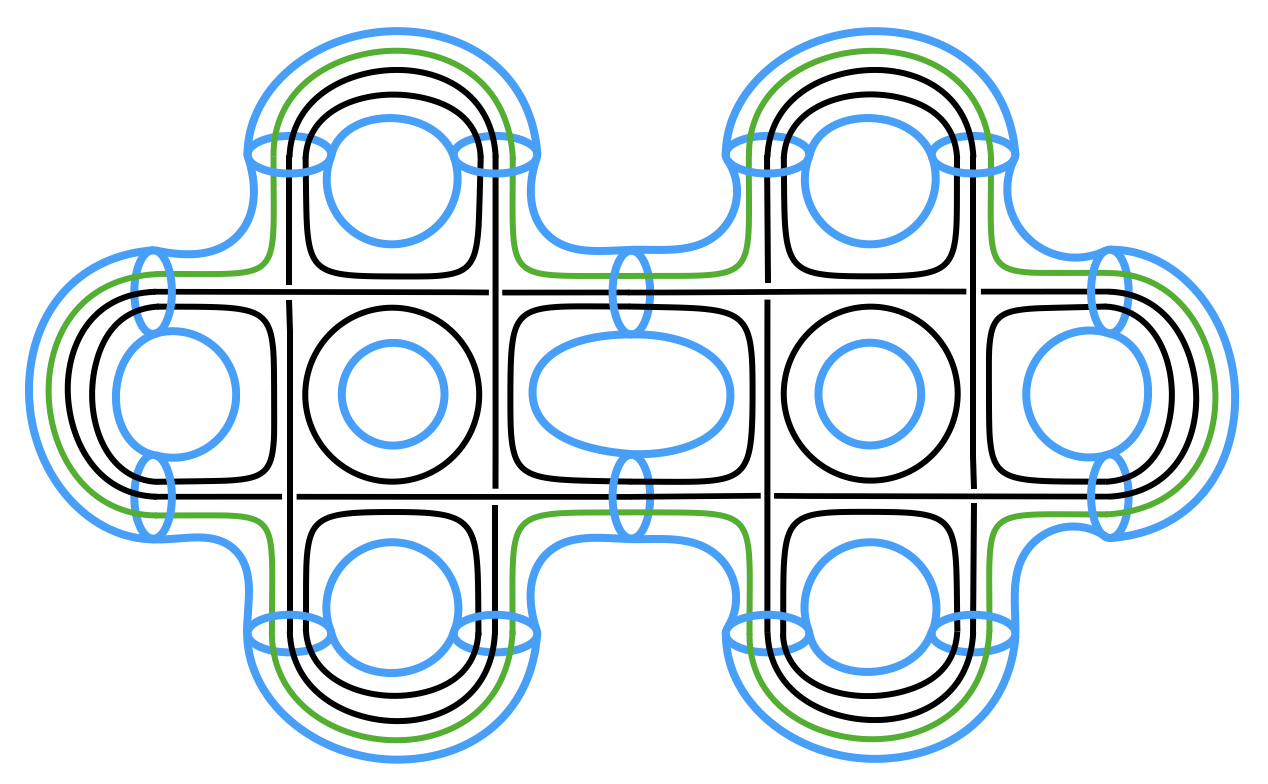}}}
\end{align}
The  non-contractible cycles in this manifold can be removed by surgery since they are wrapped by Omega loops.   This gives a link (black)  generated by the over and under crossings of the ``diagonal" lines in the 6J manifold, plus a separate unknot (colored in green) which is not entangled with the link. Surgery on this unknot would produce a manifold different from $S^3$. which we do not want. Therefore, we set this Wilson loop to be the identity.  What remains is a link in $S^3$: 
\begin{align}
\vcenter{\hbox{\includegraphics[scale=.2]{figures/fin_two6jC01.png}}} \to \vcenter{\hbox{
\includegraphics[scale=.2]{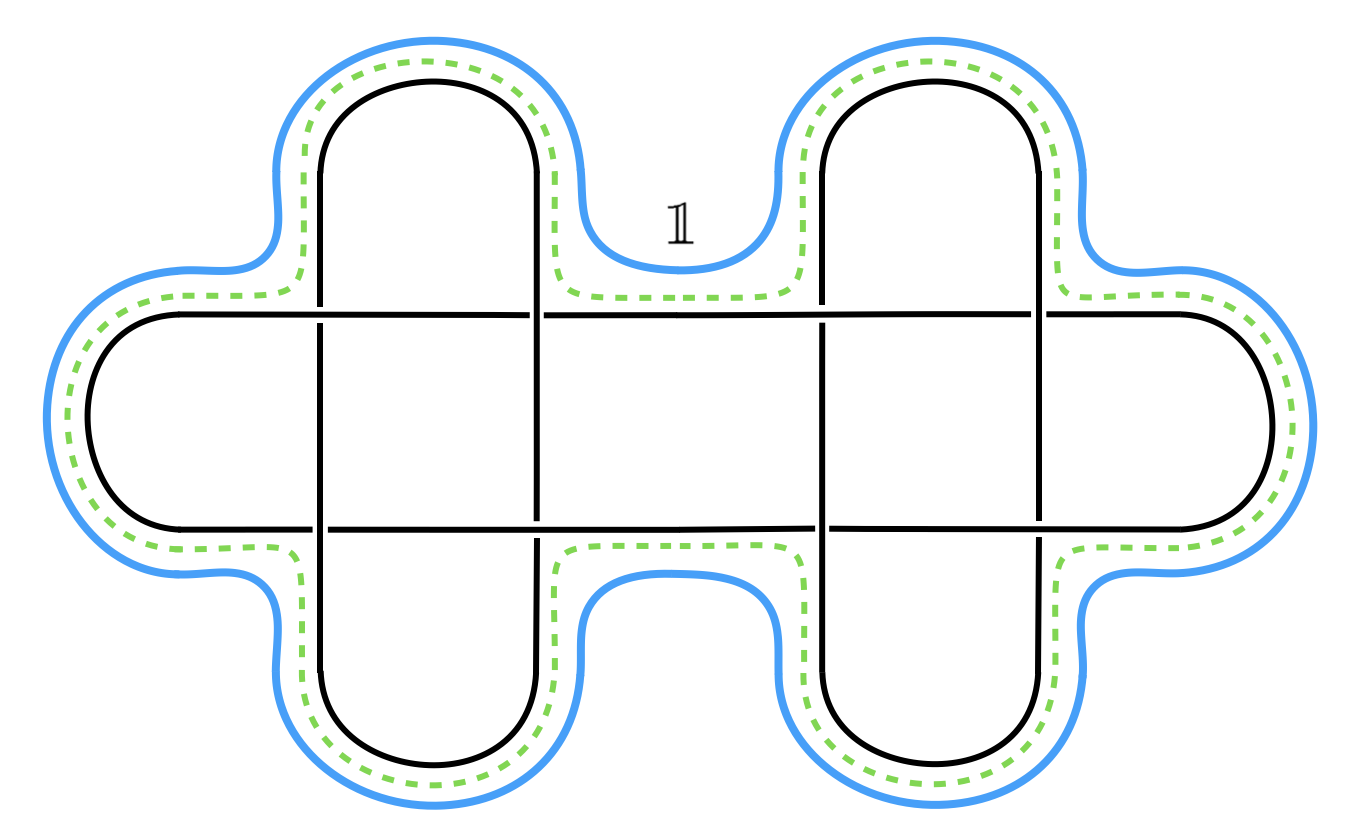}}}=\vcenter{\hbox{
\includegraphics[scale=.45]{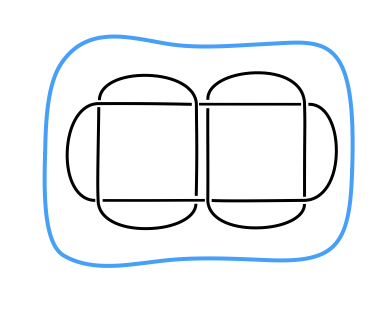}}}
\end{align}
Note that the extra (green)  Wilson loop was constructing from gluing $C_{0}$ and 6J manifolds.   Consistency with this gluing construction requires that we obtain the same link by setting segments of the Wilson loop to identity locally inside each chunk of the manifold.  This requires the following identities:
\begin{align} \label{6jID}
&\vcenter{\hbox{\includegraphics[scale=.2]{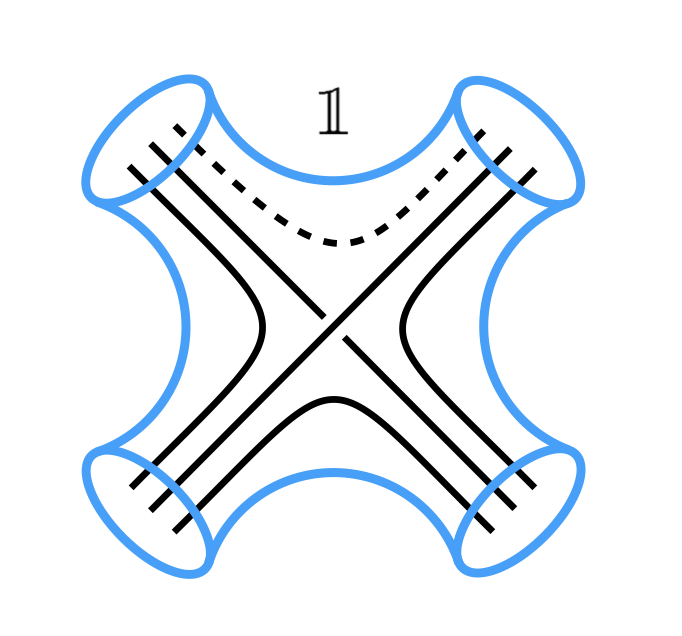} }} = \vcenter{\hbox{\includegraphics[scale=.2]{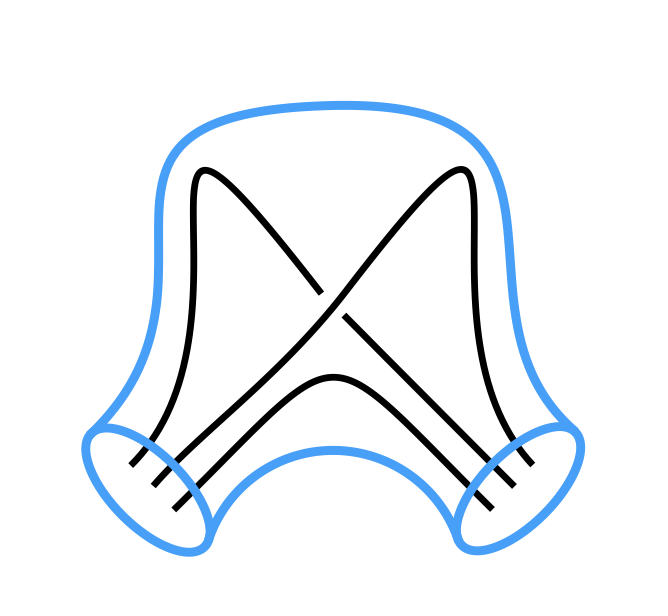} }} = \vcenter{\hbox{\includegraphics[scale=.2]{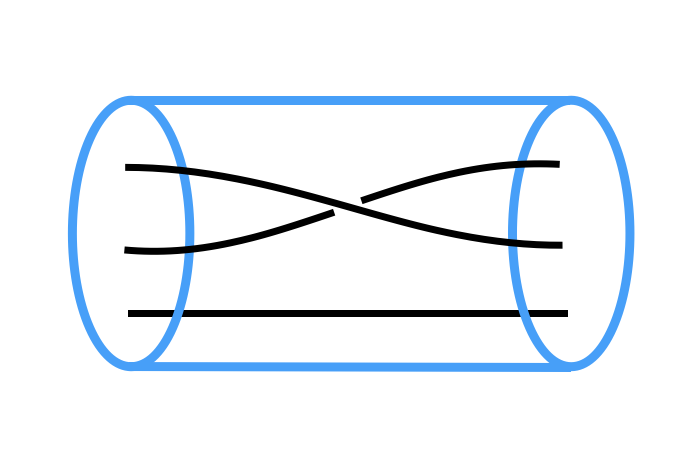} }}
\\
& \int dP_1 \, \rho_0(P_1) \lim_{P_3 \to \mathbbm{1}} \vcenter{\hbox{\includegraphics[scale=0.35]{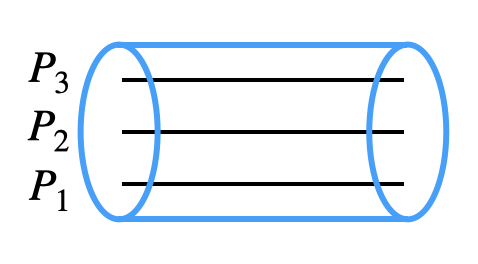}}} = \vcenter{\hbox{\includegraphics[scale=0.35]{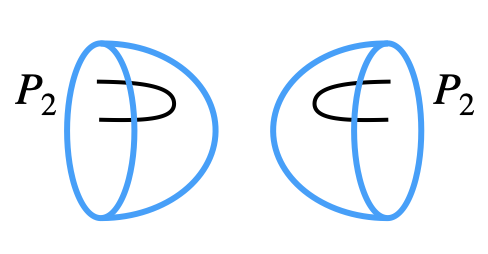}}} \, ,
\end{align}
where the second equality comes from the analytic continuation of the $C_{0}$ function
\begin{align}
    \lim_{P_{3} \to \mathbbm{1}} C_{0}(P_{1},P_{2},P_{3}) = 
    \frac{\delta(P_{1}-P_{2})}{\rho_{0}(P_{1})}.
\end{align}
Applying these identities gives the same link as above:
\begin{align}
\vcenter{\hbox{  \includegraphics[scale=.2]{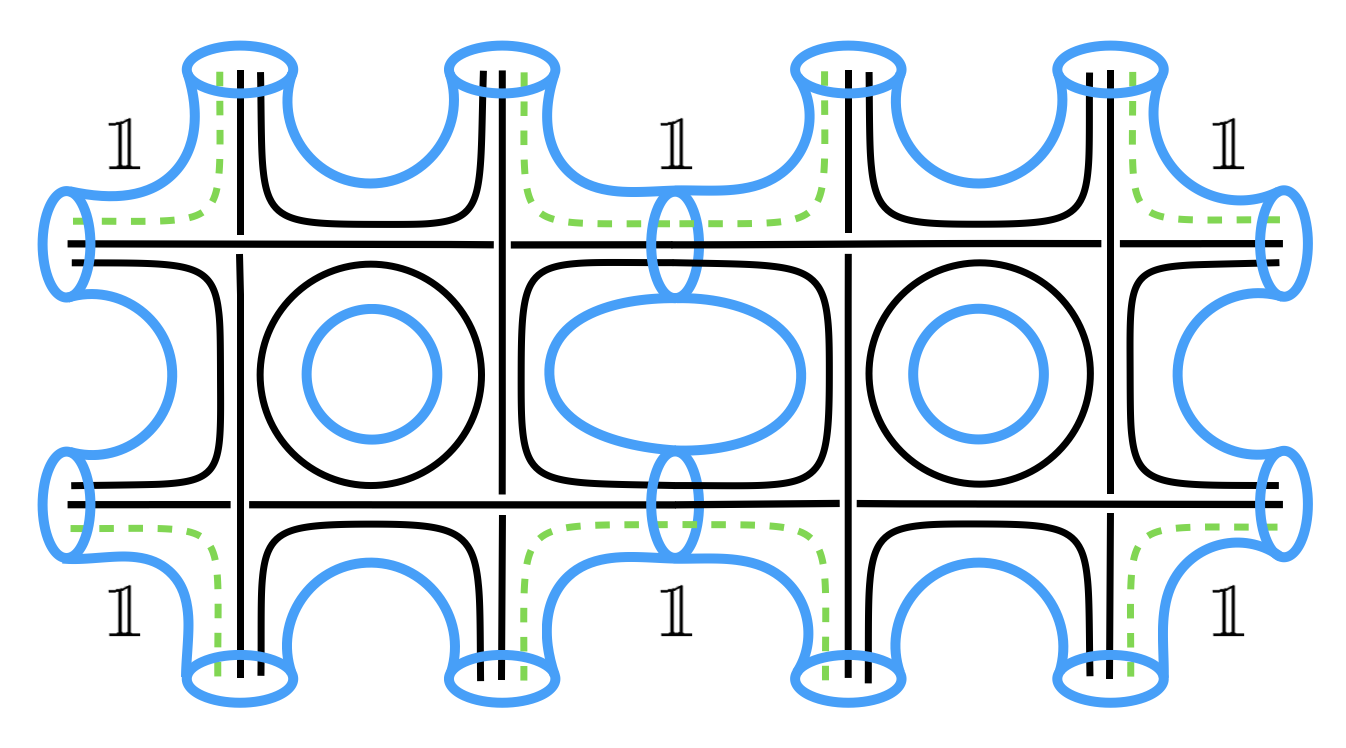}}}
=\vcenter{\hbox{  \includegraphics[scale=.45]{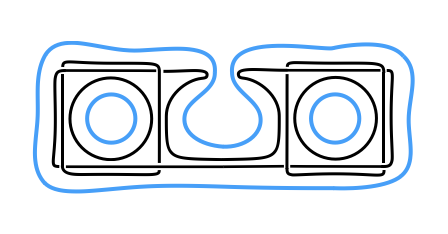}}}=\vcenter{\hbox{
  \includegraphics[scale=.45]{figures/fin_two6jlink.png}}}
\end{align}
Finally, to see that an arbitrary link is produced in this way, it suffices to observe that before closing off the manifold, we can engineer an arbitrary braid.  Then we can apply a theorem from  \cite{Birman_1976} that guarantees that an arbitrary link can be produced from the closure of an arbitrary braid.
To illustrate our construction in an non-trivial example, consider a Borromean ring in $S^3$:
\begin{align}\label{borromean_rings_link}
    \vcenter{\hbox{\includegraphics[scale=0.3]{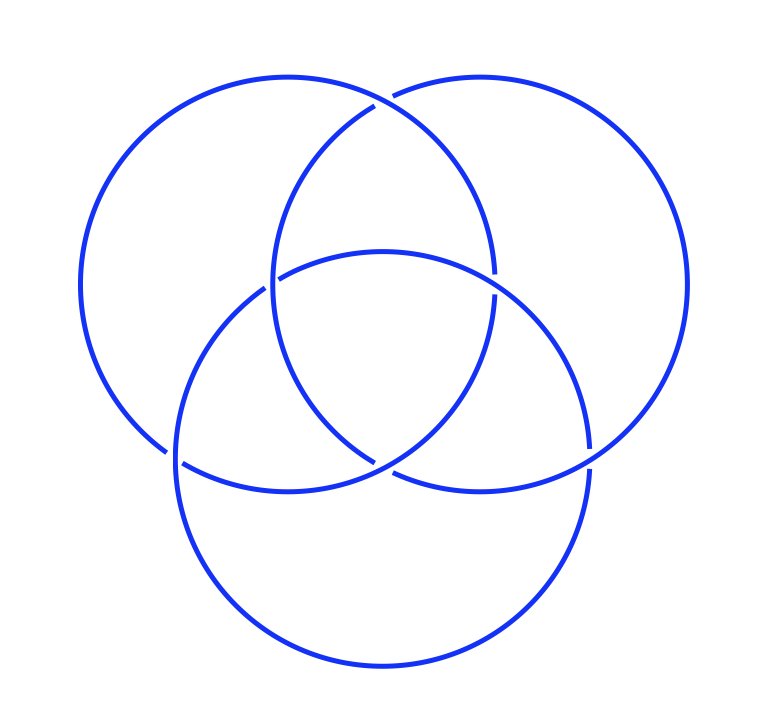}}}
\end{align}
To produce it from a gluing of 6J manifolds, we just turn each line into a triple line propagator.  This produces (orange) loops surrounding $S^2$ handles that we remove by surgery.
\begin{align}
    \vcenter{\hbox{\includegraphics[scale=0.25]{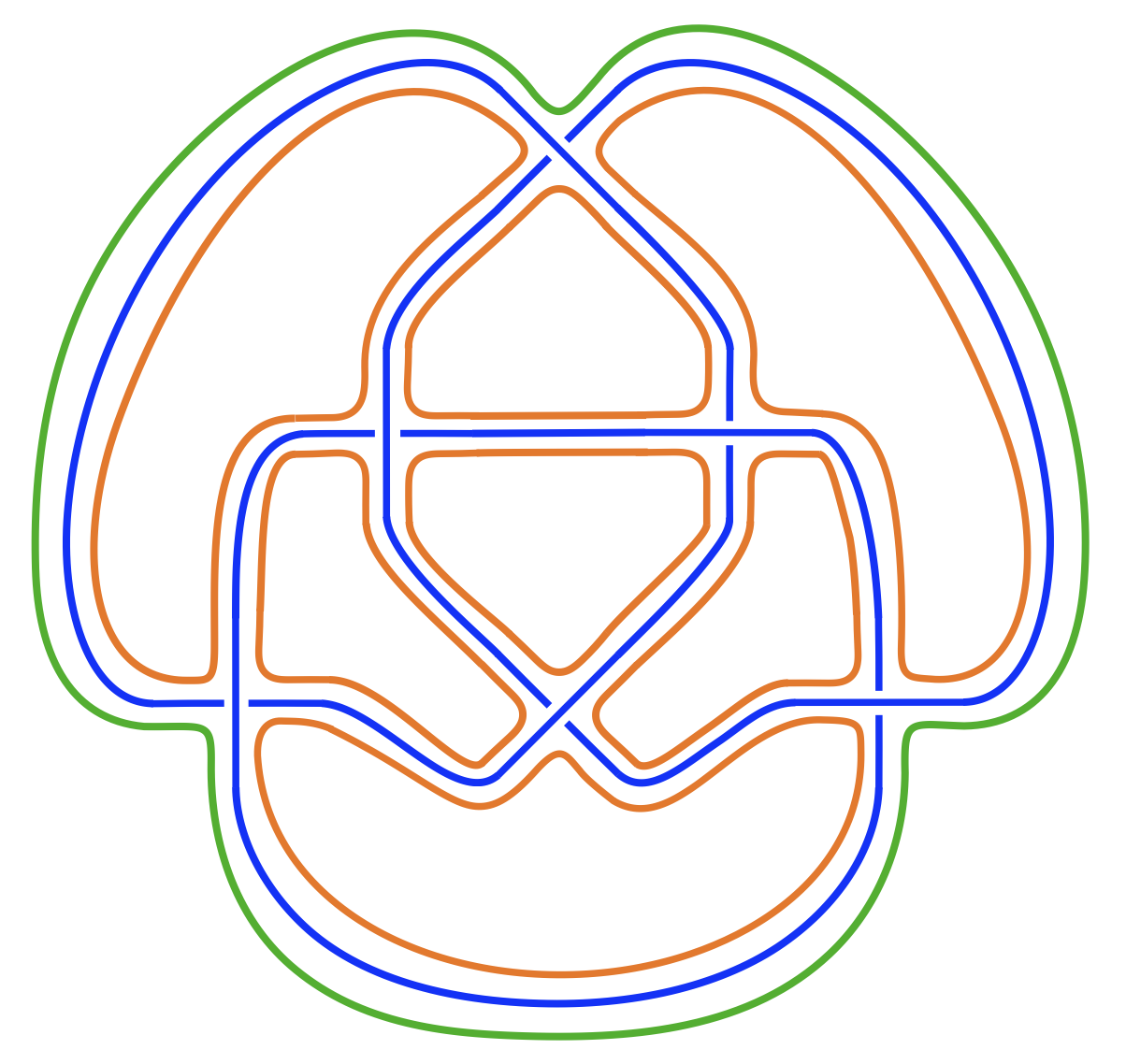}}} \rightarrow \vcenter{\hbox{\includegraphics[scale=0.25]{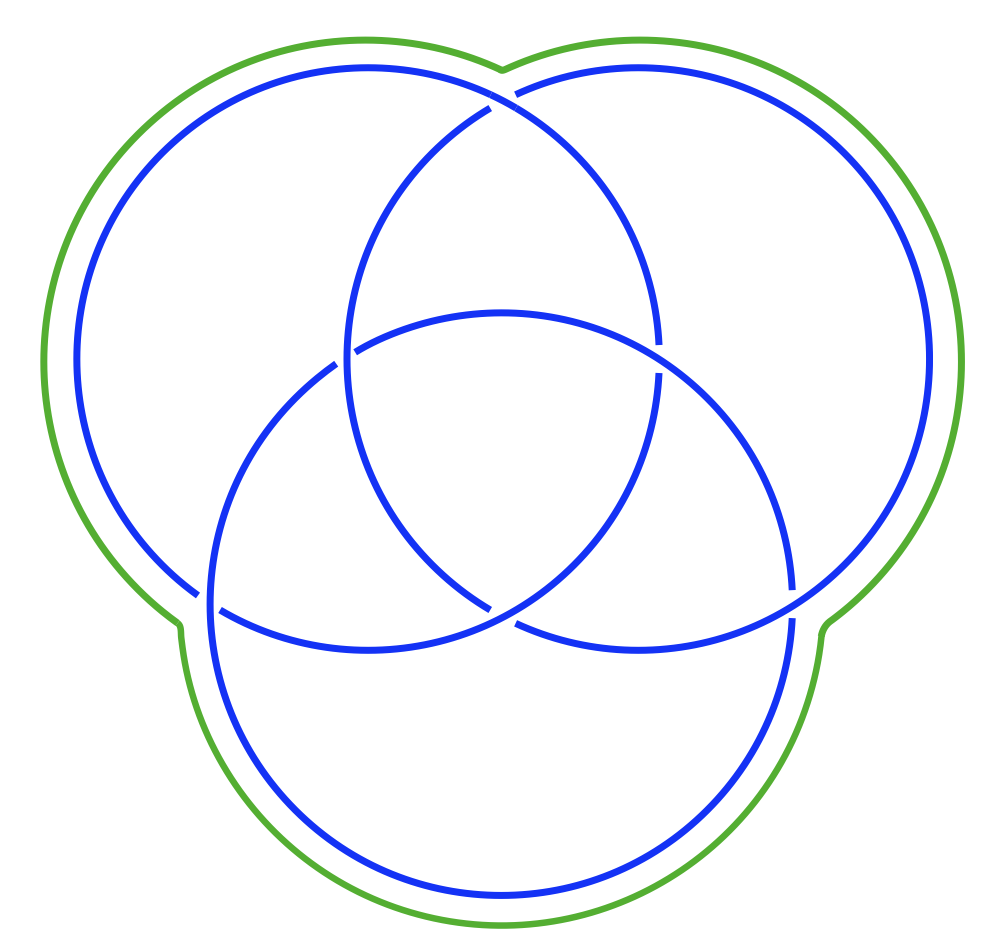}}}
\end{align}
Setting the extra green loop to identity as in \eqref{6jID} then produces the Borromean ring in $S^3$.
\paragraph{Manifolds with boundary}
There is a generalization of Lickorish's theorem, which states that any two compact, orientable 3-manifold with boundary are related by surgery on a link if their boundaries are homeomorphic \cite{Roberts1997KirbyCI}\footnote{This is explained in the 2nd paragraph of this reference.}.  This means that given a fixed boundary $\Sigma$ and a reference manifold $\mathcal{M}$ such that $\partial \mathcal{M}= \Sigma$, we can produce all 3-manifolds with boundary $\Sigma$ by performing surgery on a link in $
\mathcal{M} $ \cite{Roberts1997KirbyCI}.

We now apply this theorem to show that the CFT ensemble generates all manifolds with boundaries.   The main idea is to modify the ``boundary surgery" operation explained in \ref{sec:boundarysurgery} to include surgery over an arbitrary bulk link. 
To be concrete,  let's re-consider the expectation value of the genus 2 observable given in \eqref{genus2obs}. In section \ref{sec:boundarysurgery}, we considered a particular contribution to this expectation value given by a particular handlebody $\mathcal{M}$ bounded by a genus two: 
\begin{align}\label{genus2bdry}
 \vcenter{\hbox{\includegraphics[scale=0.23]{figures/fin_g2_bdry.png}}}= \iiint  d^2 P_i d^2 P_j d^2 P_k & |S_{\mathbbm{1} P_i} S_{\mathbbm{1} P_j} S_{\mathbbm{1} P_k}|^2 \vcenter{\hbox{\includegraphics[scale=.25]{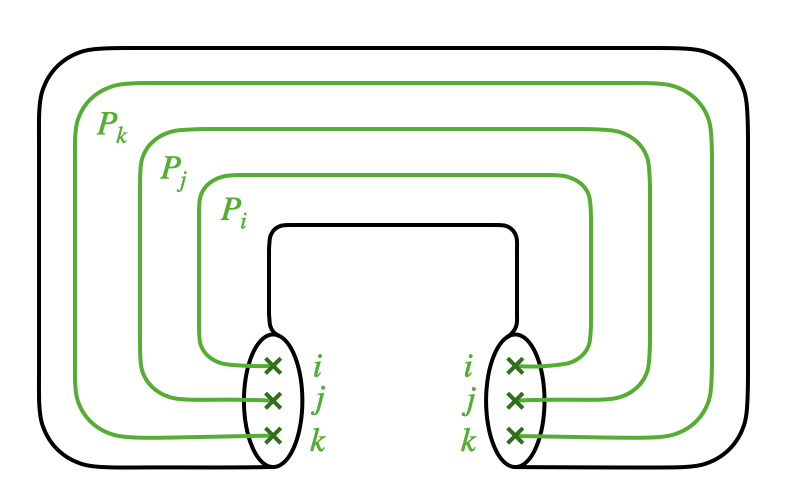}}} \nn &\times \mathcal{F}(P_i, P_j, P_k | q) \bar{\mathcal{F}} (\bar{P}_i, \bar{P}_j, \bar{P}_k | \bar{q})  
\end{align}
Here we have redrawn the $C_{0}$ manifold in a slightly different way to match with the next figure. 

With $\mathcal{M}$ as the reference manifold, we can introduce bulk links by  attaching  pillow and 6J manifolds to the thrice punctured sphere boundaries. An example with only 6J manifolds glued in is given below.
\begin{align}
    \vcenter{\hbox{\includegraphics[scale=.4]{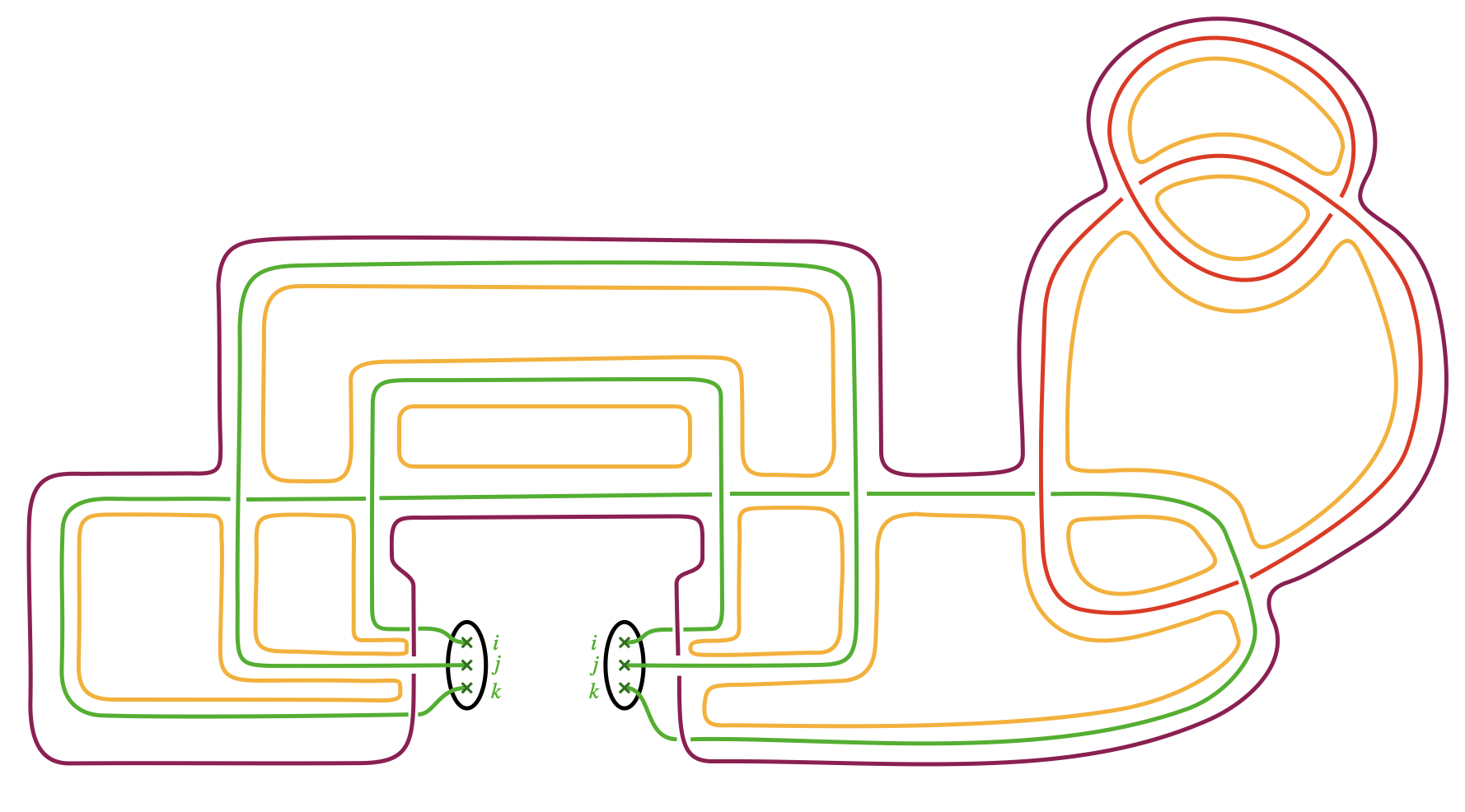}}}
\end{align}
In this diagram, surgery on the green lines that start and end on the boundary spheres produces manifolds with a genus 2 boundary as in 
\eqref{genus2bdry}. However,the bulk geometry has changed due to the insertion of the links colored in purple and red, which are produced by attaching 6J manifolds. The orange loops remove the $S^2$ handles in the bulk after surgery.   Finally, the extra purple loop can be set to identity using the same logic we applied for producing links in a closed manifolds. 

The construction of more general links for an arbitrary boundary $\Sigma$ follows the same series of steps as the above example.   We start by braiding the green lines -those that end on the boundary - so they sit in the middle of the triple line propagator.  Attaching 6J vertices then allows us to braid these green lines with an arbitrary link inside the reference manifold.     
 
\section{Schwinger-Dyson and a generalization of topological recursion }\label{SchwingerDyson}
We now address an important subtlety in the proposed relation between 3d gravity and the ensemble of approximate CFT's.  The perturbative expansion of the ensemble produces terms weighted by $\hbar$, but there is no such parameter in 3d gravity.  Indeed, the checks performed in section \ref{sec:checks} ignored these factors of $\hbar$, so the matching with 3d gravity partition functions was only done up to an overall $\hbar$ dependent coefficient. This issue was noted in \cite{belin2023}, where a solution was explained via the Schwinger-Dyson (SD) approach. In this work, we interpret the Schwinger-Dyson equation of \cite{belin2023} in a purely combinatorial/topological way. 

To see how we can obtain an exact agreement with 3d gravity, notice that due to surgery relations, the same 3-manifold appears multiple times in the expansion of the ensemble, weighted by different powers of $\hbar$.  Figure \ref{fig:ortho_mfld} illustrates this phenomenon: gluing together two 6J manifolds with propagators produces the Pillow manifold, but with a coupling that is  one higher order in $\hbar$ than the bare Pillow vertex. 

In general, by collecting together terms in the $\hbar$ expansion that produce the same manifold $M$, we can write a general tensor model  correlator in the following form:
\begin{align}\label{3dG} 
  \langle O\cdots O \rangle \left(q_{1},\cdots q_{i},\cdots \right)  =  \sum_{M} Z_{3\text{dG}}(M,c,q_{i}) f(M, \hbar),
\end{align}
where $Z_{3\text{dG}}(M,c,q_{i}) $ is the 3d gravity partition function on the manifold $M$ with boundary moduli $q_{i}$, and $f(M, \hbar)$ is a coefficient function given by a perturbative series in $\hbar$.  Note that the function $f(M, \hbar)$ is determined entirely by the combinatorial data that specifies how  $M$ can be produced by gluing together chunks of 3-manifolds.   On the other hand, all the theory dependent data is captured by $Z_{3\text{dG}}(M)$.

In the original definition of the ensemble partition function \eqref{Z}, the central charge is fixed.  However the sum over manifolds in \eqref{3dG} is naturally organized into a large $c$ expansion.  This is due to the volume conjecture, which implies the large $c$ scaling
\begin{align}
Z_{3\text{dG}}(M,c,q_{i}) \sim e^{-c \text{Vol}(M)}.
\end{align}
The volume factor in the exponent can be interpreted as the on-shell evaluation of the Einstein Hilbert action on the manifold $M$.  This gives the semiclassical  $e^{-c}$ expansion of 3d gravity.  Thus, we will interpret \eqref{3dG} as a re-organization of the perturbative expansion in which we first perform an $e^{-c}$ expansion of the ensemble at fixed $\hbar$.   To recover 3d gravity exactly, we must then take $\hbar \to 0$ term by term in $e^{-c}$, and show that $f(M,\hbar) \to 1$ for all $M$ in this limit\footnote{As explained in \cite{belin2023} achieving this limit generally requires a non-perturbative completion of the $\hbar$ expansion.  This can already be seen in a simple integral of the form $ \int dx e^{-\frac{1}{\hbar} ( x^2 - 1)^2}$.  The perturbative evaluation of the integral around $x=0$ would give a leading approximation $\braket{x^{2}}\sim -\hbar $ (i.e. the propagator), even though the non-perturbative answer gives $\braket{x^2}=1$.  This is because $x=0$ is not a minimum.}. 

Instead of showing this limit directly, we will assume that it is correct, and then insert the resulting correlators in \eqref{3dG} into the Schwinger-Dyson (SD) equations to check the consistency of our ansatz.   The SD equations are an infinite series of recursive relations for the \emph{exact} n-point functions of a theory.   
For the tensor model correlators, this is given by \cite{belin2023}
\begin{align}
     \int D \Delta_{s} \int D[C] \frac{\pd}{\pd C_{ijk}} (C_{lmn} e^{-V_{0}(\Delta_{s}) -\frac{1}{\hbar} V[\Delta_{s},C])} =0
\end{align}
For example, the exact 2 and 4 point functions must satisfy the relation shown below.
\begin{align}\label{fig:schwingerdyson}
    \includegraphics[scale=0.5]{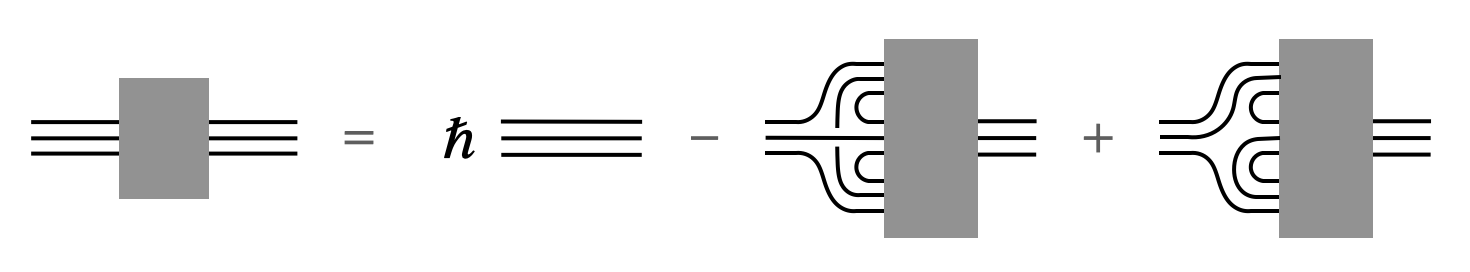}
\end{align}

The left hand side of this equation corresponds to the full 2 point function $\braket{C_{ijk}C_{jik}}$, which includes contributions from all manifolds with two boundaries given by thrice punctured spheres $\Sigma_{0,3}$: this indicated by the gray blob.   Similarly the gray blobs on the RHS represent the full 4 point function including contributions from  all manifolds with four $\Sigma_{0,3}$ boundaries. This equation seems to be a 3d analog of Mirzakhani's recursion relations \cite{Mirzakhani:2006fta, eynard2007weilpetersson} for moduli spaces of Riemann surfaces, which can be interpreted as the solution to the Schwinger-Dyson equations in 2d.

To apply the SD equation, we fix a 3-manifold -- say the $C_{0}$ manifold -- that appears on the LHS, and assume that $f(C_{0}, \hbar) \to 1 $ as $ \hbar \to 0$.  In order for this ansatz to be consistent, the RHS must produce the $C_{0}$ manifold exactly once in this limit.    
Since the bare propagator drops out when $\hbar \to 0$, the resulting equation becomes  
\begin{align}\label{SchwingerDysonC0}
\vcenter{\hbox{\includegraphics[scale=0.2]{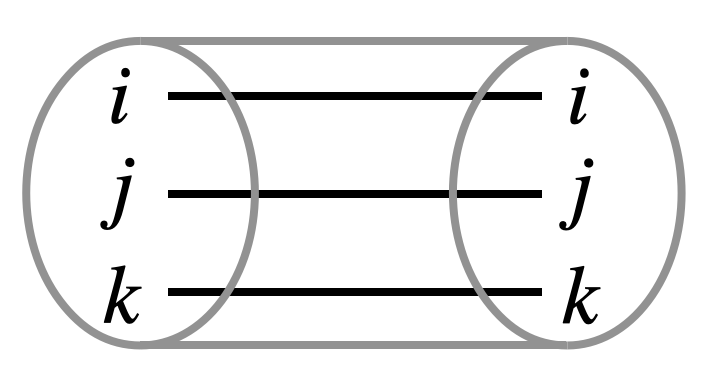}}} = \sum_{M \in \mathcal{M}_{\text{P}}} \vcenter{\hbox{\includegraphics[scale=0.2]{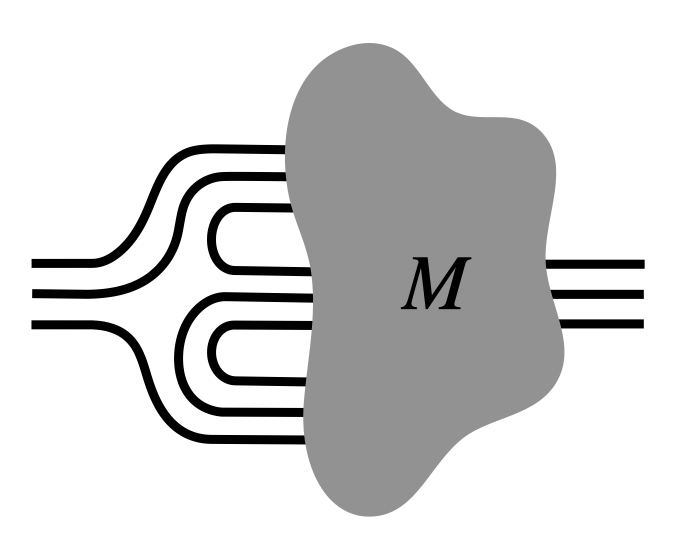}}} - \sum_{N \in \mathcal{M}_{6\text{J}}} \vcenter{\hbox{\includegraphics[scale=0.2]{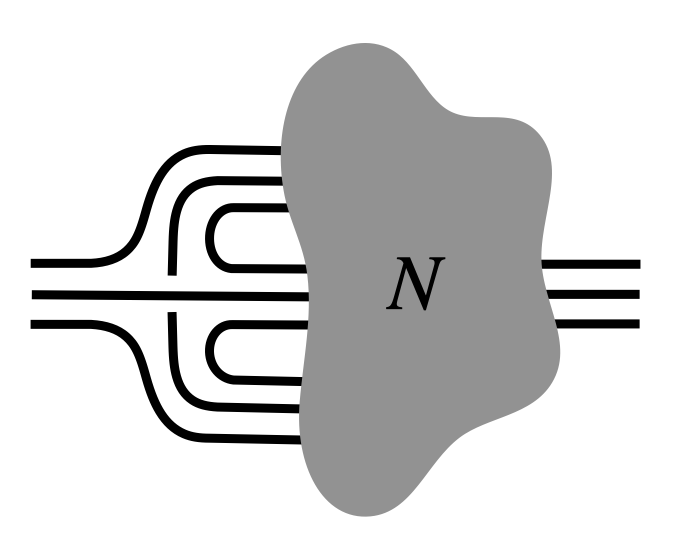}}}
\end{align}
Here $\mathcal{M}_{\text{P}}$ denotes the set of 4-boundary manifolds that produce the $C_{0}$ when glued in to the pillow, and $\mathcal{M}_{6\text{J}}$ are the 4-boundary manifolds that produce the $C_{0}$ manifold when glued into the 6J. 
Notice that strictly speaking, \eqref{SchwingerDysonC0} follows from \eqref{fig:schwingerdyson} in the $e^{-c}$ expansion because the semi classical expansion is a topological expansion. This is a consequence of Mostow rigidity, which states that for a fixed topology $M$ there is only one hyperbolic metric we can put on $M$.   This implies $\text{Vol}(M)$ is a topological invariant for hyperbolic manifolds, and equation \eqref{SchwingerDysonC0} follows from matching powers of $e^{-c}$ on both sides of the SD equation.

For instance, some examples of (connected) manifolds in $\mathcal{M}_{\text{P}}$ are shown below (drawn in green):
\begin{align}\label{exP}
    \vcenter{\hbox{\includegraphics[scale=.3]{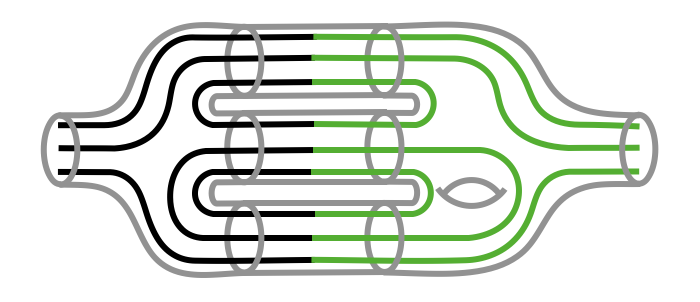}}} = \vcenter{\hbox{\includegraphics[scale=.3]{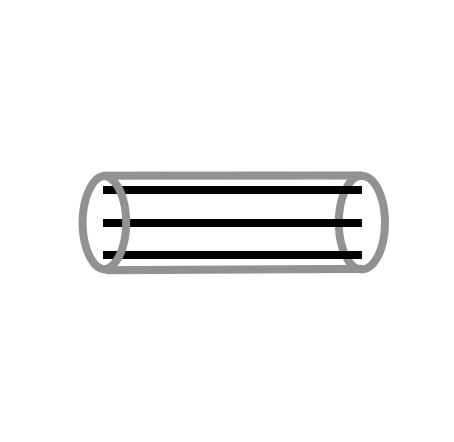}}}, \quad
    \vcenter{\hbox{\includegraphics[scale=.3]{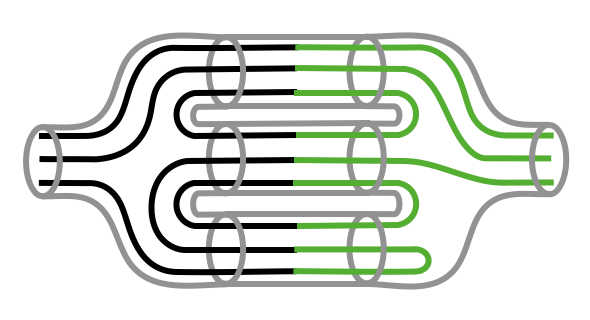}}} =\vcenter{\hbox{\includegraphics[scale=.3]{figures/fin_C_grey.png}}}
\end{align}
Similarly, examples of elements in $\mathcal{M}_{6\text{J}}$ are: 
\begin{align}\label{ex6J}
    \vcenter{\hbox{\includegraphics[scale=.3]{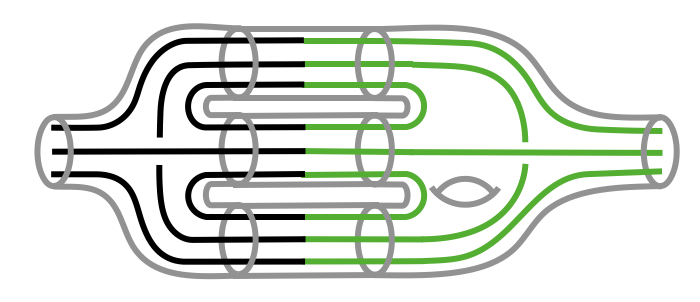}}} =\vcenter{\hbox{\includegraphics[scale=.3]{figures/fin_C_grey.png}}}, \quad
    \vcenter{\hbox{\includegraphics[scale=.3]{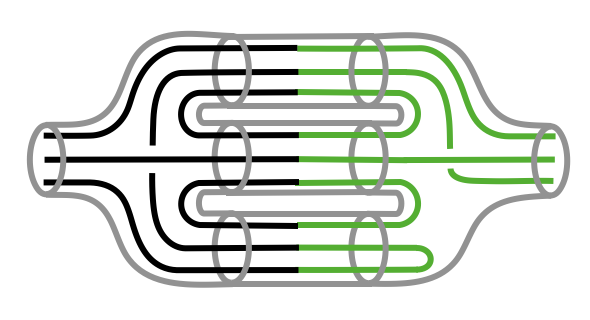}}} = \vcenter{\hbox{\includegraphics[scale=.3]{figures/fin_C_grey.png}}}
\end{align}
Some of these examples were studied in \cite{belin2023}. Note that some of the green manifolds have one $S^2 \times I$ handle -- this is in fact the maximum allowed number of handles for the manifold that we glue to the pillow or 6J. The surgeries will then close the two handles obtained from the gluing as well as the pre-existing handle in $\mathcal{M}$.
In order for the equation \eqref{SchwingerDysonC0} to be satisfied, there is must be a precise cancellation between manifolds glued into the 6J and Pillow manifolds.  In  particular, the number of elements in  $\mathcal{M}_{6\text{J}}$ and $\mathcal{M}_{\text{P}}$ must differ by 1:
\begin{align}\label{SDC0}
|\mathcal{M}_{\text{P}}|=|\mathcal{M}_{6\text{J}}|+1
\end{align}
Note that $|\mathcal{M}_{\text{P}}|$ and $|\mathcal{M}_{6\text{J}}|$ are \emph{finite}, so checking \eqref{SDC0} is a finite combinatorial question.   This is because a manifold in $\mathcal{M}_{6\text{J}}$ or $\mathcal{M}_{\text{P}}$ must connect the 9 in coming lines to 3 out going lines, and there are $\begin{pmatrix} 9 \\3 \end{pmatrix}$ ways to do this.  This is an upper bound, because nontrivial braiding patterns are not allowed as they would produce phases that correspond to 
a manifold different than $C_{0}$.   

To formulate our mathematical problem more precisely, it is useful to treat $\mathcal{M}_{\text{P}}$ and $\mathcal{M}_{6\text{J}}$ as vector spaces over the integers, with a basis 
\begin{align}
   \{\ket{M} , \, M \in \mathcal{M}_{\text{P}} \} ,\quad   \{\ket{N} , \, N \in \mathcal{M}_{6\text{J}} \}
\end{align}
 This means we can formally add and subtract these manifolds and the dimension of these vector spaces are just
\begin{align}
    \dim \mathcal{M}_{\text{P}} = |\mathcal{M}_{\text{P}}|, \qquad \dim \mathcal{M}_{6\text{J}} = |\mathcal{M}_{6\text{J}}|
\end{align}
Then we can view the Pillow and 6J manifolds as linear operators on these vector spaces that act by gluing on the left:
\begin{align}
    \text{P}= \vcenter{\hbox{\includegraphics[scale=.25]{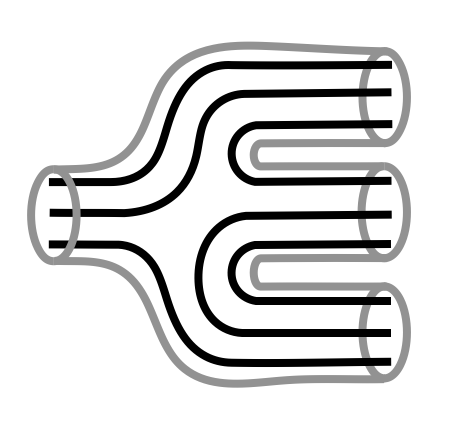}}} \qquad \text{6J}= \vcenter{\hbox{\includegraphics[scale=.25]{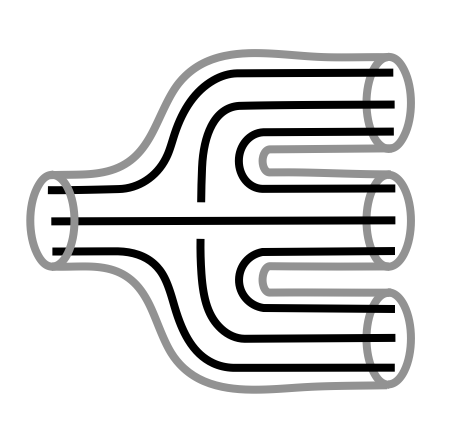}}}
\end{align}
Now we observe that $\text{P}$ and $6\text{J}$ are related by an operator we denote by $\text{E}$: 
\begin{align}
\text{E}=\vcenter{\hbox{\includegraphics[scale=.25]{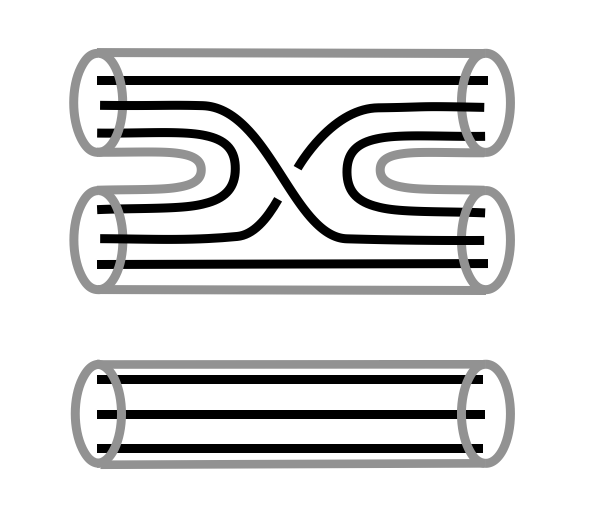}}},\qquad 6\text{J} = \text{P} \, \text{E} 
\end{align}
The means that
\begin{align}
    6\text{J} \left|M \right\rangle = \left| C_{0} \right\rangle ,  \to \text{P}(\text{E}\left|M \right\rangle) =6\text{J} \left|M \right\rangle =\left| C_{0} \right\rangle. 
\end{align}
so that gluing on the left by $\text{E}$  defines a linear map 
\begin{align}
    \text{E} : \mathcal{M}_{6\text{J}} \to \mathcal{M}_{\text{P}}
    \end{align}
  The equation \eqref{SDC0} then demands that 
  \begin{align}
      \dim \text{co-kernel} (E) - \dim \text{kernel}(E)=1
  \end{align}  We leave the proof of this and the complete SD equations as an exercise for the reader. 
 
\section{Discussion}
We have illustrated how the  topological expansion of 3d gravity arises from a random ensemble of approximate CFT's.  In this interpretation, the sum over 3-manifolds provides a mechanism to approximately implement the locality constraints of the modular bootstrap. In the $\hbar \to 0$ limit, the constraints are implemented order by order in an asymptotic $e^{-c}$ expansion, provided we can show that the SD equations are satisfied.   This is an interesting mathematical problem involving the combinatorics of 3-manifolds gluings.   The complete solution would possibly require a generalization of topological recursion in 2d. 
\paragraph{Non-perturbative completions}
An exact implementation of the bootstrap would require a non-perturbative resummation of the asymptotic series.  By analogy with the SSS model, we expect that this comes from doubly non-perturbative corrections of order $e^{-e^{c}}$.  However, whereas the non-perturbative completion of SSS is a choice of contour for the matrix ensemble, we expect the completion of our ensemble to produce an exact $\text{CFT}_2$. A general non-perturbative completion of the ensemble will be a linear combination of results from different CFT's, i.e. an average over those theories. This is similar to a general contour in the non-perturbative definition of an integral being expressed as a linear combination of steepest descent contours.  While doing the full non-perturbative sum over all 3-manifolds is a daunting challenge, the non-perturbative completion of the matrix model may be a more tractable problem.   Notice that the $\hbar \to 0$ limit of the matrix model may already be sufficient to rigidify the model, since CFT spectra that satisfy 
exact $SL(2,\mathbb{Z})$ invariance are  expected to be sparse \cite{Blommaert_2022}.
\paragraph{Loose ends}
Aside from solving the SD equations, there are some important loose ends that need to be addressed in the future.  First, there are divergences in the tensor model arising from the fact that Virasoro TQFT gives divergent answers for non-hyperbolic, ``off-shell" manifolds.  In particular, manifolds containing $S^2 \times S^1$ factors would give infinities that must be cancelled or renormalized in some way.  

Another complication comes from the fact that the same manifold can be produced by the matrix model and the tensor model diagrams.  A simple case involves $\Sigma_{0,3}\times S^1$, which arises from a tensor model loop for $C_{ijk}$: this gives the log of the tensor model propagator, summed over $i,j,k$.  $\Sigma_{0,3}\times S^1$ is topologically equivalent to  the 3-boundary matrix model manifold depicted in \eqref{3toruswh}, since $\Sigma_{0,3}$ can be viewed as a pair of pants manifold.  In this special case, we can set the tensor model contribution to zero simply by renormalizing $C_{ijk}$ because of the  logarithm in the tensor model loop.  However, in general, we need a mechanism to delete the tensor model contributions to these off-shell manifolds. The most direct solution would be to show that the function $f(M,\hbar)\to 0$ as $\hbar \to 0 $ for these manifolds.  
Finally, it is well known that the sum over $SL(2,\mathbb{Z})$ images of a manifold produces accumulations points corresponding to the hyperbolic cusp.  This leads to a divergence that must be regularized.  However, this problem has been encountered previously in \cite{Maloney_2010, Keller_2015, dijkgraaf2007black}, where zeta function regularization was applied to remove the divergence. Perhaps the matrix integral will provide a new perspective on this regularization.

\paragraph{Comparison with Simplicial gravity}
 Here we would like to compare and contrast our model to 3d simplicial gravity and the Turaev-Viro (TV )-like state sum models \cite{Turaev:1992hq,Turaev+2016}.   The Pillow and the 6J manifolds, which make up the basic building blocks of the tensor model, have natural interpretations in terms of 3-simplices:  
\begin{align}\label{simplices}
    \text{6J Simplex} = \vcenter{\hbox{\includegraphics[scale=.25]{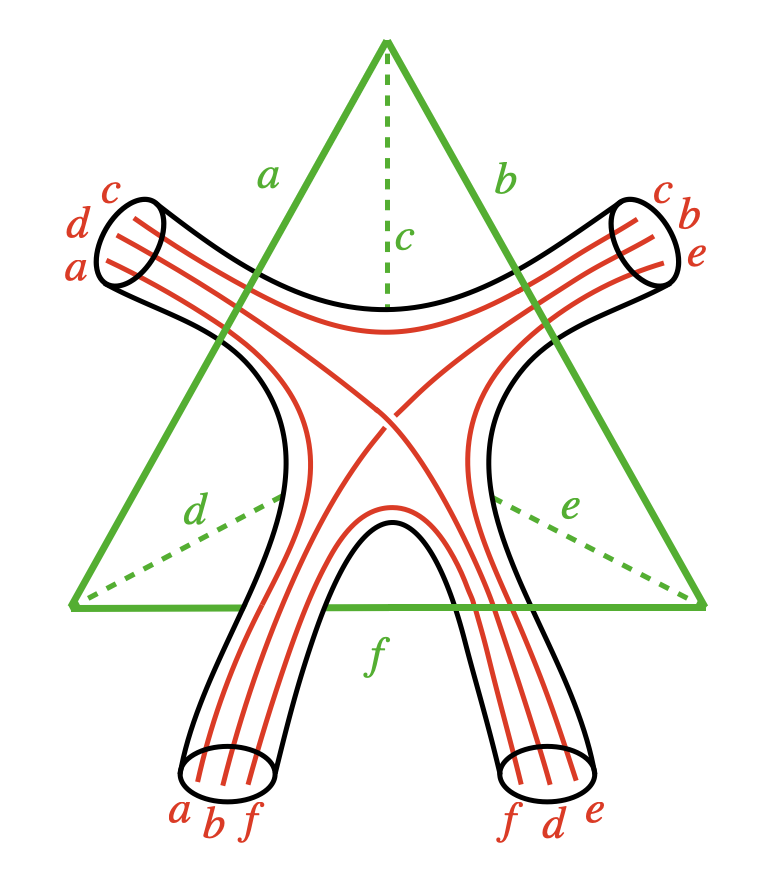}}} ,\qquad \qquad  \text{Pillow Simplex}= \vcenter{\hbox{\includegraphics[scale=.25]{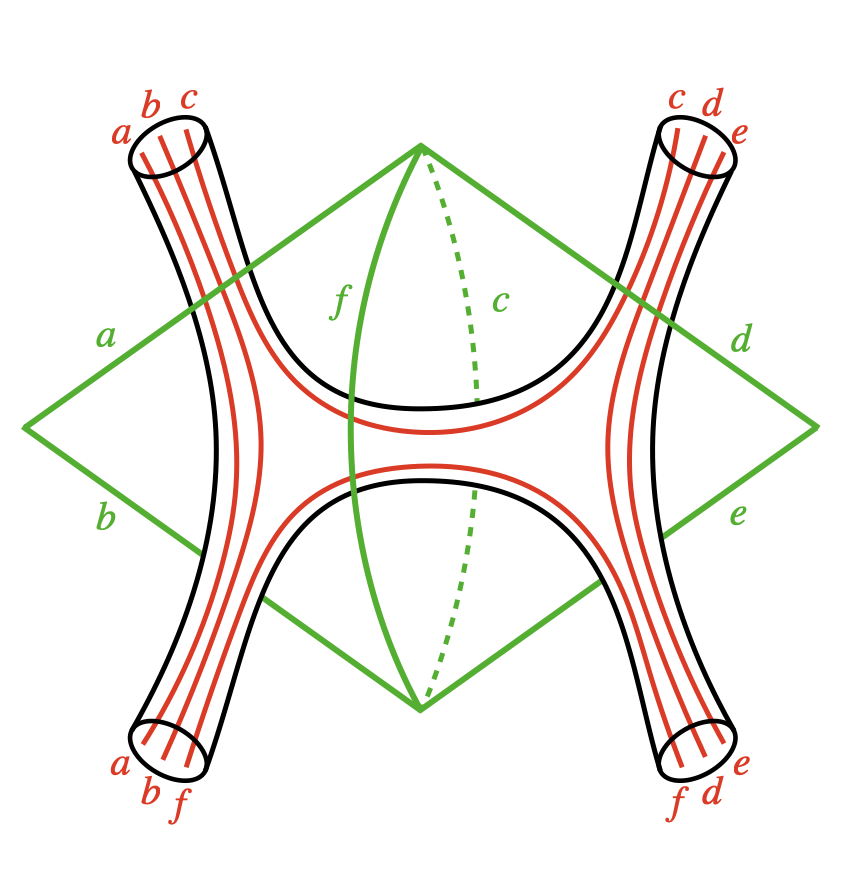}}}
\end{align}

In this mapping between 4-boundary wormholes and simplices, each thrice punctured sphere boundary corresponds to a face of the simplex, while the Wilson lines connecting the spheres specify the number of of edges shared between the faces\footnote{In the literature, the multiboundary wormholes dual to the simplices are viewed as the thickened graphs of the dual triangulation.  In that case the cross section of the wormhole geometries are disks, and the Wilson lines live on the boundary of the disk \cite{costantino2005triangulations3manifoldshyperbolicrelative, ROBERTS1995771}.}.   Given a simplicial decomposition of a manifold M in terms of the 6J and pillow simplices, the mapping produces a 3d gravity partition function from a triangulation via a rule that is similar to the TV state sum, provided that we identify the quantum dimension as $\text{dim}_{q}P \equiv S_{\mathbbm{1}P}$ and ignore factors of $S_{\mathbbm{1}P}$ in the TV model.    In this context, the surgery relations corresponds to equivalence under changing the triangulations\footnote{For example the pentagon identity corresponds to a 2-3 Pachner move.}.   Note that the diagrammatics defined by \eqref{simplices} also appears prominently in the group field theory literature, where the continuum limit of similar tensor models have been studied \cite{Gurau_2011,Ben_Geloun_2010,Gurau:2009tw,Gurau_2012,carrozza2024tensormodelsgroupfield}.  

However, not all 3d gravity partition functions defined by the CFT ensemble map to a triangulation of a smooth manifold.  This is because the ensemble produces all possible gluings of the simplices, including the cases when the ``link" of\footnote{ Here by link of a vertex, we mean set of faces the that topologically links with the vertex.     This is distinct from a set of entangled knots, which we also refer to as a link. } of a vertex is a not a sphere: this implies that the simplices do not fit together around a vertex to give a smooth manifold. 

In the context of ordinary, nonchiral TQFT's, there is an elegant way to relate the Turaev-Viro state sum  for a manifold $M$ to the computation of a link invariant associated to the Chain mail link $L(M)$ \cite{ROBERTS1995771}.   Given a triangulation $T$ of $M$, we can essentially define $L(M)$ by gluing together the Wilson lines of the 6J and Pillow geometries in \eqref{simplices}. 
 More precisely, we apply the mapping\footnote{In general the Chain mail link is defined from the any handlebody decomposition of $M$.  Here we give the construction for the special case where the handlebody decomposition comes from a thickening of the triangulation $T$.}: 
\begin{align}
  \text{6J Simplex} = \vcenter{\hbox{\includegraphics[scale=.25]{figures/fin_6jsimplexWilson.png}}}  \longrightarrow \vcenter{\hbox{\includegraphics[scale=.25]{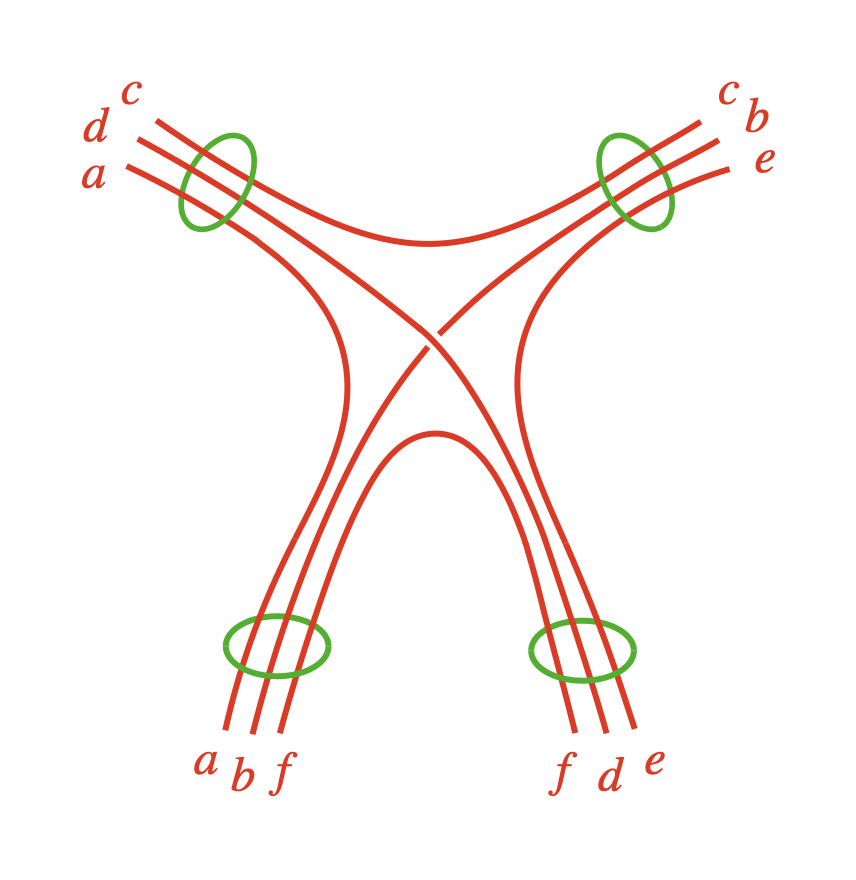}}} 
\end{align}
to produce the link $L(M)$ sitting inside $S^3$.
Here we have introduced  green circles linking with each triplet of lines connecting the 6J's and pillows.    The link invariant is then defined by putting the Omega loop on each circle of the link $L(M)$, and then evaluating the expectation value of the loops in the ``chiral half" of the TQFT.   More precisely, computing expectation value gives the Resthitkhin-Turaev-Witten link invariant for $L(M)$, and performing the surgeries associated to the Omega loops leads to the Turaev-Viro state sum via the formula:
\begin{align}\label{TVRTW}
Z_{RTW}(L(M),S^3) &=  Z_{RTW} ( M \#_{\text{all vertices}}  \bar{M} )\nn
&=Z_{TV}(M)
\end{align}
The notation $M \#_{\text{all vertices}}  \bar{M} $ refers to the connect sum of $M$ with its mirror image $\bar{M}$ over all the links of the vertices in the triangulation: this is the result of performing surgery on the link $L(M)$.  For a smooth triangulation the links of all vertices are spheres, and the partition function has a simple behavior under the connect sum: 
\begin{align}
    Z_{RTW}(M\#_{S^2} \bar{M}) = \frac{Z_{RTW}(M)Z_{RTW}(\bar{M})}{Z(S^3)}
\end{align}
from which the second equality in \eqref{TVRTW} follows. 

Due to the ill-defined nature of the Hilbert space on $S^2$, the logic of equation \eqref{TVRTW} fails in our 3d gravity model whenever the links of vertices are spheres.  However, equation  \eqref{TVRTW} could perhaps be applied to the singular triangulations in which the links of vertices are higher genus Riemann surfaces, where the 3d gravity Hilbert space is well defined.   We leave the investigation of this intriguing possibility to future work.  

\paragraph{Future directions}
There are several natural extension of our model that we plan to pursue in the future. One straightfoward extension involves conformal field theories with additional symmetries.  For example, a CFT ensemble with reflection symmetry should  lead to a sum over non-orientable 3-manifolds.  We could also consider superconformal theories, and theories with additional Kac-Moody algebras. The formalism is identical, except that the 6J symbols and torus characters and crossing kernels become those of super-Virasoro or Virasoro-Kac-Moody respectively. The operators now label primaries of the larger symmetry algebra, and the leading disk spectral density should be taken to be the $S$ transform of the identity character. In cases where there are degenerate representations, for example BPS multiplets, those must be including separately, since the formalism assumes generic characters. It would be interesting to import exactly known collections of BPS operators of an SCFT and treat the remainder of the spectrum according to the logic of the matrix/tensor integral.

Another natural extension is to introduce boundaries in the ensemble of CFT's.  Such a random BCFT ensemble would include additional data given by a set of Cardy boundary conditions, boundary OPE's and bulk to boundary OPE's which satisfy an enlarged set of bootstrap equations \cite{LEWELLEN1992654}. Such an ensemble would involve averages over the choice of local boundary conditions.   We expect this to be dual to $\text{AdS}_3$ gravity with end of the world branes\footnote{It would also be interesting to incorporate \emph{entanglement} boundaries, which were studied in \cite{Mertens:2022ujr,Wong:2022eiu}.  More recently, BCFT data associated to  these ``shrinkable" boundaries in Liouville theory was used to construct 
a novel type of simplicial 3d gravity in \cite{Chen:2024unp}.}.  
It would also be extremely interesting to consider modifying the ensemble to produce dS$_3$ gravity in the bulk. This will be associated to the bootstrap problem for the Euclidean CFT's dual to de Sitter, possessing an infinite dimensional extension of the $SL(2,\mathbb{C})$  symmetry algebra. 

 Matter can be added in the bulk by modifying the leading Cardy spectrum to include specific states below the black hole threshold.  If pure gravity belongs to the swampland, this modification would be essential to obtain a nontrivial non-perturbative completion of the ensemble.

Finally, there is a fascinating possibility that the model discussed here provides special case of matrix-tensor models for membranes. The SSS matrix model can be interpreted as a special case of the duality between matrix integrals and string theory (the minimal string or the topology string), with a particular target space for which the worldsheet theory becomes pure JT gravity\footnote{The target space is a limit of spacelike Liouville times timelike Liouville \cite{collier2023virasoro} (see also \cite{Altland_2023}).}. Similarly, our tensor model might correspond to M2 branes on a peculiar target space such that their worldvolume theory becomes pure AdS$_3$ gravity. 

A hint may come from a surprising connection between M5 branes and VTQFT. The supersymmetric partition function of a pair of M5 branes on $S^3 \times M_3$, topologically twisted over the 3-manifold and with deformed supersymmetry on the sphere, was shown to be equal, by a dimensional reduction, to $SL(2)$ Chern-Simons theory on $M_3$ \cite{Terashima:2011qi, Dimofte:2011ju, Yagi:2013fda, Lee:2013ida,  Cordova:2013cea, Dimofte:2014zga}. The precise 3d TQFT determined by the resulting unusual integration cycle was argued to be exactly VTQFT \cite{Mikhaylov_2018}. 

It would be very interesting if a more general class of matrix-tensor models resulted in membrane worldvolume theories with more conventional target spaces. 
\section*{Acknowledgements}
It is a great pleasure to thank Alex Belin, Jan de Boer, Scott Collier,  Jordan Cotler, Lorenz Eberhardt, David Gross, David Kolchmeyer, Juan Maldacena, Alex Maloney, Mark Mezei, Greg Moore, Pranjal Nayak, Julian Sonner, Douglas Stanford, Jorge Teschner, Herman Verlinde, Diandian Wang, and Edward Witten for useful discussions.  GW is supported by STFC grant
ST/X000761/1, the Oxford Mathematical Institute, and Harvard CMSA.  The work of DLJ and LR is supported in part by DOE grant DE-SC0007870 and by the Simons Investigator in Physics Award MP-SIP-0001737. GW would like to thank CERN and the Higgs Center for their hospitality during the completion of this work.  DLJ would like to thank the IAS for hospitality during the completion of this work.
\appendix

\section{Symmetries of $C_{ijk}$} \label{app:sym}
We define the OPE coefficients by the 3 point function
\begin{align}\label{3pt1App}
\braket{O_{i}(z_{1}) O_{j}(z_{2}) O_{k}(z_{3})} &=\frac{C_{ijk}}{z_{12}^{h_{i}+h_{j}-h_{k}} z_{23}^{h_{j}+h_{k}-h_{i}} z_{31}^{h_{i}+h_{k}-h_{j}} \times (\text{anti-holomorphic}) } \nn
z_{ij} &=z_{i} -z_{j} 
\end{align} 
We assume integer spin. 

%In Ginsparg, he writes $z_{13}$ in the denominator. I changed this to simplify the derivation below

To find the exchange symmetries of $C_{ijk}$, consider exchanging $i$ and $j$
\begin{align}\label{3pt2}
\braket{O_{j}(z_{1}) O_{i}(z_{2}) O_{k}(z_{3})} &\equiv \frac{C_{jik}}{z_{12}^{h_{i}+h_{j}-h_{k}} z_{23}^{h_{i}+h_{k}-h_{j}} z_{31}^{h_{j}+h_{k}-h_{i}} \times (\text{anti-holomorphic}) } 
\end{align}
Since the ordering of operators inside $\braket{}$ is immaterial this is the same as interchanging $z_{1}$ and $z_{2}$ in equation \eqref{3pt1} via a continuous process.  At the endpoint of this process we have:
\begin{align}\label{3pt3}
\braket{O_{i}(z_{2}) O_{j}(z_{1}) O_{k}(z_{3})}
&= \frac{C_{ijk}}{z_{21}^{h_{i}+h_{j}-h_{k}} z_{13}^{h_{j}+h_{k}-h_{i}} z_{32}^{h_{i}+h_{k}-h_{j}} \times (\text{anti-holomorphic}) }
\end{align}
To compare this expression to \eqref{3pt2} we see that we have to relate $z_{ij}^{a}$ to $z_{ji}^{a}$ for non integer a:  due to the branch cut we need to think of the exchange of $i$ and $j$ as a braiding process.  Defining $z_{ij}^{a}=\exp a \log z_{ij}$ with the branch cut on $z_{ij}\in \mathbb{R}^+ $ implies that 

\begin{align}
z_{ji}^a = e^{i \pi a} z^{a}_{ij} 
\end{align}
Accounting for this phase, we find 
\begin{align}\label{3pt4}
\braket{O_{i}(z_{2}) O_{j}(z_{1}) O_{k}(z_{3})}= & \frac{C_{ijk}}{e^{i \pi (h_{i}+h_{j}+h_{k})} e^{-i\pi (\bar{h}_{i}+\bar{h}_{j}+\bar{h}_{k})}} \\
&\times \frac{1}{z_{12}^{h_{i}+h_{j}-h_{k}} z_{23}^{h_{j}+h_{k}-h_{i}} z_{31}^{h_{i}+h_{k}-h_{j}} \times (\text{anti-holomorphic}) }
\end{align} 
where we have used the fact that  $z_{ij}$ and $\bar{z}_{ij}$ are conjugate variables and should have opposite phases under braiding.
Note that to get a phase  that is symmetric in all the conformal dimensions, we need to view the minus signs from  $z_{23}$ and $z_{31}$ factors as also coming from a continuous process moving $2\to 3$ and $3\to 1$. This means we do three half braids, which leads to the symmetric phase.

Then equating \eqref{3pt2} and \eqref{3pt4} gives
\begin{align}\label{sym}
C_{jik} = C_{ijk} \exp( i\pi (s_{i}+ s_{j}+s_{k}  ))
\end{align}
This implies that for integer spins or half integer, the $C_{ijk}$'s are cyclically symmetric.

\section{Reality condition on $C_{ijk}$} \label{app:real}
Here we derive the reality condition on $C_{ijk}$ which gives the phases that relate $C_{ijk}$ and $C_{ijk}^* $
First, let's recall the Hilbert space definition of $C_{ijk}$.  In radial quantization, we define states living on the unit circle    
\begin{align}
    \ket{O_{i}} = O_{i}(0) \ket{0}  \in \mathcal{H}_{S^1} 
\end{align}
obtained by the path integral on the unit disk with $O_{i}(0)$  inserted. Then the OPE coefficient with one upper index is defined by
\begin{align}
    \hat{O}_{i}  \ket{O_{j} }= \sum_{k} C_{ji}^{k} \ket{O_{k}}
\end{align}
where $\hat{O}_{i}$ acts on $\mathcal{H}_{S_{1}}$. This action can be defined by inserting $O_{i}$ at some point $z$ anywhere inside the unit disk.  By convention we pick $z=1$.  Then we can extract the OPE coefficient from the overlap with $\bra{O_{k}}$:
\begin{align}\label{cijk}
C_{jik}\equiv \braket{O_{k}|\hat{O}_{i} |O_{j} }= \sum_{l} C_{ji}^{l} \braket{O_{k}|O_{l}}   
\end{align} 
which glues the unit disk to another disk to make a 3 punctured sphere.   Here we treat the two point function $\braket{O_{k}|O_{l}}$ as a metric that we use to lower the indices.  This metric depends on a choice of operator normalization and  so does the OPE coefficient.  

Note that in terms of operator insertions,  $\bra{O_{k}}$ is an insertion of $O_{k}$ ``at infinity", which is defined by
\begin{align}\label{bra}
    \bra{O_{k}} =  \lim_{z_{k}\to \infty} (z_{k})^{2h_{k}} \bra{0}O_{k} (z_{k})  .
\end{align}
So to summarize we have
\begin{align}\label{phase}
C_{jik}= \braket{O_{k}|\hat{O}_{i} |O_{j} } \equiv \lim_{z_{k}\to \infty} (z_{k})^{2h_{k}} \braket{ O_{i}(1)  O_{j}(0) O_{k} (z_{k}) }
\end{align} 
Now consider
\begin{align}\label{C*}
C_{jik}^* = \braket{O_{k}|\hat{O}_{i} |O_{j} }^* =  \braket{O_{j}|\hat{O}^{\dagger}_{i} |O_{k} }=\braket{O_{j}|\hat{O}_{i} |O_{k} }
\end{align}
where we assumed the hermiticity of $O_{i}$.
We want to relate the last expression to $C_{ijk}$.
One way to do this is just to apply an inversion $z \to w= 1/z$ that interchanges the insertion points $z_{k} $ and $z_{j}$.  Suppressing   antiholomorphic dependence, we get:
\begin{align}
C_{jik}^* &=  \lim_{z_{j}\to \infty} (z_{j})^{2h_{j}} \braket{ O_{i} (z_{i}) O_{j}(z_{j} ) O_{k} (z_{k}) }|_{z_{i}=1,z_{k}=0}\nn
&= \lim_{z_{k} \to 0} \lim_{z_{j}\to \infty} (z_{j})^{2h_{j}} \braket{ O_{i} (\frac{1}{z_{i}}) O_{j}(\frac{1}{z_{j}}) O_{k} (\frac{1}{z_{k}}) } (-z^{2}_{i})^{-h_{i}} (-z^{2}_{j})^{-h_{j}}(-z^{2}_{k})^{-h_{k}} \nn
 &= \lim_{w_{k} \to \infty} (w_{k})^{2h_{k}}  \braket{ O_{i} (w_{i}) O_{j}(w_{j}) O_{k} (w_{k}) } (-1)^{h_{i}+h_{j}+h_{k} } |_{w_{j}=0,w_{i}=1} 
\end{align} 
In second line we applied a conformal map, and in the third line we cancelled the factors of $z_{j}^{-2h_{j}}$ before sending $z_{j} \to 0$.
Adding back in the antiholomorphic part, we get:
\begin{align}\label{C*ijk}
C_{jik}^*&=C_{jik} (-1)^{s_{i}+s_{j}+s_{k}} \nn
&=C_{jik}e^{i\pi( s_{i}+s_{j}+s_{k})} \end{align} 
where in the second line we made a choice of branch cut. 
For integer or half integer spins, this is consistent with $C^*_{jik}= C_{kij}$ from \eqref{C*} and \eqref{sym}.

\section{$SL(2,\mathbb{Z})$ symmetrizer from the matrix model expansion}
\label{app:SL2}
Here we show how the matrix model expansion produces a $SL(2,\mathbb{Z})$ symmetrizer in the $\hbar \to 0 $ limit.   To begin with, let us redefine our matrix model propagator by including the $\frac{1}{\hbar} K$ factor in the constraint square potential $V_{S}$ as part of the quadratic term of the effective action.  Thus our new propagator and vertex are:
\begin{align}
\vcenter{\hbox{\includegraphics[scale=0.35]{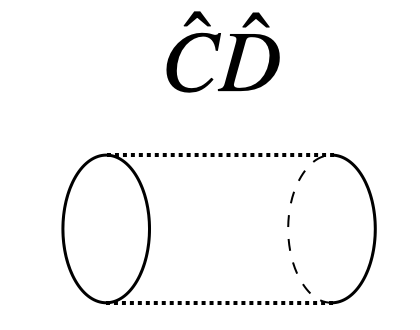}}} = (\frac{\hbar}{2+\hbar})^{-1} \hat{C} \hat{D},\qquad  \vcenter{\hbox{\includegraphics[scale=0.35]{figures/fin_VS2_obj.png}}}= \frac{2}{\hbar} \hat{K}\hat{S}
\end{align}
This definition produces a convergent geometric series, since the factors of $\frac{1}{\hbar}$ in the vertex is now cancelled by $\hbar$ factor in the propagator.  The resummed matrix model propagator  \eqref{string} is now given by 
\begin{align} \label{fullprop1}
    \sum_{n=0}^{\infty} \frac{1}{n!} \vcenter{\hbox{\includegraphics[scale=0.35]{figures/fin_pot_series2.png}}} &= \frac{\hbar}{2+\hbar} \hat{C} \hat{D} \frac{1}{1-\frac{2\hat{S}\hat{D}}{2+\hbar}} \\
    &= \hbar \hat{C} \hat{D} \frac{1}{ 2+ \hbar -2 \hat{S}\hat{D}}
    \end{align}
Ignoring $T$ transforms, i.e. setting $\hat{D}= \mathbbm{1}$
gives the projector $\frac{\mathbbm{1}+\hat{S}}{2}$ as before.  To incorporate $T$ transforms, we first regulate the operator $\hat{D}$ by writing 
\begin{align} \label{D}
    \hat{D} \to \hat{D}(l) \equiv \frac{1}{2l+1} \sum_{n=-2l}^{2l} \hat{T}^n
\end{align}
We would like to show that the full propagator takes the form: 
\begin{align}\label{fullprop2}
        \sum_{n=0}^{\infty} \frac{1}{n!} \vcenter{\hbox{\includegraphics[scale=0.32]{figures/fin_pot_series2.png}}}   = \hat{C} \sum_{\gamma \in SL(2,\mathbb{Z})}f_{\gamma}(\hbar) \hat{\gamma}
\end{align}
with the coefficients $f_{\gamma}(\hbar)$ independent of $\gamma$ as $\hbar \to 0$, thus producing the $SL(2,\mathbb{Z})$ symmetrizer.

We check this by equating  \eqref{fullprop1} with \eqref{fullprop2}:
\begin{align}\label{solvef}
    \frac{\hbar}{2+\hbar} \hat{C} \hat{D}(l) &= (1-\frac{2\hat{S}\hat{D}(l)}{2+\hbar} ) \sum_{\gamma \in SL(2,\mathbb{Z})}f_{\gamma}(\hbar) \hat{\gamma} \nn
    &=\sum_{\gamma \in SL(2,\mathbb{Z})} \hat{\gamma} \left(f_{\gamma}(\hbar)-  \frac{2}{(2+\hbar)(2l+1)} \sum_{n=-l}^{l} f_{T^{-n} S^{-1} \gamma}(\hbar) \right),
\end{align}
and then solving for  $f_{\gamma}(\hbar)$ order by order in $\hbar$. 
In the second equality above, we have relabelled the dummy variable $\gamma$ in $f_{\gamma}(\hbar)$

Expanding  $f_{\gamma}(\hbar)$ 
\begin{align}
   f_{\gamma}(\hbar)=  f_{\gamma}(0) + \hbar f'_{\gamma}(0) + \cdots 
\end{align}
one finds that to $O(\hbar^0)$ equation \eqref{solvef} implies 
\begin{align}
f_{\gamma}(0)-  \frac{1}{(2l+1)} \sum_{n=-l}^{l} f_{T^{-n} S^{-1} \gamma}(0) =0.
\end{align} 
This is solved by setting $f_{\gamma}(0)$ to be a constant independent of $\gamma$, thus giving an equal weighted sum over $SL(2,\mathbb{Z})$
\paragraph{The overall coefficient}
We have seen that the matrix model produces an equal weighted sum over $SL(2,\mathbb{Z})$, but so far we have not specified the exact value of the weight.  In 3d gravity, the different $SL(2,\mathbb{Z})$ images are summed with coefficient 1.   However the full internal propagator of the matrix model, defined by the string in  \eqref{fullprop1}, produces a projector onto the $SL(2,\mathbb{Z})$ invariant state, which should weigh each $SL(2,\mathbb{Z})$ by a normalization factor.   We illustrated this explicitly when we truncated the sum to only $S$ transforms, but we expect this to be true for the general sum.   This normalization factor is related to the normalization in the regulated sum in \eqref{D}, which is necessary to produce a projection operator on the integer spin part of the torus Hilbert space.    

This apparent tension with 3d gravity can be resolved as follows.   While the internal propagator produces an average over $SL(2,\mathbb{Z})$, the external observables of the ensemble involves a sum over $SL(2,\mathbb{Z})$ with weight 1.  For example, the  torus partition function of a CFT takes the form 
\begin{align}\label{Z'}
    Z_{CFT}(\beta,\mu)  = \sum_{s\in \mathbb{Z}} \int d \Delta  \rho(\Delta, s) e^{- \beta \Delta+ i \mu s}
\end{align}
After inserting this observable into the ensemble partition function defined in \eqref{Z0}, we replace the sum over integer spins as an integral with a Dirac comb\footnote{This is a consequence of the Poisson summation formula.}:
\begin{align}
      \sum_{s \in \mathbb{Z}} \to \int ds \sum_{n=-\infty}^{\infty} e^{2\pi i n s}.
\end{align}
This produces a sum over $T$ transformations, but with no normalization factor.   When we introduce an internal propagator connecting two such observables, the normalizations of the external observables combine with that of the  projector to give a sum over $SL(2,\mathbb{Z})$ transforms with weight 1.

\bibliography{main_biblio}

\end{document}